\documentstyle[aps,epsfig]{revtex}
\input epsf
\input rotate

\advance\topmargin by 0.5cm
\advance\oddsidemargin by 1cm
\advance\evensidemargin by 1cm

\begin{document}
\newcommand{\lamtil}{{\tilde\lambda}}
\newcommand{\ktil}{{\tilde k}}
\newcommand{\dbydrho}{{\partial_\rho}}
\newcommand{\dbydrhot}{{\partial_\rho^2}}
\newcommand{\dbyds}{{\partial_s}}
\newcommand{\dbydst}{{\partial_s^2}}
\newcommand{\nvec}{{\bf n}}
\newcommand{\tvec}{{\bf t}}

%\vskip 0.2in

\begin{center}
\Large\sc
%\bf 
Phase field under stress
\end{center}

\begin{center}
\large
Klaus Kassner$^{1}$, Chaouqi  Misbah$^2$, Judith M\"uller$^3$, Jens Kappey$^{1}$, and 
Peter Kohlert$^{1}$
\vskip 0.2 cm
$^1$Institut f\"ur Theoretische Physik
Otto-von-Guericke-Universit\"at Magdeburg
Postfach 4120,
D-39016 Magdeburg, Germany\\
$^2$Groupe de Recherche sur les Ph\'enom\`enes hors  de
l'Equilibre, LSP, Universit\'e Joseph Fourier (CNRS),
Grenoble I, B.P. 87, Saint-Martin d'H\`eres, 38402 Cedex, France\\
$^3$ Instituut Lorentz,
 Leiden University,
  P.O. Box 9506,
   2300 RA Leiden, the Netherlands\\
\vskip 0.2 cm
\vskip 0.2 cm
%8 March 2000
%27 April 2000
23 May 2000
\end{center}
%\author{(\today)}

\vskip 1cm
%\author{\parbox{397pt}{\vglue 0.3cm\small 

\noindent 
A  phase-field approach describing the dynamics of a strained
solid in contact with its melt is developed. Using a formulation
that is independent of the state of reference chosen for the
displacement field, we write down the elastic energy in an 
unambiguous fashion, thus obtaining an entire class of models.
According to the choice of reference
state, the particular model emerging from this class 
will become equivalent to one of the two
independently constructed models, on which brief accounts have
been given recently [J. M\"uller and M. Grant, Phys. Rev. Lett. {\bf 82},  
1736  (1999); K. Kassner and C. Misbah, Europhys. Lett. {\bf 46},  
217  (1999)]. We show that our phase-field approach 
recovers the sharp-interface
limit corresponding to the continuum model equations describing
the Asaro-Tiller-Grinfeld instability. Moreover,
we use our model to {\em derive} hitherto unknown sharp-interface
equations for a situation including a field of body forces. 
The numerical utility of the phase-field approach is demonstrated by 
reproducing some known results and by comparison with a sharp-interface
simulation. We then proceed to investigate the dynamics of extended 
systems within the phase-field model which contains an inherent
lower length cutoff, thus avoiding cusp singularities.
It is found that a periodic array of grooves generically
evolves into a superstructure which arises from a series of imperfect
period doublings. For wavenumbers close to the fastest-growing
mode of the linear instability, the first period doubling can be
%captured 
obtained analytically. Both the dynamics of an initially periodic
array and a random initial structure can be described as a coarsening
process with winning grooves temporarily accelerating whereas losing
ones decelerate and even reverse their direction of motion. 
In the absence of gravity, the end state of a laterally finite
system is a single groove growing at constant velocity, as long as
no secondary instabilities arise
(that we have not been able to see with our code).
With gravity,
several grooves are possible, all of which are bound to stop eventually.
A laterally infinite system approaches a scaling state in the absence
of gravity and probably with gravity, too.\\

\newpage

\section{Introduction}
Already when introducing the notion of a surface quantity Gibbs 
%has introduced the notion
implicitly entertained the idea  
of a phase field $\phi$: any density of an extensive quantity (e.g., the
mass density) %, denoted as $\phi$ here)
between two coexisting phases changes gradually (but swiftly)
from its value in one phase to its value in the other. 
%\typeout{Make figure 1!}
The existence of a transition zone, 
though microscopically of atomic extent (far enough from a second-order
phase transition), underlies the very Gibbs definition
of surface quantities.
 In phase transition phenomena,
either of first or second order, this notion has been adopted
 in Landau's spirit. Because energy is an extensive quantity,
too, there is
% transition region is known to have  
an extra energetic cost associated with the transition region,
characterized
in the appropriate thermodynamical potential density by a term
of the form $\epsilon^\ast (\nabla\phi) ^2$, 
$\epsilon^\ast$ being the stiffness of the transition region.
 
The notion of a phase field has appeared abundantly in the literature 
in the context of phase transition
 phenomena\cite{Ginzburg50,Cahn58,Halperin74}.
The transition width {\em diverges} for a second-order phase transition at 
the critical point, and thus it is essential to introduce the 
transition region. 
For a first-order transition, such as a liquid-solid interface,
confering an importance to the interface thickness 
may seem quite anecdotic if one is interested in properties which  occur
on a scale larger than the atomic one; typical examples
are dendritic patterns occuring at the scale
of a $\mu m$. Nevertheless, it is here, where phase-field modeling
has become most useful in numerical treatments.

Before phase-field models became popular, 
it seemed quite natural to treat the surface as
a geometric location on which boundary conditions are imposed
 (e.g., for a moving
front the normal velocity is proportional to the 
jump in the gradient of the temperature 
or concentration field). This is the so-called sharp interface
approach, adopted both in analytical and numerical studies in 
a variety of  contexts of
front problems.

 There has been an upsurge
of interest in the  phase-field approach 
to free-boundary problems more recently, though 
the method was actually introduced pretty early \cite{Langer75,Langer92},
as a {\em computational tool} to model solidification.   
Various studies\cite{Wheeler92,Wheeler93a,Wheeler93b,Kobayashi93,%
McFadden93,Kobayashi94} have  demonstrated the 
virtues of this
method in moving-boundary problems.

Regarding the way how to use phase-field models, 
there are two distinctly different philosophies. 
These may be best discussed considering dendritic
growth, where a set of well-established continuum equations exists
describing phenomena in terms of a sharp interface. On the basis of
this knowledge, a phase-field model can be justified by simply showing
that it is asymptotic to the correct sharp-interface description, i.e.,
that the latter arises as the {\em sharp-interface limit} of the
phase-field model when the interface width is taken to zero.
This is definitely a {\em sufficient} condition for the phase-field model
to yield a correct description of the continuum limit, providing the
interface thickness is taken small enough. Small enough sometimes 
may mean impractically small. The second approach to phase-field
modeling is to guess or derive an appropriate form for the 
free energy of the two-phase system, including the energy cost of the
transition region and to regard this as a physical model in its own 
right. In this case, one might actually forgo considering the limit of
small interface thickness and such a model would even make sense,
if the strict limit of vanishing interface width did not correspond
to sensible physics.

Of course, a problem arises, if a phase-field model obtained in the
second way gives predictions that are different from that of the
sharp-interface equations. A follower of the first philosophy would
then discard the phase-field model, whereas one of the second might
contemplate the possibility that his model contain more physics
than the sharp-interface model. In the case of dendritic growth
the situation is pretty clear: the sharp interface model gives the
right answers. However, this statement cannot be generalized  easily,
since not all sharp-interface models are as well-founded as that for
dendritic growth and because the extreme smallness of the interface
width cannot be always guaranteed (it might for example become
doubtful for a phase transition that is only weakly first order).

A related issue is the question of thermodynamic consistency,
i.e., the derivation of the model in the spirit of Gibbs from a
free-energy or entropy functional.
It is clear that with a known sharp-interface limit in mind, there
is no need at all to obtain a phase-field model this way
(which would mean to make it ``thermodynamically
consistent'') as long as one ensures its asymptotic approach to this
limit. In fact, it has turned out that in some cases where both a 
thermodynamically consistent formulation of a phase-field model
and a nonvariational formulation exist, the latter was numerically
more efficient \cite{Karma98}  and hence preferable on practical grounds.

On the other hand, thermodynamic consistency has its virtues.
This can be seen particularly well in the case considered here,
the influence of elasticity on the stability of a solid interface.
It is quite straightforward to write down the contribution
of the elastic energy to the total free energy. 
Hence, if we have a good idea about the physical origin of the
free energy to be considered, 
the corresponding phase-field model is easily obtained, 
and it is {\em bound to be right}. As a result, one may 
{\em derive} sharp-interface equations in cases where they are
not known. 

For the Grinfeld instability to be considered here,
the sharp-interface equations {\em are} well known. Nevertheless,
it is of course tremendously satisfying, if they simply pop out of the
phase-field equations as the sharp-interface limit. Not only does
this provide a natural countercheck of our ansatz for the free energy,
but it also gives us a new angle of view at the instability, leading
to the prediction of circumstances in which the Grinfeld instability
should {\em not} occur under anisotropic stress, but might appear with
{\em isotropic} stress. We shall consider this point in section II.

Let us return to the advantages of the phase-field method.
The first virtue of phase fields is pretty obvious: instead
of tracking permanently the {\em a priori} unknown interface position in the
sharp-interface limit, and imposing nontrivial boundary conditions 
for the discontinuity of the fields, the
interface in the phase-field approach is nothing but the location
of a rapid variation of the  field $\phi$, while the two
phases are  treated as the same entity. Thus there is no boundary condition
to be imposed in the transition region, 
a fact which greatly facilitates both the
numerical implementation and analysis. 
This is done at some price: one must,
in principle,  mix disparate length (and thus time) scales: 
the pattern length
and the interfacial width, whose ratio may range over  many
orders of magnitude. This may render the numerical 
procedure excessively expensive, a fact which would
quickly take us back to the sharp-interface problem, where
the small length  scales out of the problem. 

As discussed recently\cite{Caroli97},
the sharp-interface limit that one would like to represent %has in mind 
when writing the phase-field
equations, only makes physical sense on ``outer'' length scales much
larger than the physical extent  $l_c$ 
of the transition region, and thus does not
depend structurally on the details of the interface shape on the inner $l_c$ 
scale.
The mathematical question, formulated in the framework  of 
phase-field models, of formally
recovering the sharp interface description via an asymptotic (multi-scale)
expansion in the limit $l_c\rightarrow 0$  might,
from this point of view, seem irrelevant. 

It should however be kept in
mind that the actual matching conditions are imposed for the limit,
where an ``inner'' variable (defined in the transition region) goes to
{\em infinity}. This entails a certain amount of liberty in the choice of
functions defining the free-energy density, because the precise behavior
of these functions on the inner scale does not matter. Hence, the 
{\em validity} of a phase-field model can indeed be judged by simply showing
that it asymptotically reproduces the correct sharp-interface description. 
Whether the additional information encoded in the structure of the phase-field
model on the inner scale is physically relevant, is a question to be decided
on a case by case basis (as implied by our discussion above).

An example where this is relevant, is provided by
 the Young condition for the contact angle
of a droplet on a substrate. This is a condition
on ``outer'' scales, while the inner scale is rather governed
by van der Waals interactions of a thin liquid film with the substrate, 
leading to some nontrivial corrections of the wetting profile at small
atomic length scales. In other words, what matters in the phase-field
description is not that the width of the interface be of atomic
extent, but rather that it be small in comparison with
 the scale of the pattern of interest. 

This physical argument has been cast in a mathematical form in \cite{Karma98},
where the {\it thin-interface limit} was considered, arising from
an alternative asymptotic procedure. This has to be 
contrasted with the sharp-interface limit with the small
parameter being the ratio of the interface thickness  to the 
%scale of the pattern of interest.
capillary length. 
Making the width of the interface 
small only in comparison with the scale of the pattern leads to a rather
important enhancement of the computing speed, thus rendering the 
phase-field approach attractive with regard to
numerical efficiency as well.
Unfortunately, in most systems the thin-interface limit is 
not as easily accessible as for dendritic growth in the thermal model 
\cite{Karma98}. Hence, it is difficult in general to make our argument
mathematically rigorous. However, we will give a consideration of
length scales in section II that clarifies what are the
ratios that have to be kept small for our model to be a good description.
Moreover, we have checked that the dependence of the results on the
interface width becomes weak for the small values that we use for the latter.

Finally the phase-field approach has the additional virtue of regularizing
instabilities, such as the development of cuspy structures, often
setting severe limitations in numerical studies. In the elastic system,
we are led to the question whether the sharp-interface models 
still make sense in the limit where they predict finite-time singularities.
No such singularities arise in the phase-field model, thus possibly
extending the range of validity of the latter beyond that of the former.
This will be discussed in more detail in
section IV.

In the context of growth phenomena phase-field approaches have been
introduced in problems involving temperature or concentration fields.
There are myriads of situations, however, where the corresponding transition is
monitored, or at least affected, by strain. A typical situation is
a solid  under uniaxial stress. This leads to the 
Asaro-Tiller-Grinfeld (ATG)  instability \cite{Asaro72,Grinfeld86} 
(see Ref. \cite{Cantat98}
for a recent review. 
A surface corrugation allows to lower the stored elastic energy.
Other examples of particular interest include solid-solid
transformations, phases in nonequilibrium gels, molecular beam
epitaxy, solidification of lava, etc.
It is thus highly desirable to develop a phase-field
approach including the stress as an active variable. Very recently,
two groups, 
M\"uller-Grant (MG) and Kassner-Misbah (KM) \cite{Mueller99,Kassner99}, have 
independently developed such an approach
and have given a brief account on it.

This paper will present extensive discussions of this question and
give new results. We shall also provide a  comparison
between the MG and KM models, and point out similarities and dissimilarities.
As already known from sharp-interface simulations\cite{Yang93,Kassner94a}
a solid under stress presents a somehow stringent behavior in that 
no stable steady-state solution seems to exist. This is also
the case from analytical studies in the long-wavelength 
limit\cite{Spencer93}.
Nozi\`eres had shown \cite{Nozieres93} that the bifurcation from 
the planar front to the deformed
one is subcritical (the analogue of a first-order transition). 
The study of Spencer and Meiron \cite{Spencer94} focused on structures with a
given basic wave number in the absence of gravity (where the instability does
not have a threshold) and on systems, in which the transport mechanism
necessary for the instability to manifest itself is surface diffusion. 
They find that in the unstable range of wavenumbers
(i.e., for wavenumbers below the marginal value, above which surface
tension stabilizes the planar interface)
there exist finite-amplitude steady-state solutions, {\em if\/} the wavenumber
is close enough to  marginal. This branch of steady-state solutions
terminates by structures developing cusp singularities, despite the
stabilizing influence of surface tension. It cannot be overemphasized
that this result is {\em not} an artifact of their numerics. Indeed, they
investigate carefully the effects of numerical fine-graining 
using a code with spectral accuracy, and their
discretization sequence seems to get fine enough to render their extrapolation
valid. The evidence for the appearance of true cusps in 
the sharp-interface continuum model 
becomes compelling, if one takes into account the work of 
Chiu and Gao \cite{Chiu93} who found {\em analytical} solutions
%it is known that these cusp singularities
developing cusp singularities in finite time. 
The conclusion of Spencer and Meiron is that
for generic initial conditions, including sufficiently small
wavenumbers, finite-time singularities will always occur.
Moreover, they state that this is also true in the presence of gravity
(beyond the threshold) and under fairly general conditions.

For a physical system,  finite-time singularities will be prevented 
by intervening effects that are not considered in the model.
This would mean that nonlinear elasticity and plasticity have to
be taken into account. For example, the formation of dislocations (plasticity)
could blunt cusps again.

%on the tracking of the steady-state branch, and have shown that the branch
%ceases to exist below some value of the control parameter (Fig. 2)
%whereby the cellular structure exhibit a slope divergence.
 Two questions then arise. What kind of structures can be expected
before cusp formation and what kind of structures will prevail eventually? 
Our previous study\cite{Kassner94a}
has shown that the initial cellular pattern may develop into a super-structure,
where a groove or several of them accelerate in a spectacular fashion,  thus
relieving the stresses in their surroundings significantly and 
causing nearby grooves to
recede. Further evolution of the structure
was difficult to handle numerically, due to the development
of cusps which appear in the numerics as they should according
to the analytic results \cite{Chiu93}. %in the sharp interface limit.
In the phase-field approach presented here, no cusps can arise.
The question is of course legitimate, whether the model, which allows
to track the   dynamics of the structure much beyond
the times where earlier studies had to fail numerically, still gives
a faithful description of the physics. Here we take the point of view,
that the details of the description in the locations, where stresses
become large, may not be correctly captured by the model, since we neither
include nonlinear elastic effects  nor other effects
such as capillary overpressure explicitly \cite{footnote1}.
 If they have the right sign, these might
prevent cusp formation \cite{Spencer94}. However, the result of any such
effect must be to blunt cusps, which the phase-field model does. 
As we shall see, it does so in a non-obtrusive way by introducing
a cutoff for interface curvature.
Moreover,
away from the cusps, stresses are low enough for linear elasticity to apply.
Hence, we believe that the development of the overall morphology is still
correctly described by the phase-field model.

A more detailed justification would point out that nonlinear elasticity
will first  make itself felt by stresses increasing more slowly 
as a function of strains than
in the linear case and that next plasticity will act to introduce an
upper cutoff for stresses, where the material will yield. Now the
effect of any resulting modification in the stress-strain relationship
on the remaining body
can be reproduced by cutting out the piece of material where linear
elasticity ceases to hold and by requiring boundary conditions at 
the edge of the cut-out piece that correspond to the correct stresses.
In the case of  material yielding near a singularity  
 of the stress
field (obtained assuming linear elasticity), these boundary
conditions will essentially be that the stresses are close to the 
yield stress at the boundary. This is mimicked by the phase-field
model in which the maximum supportable stress is, for a given
geometry, determined by the interface width.   

Therefore, we think this is one case, where the 
phase-field model can do more in
the description of the physical system than its sharp-interface limit.

%allows one to go much
%beyond previous studies and to track the dynamical evolution 
%of the structure.
 It will emerge that usually one leading groove continues to deepen
while neighboring ones recede
after  the winner has started to relieve the stress that kept them 
growing. %continuously in time. 
Finally the surface shows a single deep
groove evolving in time and becoming a location of a strong stress
accumulation, possibly until the fracture threshold is reached. 
Presumably, before that 
stage is reached the validity of the model will break down. We shall
make some speculation on future directions to elucidate the physical
 behavior in real systems. For generic initial conditions,
we may consider the dynamics a continual 
coarsening process
which initially develops as described in \cite{Mueller99}
and later is dominated by groove growth and shrinkage.

 The  paper is organized as follows. In section II we give the
 continuum equations ordinarily used in the description of the Grinfeld
 instability. This is mainly done in order to introduce appropriate
 length and time scales in nondimensionalizing the equations; then we
 present our 
 phase-field approach and discuss how the
 interface width has to be chosen in comparison with the
 other length scales. We demonstrate how the phase-field model
can be employed to derive new sharp-interface equations in the 
presence of body forces breaking rotational invariance.

%In section III, we shall derive the sharp-interface limit of the model
% via a mathematically rigorous asymptotic expansion. 
 Section III presents validation results, a comparison  of the 
MG and KM models and
  describes the main findings of our simulations.
%  and how further assumptions on the technical level
%  allow to optimize the computing time. 
Section IV sums up the results and discusses perspectives. 
The mathematically rigorous asymptotic expansion used to derive
the sharp-interface limit has been relegated to an appendix, as the
calculation is somewhat lengthy and would interrupt the flow of the 
text.
 Since we use the MG model in a slightly different form from that presented
originally \cite{Mueller99}, we give the connection between the 
two formulations in a second appendix.
Finally, appendix \ref{APPCONFMAPP} 
contains the analytic derivation via 
conformal mapping of the stresses for a particular interface shape
to be compared with the numerics.

 \section{The Grinfeld instability} %Elastic energy and reference state}
 \subsection{Sharp-interface equations \label{sharpinterface}}
A description of the basic ingredients of the Grinfeld instability
has been given elsewhere \cite{Cantat98}. Therefore,
 we may restrict ourselves
to explaining the physical mechanism and giving the equations.

We wish to describe the behavior of a solid submitted to uniaxial 
stress, at the surface of which material transport is 
possible. Consider the example of a solid in contact with its melt.
An accidental corrugation of the surface will act to reduce
the stress at its tip and increase it in the valleys next to it.
That is, the solid can decrease its elastic energy density by growing 
tips (where the stress is lower) and by increasing the depths of
valleys (where it gets rid of material having a higher density of elastic
energy due to larger stresses). This tendency is most easily cast into
equations by writing down the chemical potential difference 
$\Delta \mu = \mu_s -\mu_\ell$ at the interface (the subscripts refer
to the solid and liquid or nonsolid phases, respecively) \cite{Kassner94a}:
\begin{equation}              
  \Delta \mu = {1-\nu^2\over 2E\rho_s}     
   (\sigma_{tt}-\sigma_{nn})^2+  \frac{1}{\rho_s}\gamma \kappa    
    +   
      \frac{\Delta\rho}{\rho_s} g \zeta(x)   
         \label{dmu1}
  \end{equation}
Herein, the first term is of elastic origin; $\sigma_{tt}$ and $ \sigma_{nn}$
are the normal stresses tangential and perpendicular to the interface,
$E$ is Young's modulus, $\nu$ the Poisson number, and $\rho_s$ the density
of the solid. The second term describes the stabilizing influence of the
surface stiffness $\gamma$, taken isotropic here (so it becomes identical 
to the surface energy). 
 $\kappa$ is  the curvature of the interface; 
for simplicity, we consider the two-dimensional case only. Finally, the
third term is the contribution of gravity ($g$), where $\Delta\rho=
\rho_s-\rho_\ell$ is the density contrast between the solid and the 
liquid (or vacuum) and $ \zeta(x)$ is the interface position, given by
its $z$ coordinate (the $z$ axis is oriented antiparallel to the gravitational
force). Equation (\ref{dmu1}) holds for plane strain. For plane stress,
the prefactor $1-\nu^2$ has to be dropped.
        
The dynamics is then described by giving the normal velocity in terms of
the chemical potential difference. For a solid in contact with its melt
this would simply be
\begin{equation} 
v_n = -\frac{1}{k} \Delta \mu   \>, \label{vn}
\end{equation}
where $k$ is an inverse mobility with the dimension of a velocity.
 In the case of a solid in contact with vacuum and surface diffusion
as the prevailing transport mechanism we would have
$ v_n = D \nabla^2 \Delta \mu$ instead.

Of course, in order to compute $v_n$, we must first obtain the 
stresses entering (\ref{dmu1}). This involves solving an elastic
problem ($\partial_j \sigma_{ij} = 0$)
with a prescribed external stress and boundary conditions on the interface,
assuming an appropriate constitutive law. Ordinarily, Hooke's law for isotropic
elastic bodies is assumed [therefore we have only two elastic constants in
(\ref{dmu1})]. Neglecting the capillary overpressure, which usually
is a good approximation, we have as 
boundary conditions at the interface $\sigma_{nn} = -p$,
where $p$ is the pressure in the second phase, 
and $\sigma_{nt}=0$, i.e., the shear stress vanishes.

A linear stability analysis of a planar interface under the dynamics
given by equations (\ref{dmu1}) and (\ref{vn}) yields the following 
dispersion relation ($\omega$ is the growth rate, $q$ the wave number)
\begin{equation}   
  \omega=\frac{1}{k\rho_s}\left\{{2 \sigma_0^2 ({1-\nu^2})\over  E}q
  -\gamma q^2 -
  \Delta \rho g\right\} \>.    \label{dispers} 
  \end{equation}   
$\sigma_0$ is the uniaxial external stress. Eq.~(\ref{dispers}) provides
us with a critical wave number $q_c = \sqrt{\Delta\rho g/\gamma}$ (an
inverse capillary length) and a critical stress 
$\sigma_{0c}= \left[\gamma q_c E/(1-\nu^2)\right]^\frac12$,
below which the planar interface is stable. The  
wave number of the fastest-growing mode can be inverted to  give a length
\begin{equation}
\ell_1 = \frac{\gamma E}{\sigma_0^2(1-\nu^2)} \>. \label{ell1}
\end{equation}
Apart from a prefactor of $2/\pi$, this length is identical to the
so-called Griffith length.

Note that  a  planar interface
will not remain at its original equilibrium position, even  when only
 a subthreshold stress is applied, i.e., when it is ``stable''.
 Our dynamical equations predict that it has a
nonzero velocity as long as $\Delta\mu$ is different from zero.
Hence it will recede to smaller values of $\zeta$.  If the density of
the solid is bigger than that of the liquid, the chemical potential
on the solid side of the receding interface decreases faster than
that on the liquid side and there is a new equilibrium position, which can
be computed directly from (\ref{dmu1}) and which evaluates to
\begin{equation}
\Delta \zeta \equiv -\ell_2 = - \frac{1-\nu^2}{2\Delta\rho g E} \sigma_0^2 \>.
\label{ell2}
\end{equation}

Equations (\ref{ell1}) and (\ref{ell2}) provide us with two independent
length scales of the problem, the first of which is due to a competition
between stress and surface energy, while the second arises from the 
competition between stress and gravity. For the purpose of nondimensionalizing 
equations, $\ell_1$ is more appropriate, as this length does not diverge
in the limit of vanishing  density difference (or gravity).
 To obtain a natural time scale $\tau$, 
we can replace $q$ in either of the two wave-number dependent terms of 
(\ref{dispers}) by $1/\ell_1$. This leads to 
\begin{equation}
\tau = \frac{k \rho_s \gamma E^2}{\sigma_0^4(1-\nu^2)^2}\>.
\end{equation}
The nondimensional version of the dispersion relation then reads
($\tilde\omega = \tau\omega$, $\tilde q = \ell_1 q$):
\begin{equation}
\tilde{\omega} = 2 \tilde{q} - \tilde{q}^2 - \frac{\ell_1}{2\ell_2} \>,
\label{disprelnondim}
\end{equation}
which shows clearly that the problem {\em without} gravity (when 
$\ell_2$ becomes infinite) can be made
parameter free, i.e., elastic and other parameters only set the time
and length scales; apart from that we should expect the same dynamics
for all systems. %, at least in the linear regime.
 {\em With} gravity, the dynamics is essentially
determined by the ratio of the two length scales introduced. 

\subsection{Elastic energy and state of reference}

Let us now proceed to investigate the contributions to the free energy
of the same system.
The phase-field  model will then consist in writing 
the free-energy density that takes
into consideration the global elastic energy in both phases. 
%We will treat the liquid phase essentially as a solid that is 
%free of shear stresses.

 As is usual with elastic problems, it is important to specify
 the state of reference defining the positions of material particles with 
respect to which displacements are measured. This is crucial whenever the
reference state is not that of an undeformed body but one that is subject
to prestraining (which will turn out useful later).
 In order to make this point clear, and in the hope
 of helping subsequent discussions, we would like first to dwell on this 
issue. 

If the only energy present is elastic energy and Hooke's law holds true,
the free energy per unit volume can be written as 
%to stress on this point.  
%The energy per unit volume obeys Hooke's law, and
% can be written as 
 %The traditional relation between the stress tensor $\sigma_{ij}$
 %and the strain tensor $u_{ij}$ is given by Hooke's law which can be
 %written as
 \begin{equation}
 f= \mu \, u_{ij}  u_{ij}+ {\lambda\over 2} u_{ii}^2 \>,
 \label{Hooke1}
 \end{equation}
where summation over double subscripts is implied.
 $\lambda$ and  $\mu$ are the Lam\'e constants, $\mu$ being better known
%is the compression modulus, while $\mu$ is
as the shear modulus. For plane strain, these elastic constants are related to 
Young's modulus and the Poisson ratio via $\mu = E/[2(1+\nu)]$ and
$\lambda = E\nu/[(1+\nu)(1-2\nu)]$.
%\typeout{possibly appendix here with plane stress relations}
 The stress tensor $\sigma_{ij}$ is 
%related to the strain tensor by the relation
then 
 \begin{equation}
 \sigma_{ij}=\frac{\partial f}{\partial u_{ij}} = 
 2\mu u_{ij}+\lambda u_{kk}   \delta_{ij}
 = 2\mu (u_{ij}- u_{kk}  \frac{\delta_{ij}}{d}) + K u_{kk}   \delta_{ij} \>.
 \label{stress1}
 \end{equation}
$K=2\mu+\lambda/d$ is the bulk modulus, 
and the last relation has the advantage of 
making explicit the  parts of the stress tensor
causing pure shape and pure volume changes, respectively.
 ($d$ is the spatial dimension.)
We will nevertheless mainly use the relations containing $\mu$ and $\lambda$, which are
more compact. 
The implied reference state here is $u_{ij}=0$, for which $\sigma_{ij}=0$.

However, if we choose a reference state given by a different strain
tensor $u_{ij}^{(0)}$, setting $\tilde{u}_{ij} = u_{ij} - u_{ij}^{(0)}$,
then we should {\em not} simply replace $ u_{ij}$ by   $\tilde{u}_{ij}$ 
in (\ref{stress1}), as the stress tensor and the elastic energy are,
in principle, measurable quantities and should thus be unaffected by
a change of strain reference state. 
Hence, we have to write
 \begin{equation}
 \sigma_{ij}=
 2\mu \left(\tilde u_{ij}+ u_{ij}^{(0)}\right)
 +\lambda\left( \tilde u_{kk} +  u_{kk}^{(0)}\right)  \delta_{ij} \>,
 \label{stressa}
 \end{equation}
and change definition (\ref{Hooke1}) accordingly, i.e.~replace 
$u_{ij}$ by $\tilde{u}_{ij}+u_{ij}^{(0)}$. In this situation, the
zero-strain state would not be  stress-free. An alternative way to 
specify a reference state would then consist in giving the stress of
the  zero-strain state.

In general, the thermodynamic system under consideration will not only
contain elastic contributions. Then the {\em equilibrium state}, corresponding
to a minimum of the free energy, may not be a state of vanishing strain.
A trivial example is a solid in equilibrium with its melt, 
where the equilibrium state in the solid 
corresponds to the strain produced by the equilibrium pressure $p$ of 
the liquid (the equilibrium stress tensor of the solid is $-p \delta_{ij}$).
The form of the free-energy density accounting for such a situation is not
(\ref{Hooke1})
 [which does not exhibit a minimum at $u_{ij}^{({\rm eq})} \ne 0$]
but
% The free energy would
% then take the form
 \begin{equation}
  f =  \mu \left(u_{ij}-u_{ij}^{({\rm eq})}\right)\left(u_{ij}-u_{ij}^{({\rm eq})}\right)
+ {\lambda\over 2} \left(u_{ii}-u_{ii}^{({\rm eq})}\right)^2  \>.
   \label{Hooke2}
   \end{equation}
This is manifestly minimum at $u_{ij}^{({\rm eq})}$ and the nonzero value of
 the latter quantity takes into account nonelastic contributions to the
free energy. If we now define the stress from the first relation in 
Eq.~(\ref{stress1}), i.e., $\sigma_{ij} = \partial f/\partial u_{ij}$, it will
be nonzero only, if there are  forces driving the system away from 
equilibrium. If we  rather define it via the second equality of
Eq.~(\ref{stress1}), meaning that we set 
$\sigma_{ij}=2\mu u_{ij}+\lambda  u_{kk} \delta_{ij}$ (which is now different
from $\partial f/\partial u_{ij}$), it will describe, in addition to 
these forces the prestress necessary to {\em keep} the system in equilibrium.
An invariant relation between stresses and strains follows from 
requiring
 \begin{equation}
f = \int_{u_{ij}^{({\rm eq})}}^{u_{ij}} \left(\sigma_{ij}-\sigma_{ij}^{({\rm eq})}\right) du_{ij} \>, 
\label{freeint}
\end{equation}
which leads to 
 \begin{equation}
 \sigma_{ij}-\sigma_{ij}^{({\rm eq})} =
 2\mu \left(u_{ij}-u_{ij}^{({\rm eq})}\right)
 +\lambda\left( u_{kk} -u_{kk}^{({\rm eq})}\right)  \delta_{ij} \>.
 \label{stressb}
\end{equation}
Once both $\sigma_{ij}^{({\rm eq})}$ and $u_{ij}^{({\rm eq})}$ are specified,
this equation gives us an unambiguous relationship between stresses and
strains. Depending on which variables we choose to define the reference
state, we obtain the conjugate variables of the same state from
(\ref{stressb}). If we choose, for instance, a strain-free state as reference,
this equation will provide us with the corresponding stress of reference, if we choose a
stress-free state of reference, it will yield the strain of reference.

As an example we can look at a case where
the equilibrium stress is 
$\sigma_{ij}^{({\rm eq})}=-p_0 \delta_{ij}$, and ask what should
the strain be. 
Hooke's law  -- written in a such a way
that the absence of strain implies the absence
of stress as well [see (\ref{stress1})] --
then gives us an equilibrium strain
\begin{equation}
u_{ij}^{({\rm eq})} = - {p_0\over 2\mu +\lambda d} \delta_{ij} \>. \label{str0}
\end{equation}

Since the free energy must not depend on the choice of reference state,
it is clear that it does not matter whether we use ${u}_{ij}$
or  $\tilde{u}_{ij}$
in (\ref{stressb}), providing that we use the 
correct values of $u_{ij}^{({\rm eq})}$ and $\tilde{u}_{ij}^{({\rm eq})}$,
respectively.
Suppose that we choose another reference state characterized by
the strain tensor $\tilde u_{ij}$ in such a way that
when the strain is zero, 
the stress is equal to $-p_{0s}\delta_{ij}$.
A vanishing strain then corresponds to  
a pre-stressed situation.
If the equilibrium stress is again $-p_0\delta _{ij}$ as above, we must
have a new equilibrium strain $\tilde u_{ij}^{({\rm eq})}$ obeying, 
according to our invariant relation (\ref{stressb}), 
$-p_{0s}\delta _{ij}+p_{0}\delta _{ij} = 
- 2\mu \tilde u_{ij}^{({\rm eq})}- \lambda  
\tilde u_{kk}^{({\rm eq})} \delta_{ij}$. After a simple manipulation we obtain
\begin{equation}
\tilde u_{ij}^{({\rm eq})}= {p_{0s} -p_{0}\over 2\mu +\lambda d} \delta _{ij}
\label{strain2}
\end{equation}
%This defines another equilibrium strain reference. Obviously the energy
%must not depend on the choice of the reference state. This means that
%we must have 
Of course, Eq.~(\ref{str0}) is a special case of (\ref{strain2}) for
$p_{0s} = 0$.
The free energy, expressed by $\tilde u_{ij}$, then reads
 %is independent of the  choice of the reference state, we have
\begin{eqnarray}
f&=&
 \mu (\tilde u_{ij}-\tilde u^{(\rm eq)}_{ij}) 
(\tilde u_{ij}-\tilde u^{(\rm eq)}_{ij})
+
{\lambda\over 2} (\tilde u_{ii}-\tilde u^{(\rm eq)}_{ii})^2 
 \nonumber\\
  & =&\mu \tilde u_{ij} \tilde u_{ij} + {\lambda\over 2} \tilde u_{ii} 
     \tilde u_{jj}+(p_0-p_{0s})\tilde u_{ii} + {d \over 2 (2\mu +\lambda d)}
     (p_0-p_{0s})^2
\label{hooke3}
\end{eqnarray}
It should be realized that Eqs.~(\ref{Hooke2}) and (\ref{hooke3}) describe
exactly the same situation, if {\em corresponding} values for the strain fields
without and with a tilde are inserted. At this point we have said nothing
about applied external stresses or so. However, if we choose, say, 
vanishing displacement as boundary
condition at a planar interface, directed along the $x$ direction,
then this will correspond to two different physical situations for the
two different choices of the state of reference.
 Let us for simplicity assume $p_0=0$. According to (\ref{str0}),
 we  then have $u^{(\rm eq)}_{ij}=0$, and (\ref{Hooke2}) implies (\ref{Hooke1}).
Setting $u_{xx}=0$ in Eq.~(\ref{stress1}) we  obtain, because of the
boundary condition $\sigma_{zz}=0$ that also $u_{zz}=0$, and there is no
stress at all. On the other hand, if we set $\tilde{u}_{xx}=0$, we have to
use (\ref{stressb}) and (\ref{strain2}) to obtain the elastic state
of the solid. The boundary condition for $\sigma_{zz}$ implies
$u_{zz}=p_{0s}/(2\mu+\lambda)$, which in turn leads to 
$\sigma_{xx}=-2\mu p_{0s}/(2\mu+\lambda)$; hence vanishing displacement
along our planar interface means a solid that is homogeneously strained
in the $x$ direction with
a prestress $\sigma_0=\sigma_{xx}$.

As we shall see later, the latter choice of the state of reference 
has been made in the phase-field
model discussed in \cite{Mueller99}, the former  (setting $p_{0s}=0$)
 in \cite{Kassner99}.
These are the most natural choices, although an infinity of (less natural)
alternatives is available.

\subsection{The phase-field model \label{secphase}}

The total (solid+liquid)
free energy of the system can be written as
\begin{equation}
F[\phi ,\; \{u_{ij}\}]=\int dV \left[f(\phi ,\, \{u_{ij}\})
 + {1\over 2} \Gamma \epsilon^2
({\bf \nabla} \phi)^2\right]
\label{F}
\end{equation}
where %$f$ is the free-energy density, $u_{ij}$ is the strain tensor, and 
$\epsilon$ is a length parameter controlling
the order of magnitude of the transition region described by the
phase field.  $ \Gamma = 3 \gamma / \epsilon$
is the energy density corresponding to 
the surface energy $\gamma$ being distributed over
a layer of width $\approx \epsilon$. (The factor 3 is just a convenient
choice, simplifying later derivations.)

If we start from the invariant form (\ref{stressb}), we can set up a whole 
class of phase-field models at once and specify the reference state later.
In order to be able to write a single elastic energy expression for
the two-phase system, we formally treat the liquid as a shear-free solid
(not including hydrodynamics). 
%This is possible providing that
%tensile stresses do not exceed the original equilibrium pressure of 
%the liquid; ordinarily, a Young's modulus cannot be assigned to a
%liquid, since it does not resist dilation, but for deviations from
%the equilibrium pressure this is allowed  
%as long as the total pressure does not become negative.
We will discuss some  consequences of this approach later.
%regarding the liquid as a ``particular'' solid later.

A straightforward ansatz for the elastic energy density %in the two phases 
is then
\begin{eqnarray}
f_{{\rm el}}(\phi,\{u_{ij}\})  &=& h(\phi) f_{{\rm sol}}(\{u_{ij}\}) 
                            + [1- h(\phi)] f_{{\rm liq}}(\{u_{ij}\}) \nonumber \\
&=& h(\phi)\left\{\mu \left(u_{ij}-u_{ij,s}^{({\rm eq})}\right) 
    \left(u_{ij}-u_{ij,s}^{({\rm eq})}\right)
   + {\lambda\over 2} \left(u_{ii}-u_{ii,s}^{({\rm eq})}\right)^2 
                \right\} \nonumber \\
&& \mbox{} + \left[1-h(\phi)\right] {\tilde\lambda \over 2}
         \left(u_{ii}-u_{ii,\ell}^{({\rm eq})}\right)^2  \>, 
\label{felas}
\end{eqnarray}
where $f_{{\rm sol}}(\{u_{ij}\})$ and $ f_{{\rm liq}}(\{u_{ij}\})$
 are the densities
of elastic energy in the solid and in the liquid, respectively, and
where $ h(\phi)$ may be interpreted as a ``solid fraction'', which
must be equal to one in the solid and equal to
zero in the liquid. We choose  $h(\phi)=\phi^2(3-2\phi)$ for reasons
of convenience: with this choice $h'(\phi)=0$ for $\phi=0$ and for 
$\phi=1$, i.e., in the bulk phases. This leads to the advantage
(see appendix \ref{APPSHARPINT}) 
that the zeroth-order solution of the asymptotic
expansion in powers of $\epsilon$ is valid to all orders in the outer
region considered. 

Since different reference states may be chosen in the solid and in
the liquid, the equilibrium strains carry a subscript $s$ or $\ell$, 
respectively.
This would not be necessary here, because the prefactor [$h(\phi)$ or $1-h(\phi)$]
decides whether the equilibrium expression for the strain in the liquid or in the
solid has to be taken. However, as soon as we take derivatives with respect to 
$\phi$, this criterion of distinction becomes ambiguous, so we prefer to make the 
difference explicit from the outset.
$\tilde \lambda$ is the bulk modulus of the liquid.

To account for the possibility of a phase transition, we introduce
a double well potential
\begin{equation}
f_{{\rm dw}}(\phi) = 2 \Gamma g(\phi)  \>, 
\label{fdw}
\end{equation}
where $g(\phi)=\phi ^2 (1-\phi)^2$. The minimum at $\phi=1$ corresponds to the 
solid phase, the one at $\phi=0$ to the liquid phase. Note that while this potential
looks similar to the fourth-order polynomial used in the Landau theory of 
second-order phase transitions, it is employed 
in quite a different manner here.
The two minima correspond to the two phases and the symmetry of the potential
is of secondary importance; in Landau's approach,  symmetry considerations are
 at the heart of the theory, the symmetric minima describe the {\em same} phase,
 and the second phase corresponds to the unstable maximum in between.
Since in our case both phases sit at a minimum, the transition described by the
double well potential is of first order. We do not need a sixth-order polynomial
as would be necessary in Landau's theory for first-order phase transitions.

Gravity will be included in essentially the same way as in the sharp-interface
equations discussed above; i.e., its effect as a body force in 
the mechanical equilibrium condition is neglected but its
influence on the chemical potential is taken into 
account. This is a good approximation  usually (one can estimate the 
cross-effect of gravity on the elastic energy to be on the order
of $\rho_s g H/\sigma_0 \ll 1$, for typical heights $H$ of the sample).
Then the contribution of gravity to the free-energy density becomes
\begin{eqnarray}
f_{{\rm grav}}(\phi,z) & = & (z-z_0)\bigl\{\rho_s h(\phi) + \rho_\ell  
\left[1-h(\phi)\right] \bigr\} g  \nonumber \\
& = &   (z-z_0) h(\phi) \Delta \rho g  +   (z-z_0)\rho_\ell  g \>,
\label{fgrav}
\end{eqnarray}
where we have taken the zero point of this potential energy at $z=z_0$.
Note that in taking fixed values for $\Delta\rho$ and for $\rho_s$,
we also neglect the second-order effect caused by density changes 
due to strain.

Finally, we wish to be able to control the equilibrium position of the
interface ``by hand'' via addition of a constant to the free-energy density of
one phase; this phenomenogical contribution to the total free-energy density
may be conveniently written as
\begin{equation}
f_{{\rm c}}(\phi) = - h(\phi) \frac{1-\nu^2}{2E} \sigma_{00}^2 
            = - h(\phi) \frac{2\mu+\lambda}{8 \mu (\mu+\lambda)} \sigma_{00}^2
\label{fc} 
\end{equation}
and it is normalized such that setting $ \sigma_{00} = \sigma_0$,
we can keep the equilibrium position of the planar interface at 
the fixed value $z_0$, independent of $\sigma_0$. This is useful, for
example, if one wishes to assess the relative position of the maxima
or minima of an evolving structure with respect to a planar interface at the
same external stress. Because of the recession of a planar interface
according to (\ref{ell2}), such a comparison would otherwise be difficult.

Collecting all contributions, we obtain for the total free-energy density
\begin{eqnarray}
f(\phi\; , \{u_{ij}\},z) &=& f_{{\rm dw}}(\phi) + f_{{\rm el}}(\phi,\{u_{ij}\})
+ f_{{\rm grav}}(\phi,z) + f_{{\rm c}}(\phi) \nonumber \\
 &=& \Gamma \Biggl( 2 g(\phi)
 + \frac{\epsilon}{3\gamma} \biggl\{h(\phi)\left[ 
           \mu \left(u_{ij}-u_{ij,s}^{({\rm eq})}\right)
                \left(u_{ij}-u_{ij,s}^{({\rm eq})}\right)
              + {\lambda\over 2}
               \left(u_{ii}-u_{ii,s}^{({\rm eq})}\right)^2
 \right] \nonumber \\
&& \phantom{\Gamma \Biggl( 2 g(\phi)+ \frac{\epsilon}{3\gamma} \biggl\{ }
\mbox{}   + \left[1-h(\phi)\right] {\tilde\lambda \over 2}
                \left(u_{ii}-u_{ii,\ell}^{({\rm eq})}\right)^2 \nonumber \\
&&  \phantom{\Gamma \Biggl( 2 g(\phi)+ \frac{\epsilon}{3\gamma} \biggl\{ }
\mbox{} +  h(\phi) \left[ \Delta \rho g  (z-z_0) 
- \frac{2\mu+\lambda}{8 \mu (\mu+\lambda)} \sigma_{00}^2\right] +  (z-z_0) \rho_\ell  g  \biggr\} \Biggr) \>.
\label{genphasefield}
\end{eqnarray}
Note that here the  terms in braces, in particular the elastic term,
 have acquired a prefactor $\epsilon$. This $\epsilon$ dependence is spurious,
as we have taken the prefactor $\Gamma \propto 1/\epsilon$ in front
of everything, and the factor $\epsilon$ just serves to cancel this.
 In fact, the only contribution to the free energy that
can depend on $\epsilon$ explicitly is the double well potential, which
must ensure that in the limit $\epsilon\to0$ the only possible 
states are the bulk phases and must therefore become infinite for all
values of $\phi$ different from 1 or 0. All the other energies can
depend on $\epsilon$ only implicitly via $h(\phi)$, the
local solid fraction of the two-phase system. 

We then require $\phi$ to satisfy a relaxation equation for a non-conserved
order parameter. This equation takes the form
\begin{equation}
 \frac{\partial \phi}{\partial t} =-R\frac{\delta F}{\delta \phi} \>,
\end{equation}
and the prefactor $R$ should contain the mobility $1/k$ defined in 
(\ref{vn}). The dimension of $R$ must be 
(energy~density\ $\times$~time)$^{-1}$, which leads us to choosing
$R = 1/(3k\rho_s\epsilon)$. This essentially amounts to setting the 
time scale for the evolution of $\phi$ (which must be related to the
width of the transition region, because it is only in this region where
$\phi$ has an appreciable dynamics).

We arrive at
\begin{eqnarray}
 \frac{\partial \phi}{\partial t} &=& 
\frac{\gamma}{k\rho_s} \biggl[ \nabla ^2 \phi 
-  {1\over\epsilon^2} \biggl(2 g'(\phi)
 + \frac{\epsilon}{3\gamma} h'(\phi) \Bigl\{
         \mu \left(u_{ij}-u_{ij,s}^{({\rm eq})}\right)
                \left(u_{ij}-u_{ij,s}^{({\rm eq})}\right) 
         + {\lambda\over 2}
               \left(u_{ii}-u_{ii,s}^{({\rm eq})}\right)^2 
 \nonumber \\
&&\phantom{\frac{\gamma}{k\rho_s} \biggl[ \nabla ^2 \phi 
-  {1\over\epsilon^2} \biggl(2 g'(\phi)} 
\mbox{} -  {\tilde\lambda\over 2}
               \left(u_{ii}-u_{ii,\ell}^{({\rm eq})}\right)^2  +  (z-z_0) \Delta \rho g 
- \frac{2\mu+\lambda}{8 \mu (\mu+\lambda)} \sigma_{00}^2 \Bigr\} \biggr)\biggr]
\label{phirelax1}
\end{eqnarray}
Herein, $g'(\phi)
%=2\phi(1-\phi ) (1-2 \phi )
$ and $h'(\phi)$ are the derivatives of $h(\phi)$ and $g(\phi)$ with respect to their argument.
%=6\phi(1- \phi)$. 
As we have mentioned before,
$h'(\phi)$ vanishes in the solid as well as in the liquid  phases
[see eq.~(\ref{useful2})]. 
%which is one of the reasons why this choice of $h(\phi)$ is convenient.
%This has constituted a convenience for  that choice.

In writing down an equation for the evolution of 
the elastic variables, we have to be careful about the
fact that the strains $u_{ij}$, $i,j=1,..., d$ are not independent
quantities. Therefore, the variational derivatives $\delta F/\delta u_{ij}$
are not independent. Instead of introducing Lagrangian multipliers, we
can however exploit the fact that the components $u_i$, $i=1,..., d$
of the {\em displacement}, related with the strains via Eq.~(\ref{straindef}),
%$u_{ij} = \frac{1}{2} (\partial_i u_{j}+\partial_j u_{i})$, 
{\em are} independent variables. Assuming that
the time scales of our problem are large in comparison with sound
propagation times, we conclude that the variational derivatives
 $\delta F/\delta u_i$ are equal to zero. This is is an adiabaticity assumption.
Hence, we obtain
\begin{eqnarray}
0 = \frac{\delta F}{\delta u_i}
 &=& \frac{\partial}{\partial x_j}  \frac{\delta F}{\delta u_{ij}}
\nonumber \\
&=& \frac{\partial}{\partial x_j} \left\{ h(\phi) \left(\sigma_{ij}-\sigma_{ij}^{({\rm eq})}\right) - [1-h(\phi)] \left(p-p^{({\rm eq})}\right) \delta_{ij} \right\} \>.
\label{stressrelax1}
\end{eqnarray}
This is nothing but a generalized elasticity problem, with the generalized stress
tensor given by the quantity in braces.

%Below, we  will show that the sharp-interface limit of these
%equations becomes, given the usual equilibrium state before straining
%the solid, identical to the dynamical laws (\ref{dmu1}, \ref{vn}) plus 
%the elastic equations governing the Grinfeld instability. We will also
%see that unconventional equilibrium states, e.g., a nonisotropic state
%of a piezoelectric crystal under the influence of a homogeneous
%electric field, lead to {\em different} dynamic equations, allowing the
%presence of the
%Grinfeld instability for {\em isotropic}  and its absence for
%{\em anisotropic} strain.

Before moving to a demonstration of the sharp-interface
limit,  let us  
discuss scales. Since the elastic
problem (\ref{stressrelax1}) is formally
linear in the strains, rendering it nondimensional
is straightforward and unenlightening. On the other hand, trying
to cast (\ref{phirelax1}) into nondimensional form, we realize that
besides the length and  time scales discussed in 
subsection \ref{sharpinterface}, we need a {\em third}
length scale $\ell_3 = \gamma/K$, 
{\em apart from} the width of the transition region $\epsilon$.
So the phase-field model contains four length scales altogether.

%Introducing $\ell_3$ and 
Normalizing  elastic moduli and stresses by the bulk modulus, 
i.e., setting $M=\mu/K$, $\Lambda = \lambda/K$, $\tilde\Lambda
= \tilde\lambda/K$, $\Sigma_{00} = \sigma_{00}/K$, we obtain 
\begin{eqnarray}
 \frac{\partial \phi}{\partial \tilde t} &=& 
 \tilde\nabla ^2 \phi 
-  {\ell_1^2\over\epsilon^2} \biggl(2 g'(\phi)
 + \frac{\epsilon}{3\ell_3} h'(\phi) \Bigl\{
         M \left(u_{ij}-u_{ij,s}^{({\rm eq})}\right)
                \left(u_{ij}-u_{ij,s}^{({\rm eq})}\right) 
         + {\Lambda\over 2}
               \left(u_{ii}-u_{ii,s}^{({\rm eq})}\right)^2 
 \nonumber \\
&&\phantom{\tilde \nabla ^2 \phi 
-  {\ell_1^2\over\epsilon^2} \biggl(2 g'(\phi)} 
\mbox{} -  {\tilde\Lambda\over 2}
    \left(u_{ii}-u_{ii,\ell}^{({\rm eq})}\right)^2 + \frac{2\ell_3}{\ell_2} (\tilde z-\tilde z_0) 
- \frac{2M+\Lambda}{8 M (M+\Lambda)} \Sigma_{00}^2 \Bigr\} \biggr) \>,
\label{phirelaxnrm}
\end{eqnarray}
where $ \tilde t = t/\tau$ is the nondimensional time, and 
$\tilde z = z/\ell_1$, $\tilde \nabla = \ell_1\nabla$ are nondimensionalized
spatial operators.
Physically, $\ell_3$ represents an atomic scale. For many materials,
$\gamma/K$ is on the order of the lattice constant.

For the phase-field model to work properly, 
we must impose some conditions on the
length scale $\epsilon$. We  definitely need $\epsilon/\ell_1 \ll 1$ to have
a decently sharp interface. Moreover, the $h'(\phi)$ term must not become
too large in comparison with the $g'(\phi)$ one, otherwise the minima
of the double well  move away from the positions
$\phi=0$ and $\phi=1$. This appears to suggest that we also need
 $\epsilon/\ell_3 \ll 1$. We compute some typical values.
Using the material parameters of solid He \cite{footnote2}, 
the system for which the
Grinfeld instability has been unambiguously demonstrated by Torii and Balibar 
\cite{Torii92}, we obtain the estimates $\ell_1 \approx 0.1$ cm,
$\ell_2  \approx 0.1$ cm, $\ell_3 \approx 10^{-9}$ cm, and 
$\tau  \approx 1$ s. If we had to require $\epsilon \ll \ell_3$,
we would have a problem with very disparate length scales, as our
numerical grid would have to be smaller than $\ell_3$, whereas the
length scales governing pattern formation are $\ell_1$ and $\ell_2$.
Fortunately, the quantity $\epsilon/3\ell_3$ appearing in (\ref{phirelaxnrm})
is multiplied by squared strains, and the $u_{ij}$ are on the order of
$10^{-4}$. Moreover, we have $2\ell_3/\ell_2 \approx  2\times 10^{-8}$ 
and the last term in braces can be estimated by 
$\frac14 \Sigma_{00}^2\approx 2\times 10^{-8}$. Therefore,
the actual condition for our model to be useful is
$10^{-8} \times \epsilon \ll 3\ell_3$, which is much easier to achieve.
In our simulations, we typically had  
$10^{-8} \times \epsilon/3 \ell_3 \approx 0.1$.

Equations (\ref{phirelax1})-(\ref{stressrelax1}) constitute the basic 
phase-field equations for the phase transformation under stress.

To  specify our model completely, we have to indicate
the equilibrium stresses and strains.
Let us assume the following forms for the stress-strain relationships in the
two phases,
\begin{eqnarray}
\sigma_{ij} &=& -p_{0s} \delta_{ij} 
              + 2\mu u_{ij} + \lambda u_{kk} \delta_{ij} \>,
              \label{hooke1} \\
  p         &=& p_{0\ell} - \lamtil u_{kk} \>, \label{hooke2}
\end{eqnarray}
and require the equilibrium pressure to be $p_0$.  
For a planar interface, this fixes the normal stress in $z$ direction 
to be $\sigma_{zz}^{({\rm eq})} = -p_0$. If we assume the equilibrium
stress tensor to be isotropic (a very natural assumption in most cases),
we have $\sigma_{ij}^{({\rm eq})} = -p_0 \delta_{ij}$, and 
$u_{ij,s}^{({\rm eq})}$ in the solid is given by (\ref{strain2}).
 In the liquid,  we have
$u_{ii,\ell}^{({\rm eq})} = (p_{0\ell}-p_0)/\tilde\lambda$. Note that only
the displacement divergence $\nabla {\bf u} = u_{ii}$ appears in the elastic energy 
of the liquid. This gives us a degree of freedom (neither 
$u_{xx}$ nor $u_{zz}$ are fixed separately in the liquid, only their sum is) that will turn out
important later. (Without this degree of freedom, it would not be feasible to 
treat the liquid as a shear-free solid,
 as will be discussed in appendix \ref{APPSHARPINT}.)

Inserting these equilibrium values into (\ref{phirelax1})
and (\ref{stressrelax1}), we obtain as basic
equation of motion for the phase field (introducing the abbreviation $\ktil = k\rho_s$)
\begin{equation}
{\partial \phi \over \partial t} =\frac{\gamma}{\ktil} \biggl\{ \nabla ^2 \phi 
-  {1\over\epsilon^2} \biggl[2 g'(\phi)
 + \frac{\epsilon}{3\gamma} h'(\phi) \Bigl( \mu \, u_{ij} u_{ij}
 + \frac{\lambda-\tilde\lambda}{2}  u_{ii}^2 
+ \Delta p \, u_{ii} + \Delta W 
 +\Delta\rho g\,(z-z_0)\Bigr)\biggr] \biggr\}
\label{phasebas}
\end{equation}
where we have %set $z_0=0$ for convenience and 
defined further abbreviations
\begin{equation}
\Delta p = p_{0\ell}-p_{0s} \label{delp}
\end{equation}
and 
\begin{equation}
\Delta W = \frac12 d \frac{(p_0-p_{0s})^2}{2\mu+\lambda d} 
- \frac12  \frac{(p_0-p_{0\ell})^2}{\tilde\lambda} 
- \frac{2\mu+\lambda}{8 \mu (\mu+\lambda)} \sigma_{00}^2
\>. \label{delw}
\end{equation} 
The elastic problem can be cast into the suggestive form
\begin{equation}
0=\frac{\partial}{\partial x_j} \biggl \{h(\phi) \sigma _{ij} -
[1-h(\phi)]  p\delta_{ij} \biggr\} \>,
\label{stressbas}
\end{equation}
from which it is even more transparent
 that the expression in braces is nothing but a generalized stress tensor.
Note that the phase-field model always guarantees exact mechanical
equilibrium with respect to this stress tensor, but that the validity
of a linear  relationship between strains and generalized stresses
is only warranted outside the
interface region, where the values of $\phi$ cease to depend on the 
$u_{ij}$. (This means that in the vicinity of sharp groove tips
we will automatically have deviations from Hooke's law, albeit they 
are not modeled to satisfy a particular nonlinear constitutive relation.)

These equations are to be solved subject to the conditions that
the phase field approaches its limiting values 
in the bulk phases. To make them closed equations,
we have to replace $\sigma_{ij}$ and $u_{ij}$ by
the field variables $u_i$ using 
 the definition of the strain tensor,
\begin{equation}
u_{ij} = \frac12 \left(\frac{\partial u_i}{\partial x_j} 
                      +\frac{\partial u_j}{\partial x_i}\right)\>,
                      \label{straindef}
\end{equation}                      
and Hooke's law.

It remains now to be shown
that this model reproduces the sharp-interface limit when
the width of the interface is small. This calculation is given in the 
appendix. Its central result is formula (\ref{finres2}), which we rewrite
here in  nondimensional form (for $\sigma_{00}=0$): 
\begin{equation}
\tilde{v}_n = -\left\{ \frac12 \frac{(\sigma_{tt}-\sigma_{nn})^2}{\sigma_0^2}
+\tilde\kappa + \frac{\ell_1}{2\ell_2} (\tilde\zeta-\tilde\zeta_0) \right\} \>.
\label{vntil}
\end{equation}

The ansatz proposed in \cite{Mueller99} is slightly different.
One difference mainly concerns the {\em interpretation\/} or philosophy 
of the approach. 
%and may may be characterized as follows. 
In the MG model, the phase field is
considered the variable determining the 
shear modulus. The shear modulus is {\em the\/} macroscopic quantity
deciding whether a piece of condensed matter is solid or fluid.
Hence, the phase-field order parameter
differentiates between liquid and fluid and has a transparent
meaning in the context of liquid-solid transitions. 
Of course, the model can be extended easily to the case of two solids
with non\-vanishing shear moduli on both sides of the interface. 
In the KM approach, the traditional and more conventional view is
taken that the phase field decides between two phases characterized
by their respective free energy densities. That one of these phases
is a liquid is of secondary importance, as it were. 
Again, in principle, it might be another solid. 
% (but then a different transport mechanism
%for the Grinfeld instability would have to be invoked).
Of course, {\em if\/} the second phase is chosen a liquid, then its
shear modulus must vanish. And indeed, this is guaranteed in 
the current form of both models by construction. 
%Both approaches discussed guarantee by construction that the shear 
%modulus in the soft liquid phase is zero, whereas it stays nonzero and
%constant in the hard solid phase. Hence the solid phase supports shear
%whereas the liquid phase does not. In the original form of the MG model
%this is very conspicuous as the only elastic modulus multiplied by a
%function of the phase field (which vanishes in the liquid)
%is the shear modulus. Therefore, the phase-field order parameter has 
%a transparent meaning in the context of liquid-solid phase transitions.
For ease of further
comparison, we give the phase-field equations of \cite{Mueller99}
in appendix \ref{APPMGMAPP} and show how they are mapped onto the 
 form (\ref{phasebas},\ref{stressbas}).

In concluding this section, we would like to comment briefly on the consequences of an
anisotropic equilibrium strain. Suppose we submit a body consisting of piezoelectric material
to a homogeneous electric field. 
(Alternatively, we could consider some magnetrostrictive
material under the influence of a magnetic field.)
This body will contract or expand until it reaches a new equilibrium state compatible with
the body forces exerted by the field. 
The new state will have anisotropic strain and,
assuming isotropic elastic properties, an anisotropic stress tensor as well\cite{footnote5}.
 What will the
surface dynamics of such a body be, if uniaxial stress is applied in addition, as in the
setup of the Grinfeld instability? Of course, the assumption that the equilibrium stress
remain constant is an oversimplification now, since the dielectric properties of the 
solid and its melt will usually differ, hence the electric field would become inhomogeneous
as soon as an interfacial shape change occurs. Let us nevertheless assume the simplest
possible situation, an anisotropic but constant equilibrium state
\begin{equation}
\sigma_{ij}^{({\rm eq})} = -p_0 \delta_{ij} + \chi_0 \delta_{ix} \delta_{jx} \>. \label{anisoeq}
\end{equation}
Using the stress-strain relationship (\ref{hooke1}),
this can be inserted into our expression for 
the elastic energy density of the solid,
which then becomes
\begin{equation} 
%\begin{eqnarray}
f_{{\rm sol}}(\{u_{ij}\}) = \mu u_{ij} u_{ij} + {\lambda\over 2}  u_{ii}^2 
     +\Delta p\> u_{ii} + \frac12 \frac{\Delta p^2}{\mu+\lambda} 
%\nonumber \\ && \mbox{}
 - \chi_0 u_{11} - \chi_0  \frac{\Delta p}{2(\mu +\lambda)} 
 + \chi_0^2 \frac{2\mu +\lambda}{8\mu(\mu+\lambda)} \>, \label{enaniso}
%\end{eqnarray}
\end{equation} 
where we have set $p_{0\ell}=p_0$ for simplicity.
% $\leadsto$ $p_0-p_{0s}=\Delta p$. %[compare (\ref{hooke3})].

It is then straightforward to derive the sharp-interface limit for this modified phase-field
model. The result reads (on setting $\sigma_{00}=0$)
\begin{eqnarray}
v_n &=& - \frac1{k\rho_s}\Biggl\{ \gamma \kappa + \Delta\rho g z +\frac{1-\nu^2}{2E}
\left(\sigma_{tt}-\sigma_{nn}-\chi_0\right)^2 \nonumber\\
&&\mbox{} + \left[\frac{2(1-\nu^2)}{E} \chi_0 (\sigma_{tt}-\sigma_{nn})
                  - \frac{1+\nu}{E}  \chi_0^2\right] n_x^2 + \frac{2\nu(1+\nu)}{E} \chi_0^2 n_x^4
 \Biggr\} \>, \label{vnaniso} 
\end{eqnarray}
where $n_x$ is the component of the interface normal in $x$ direction. Rotational invariance
is broken. 

We are not aware of any previous mention of this equation in the literature, nor do we 
think this interesting case has been treated. In fact, what we have demonstrated here, is
how the phase-field model can be
 used to {\em derive} hitherto unknown sharp-interface
equations in a transparent way. 

It is clear from (\ref{vnaniso}) that an isotropic stress tensor, i.e.,
 $\sigma_{tt}=\sigma_{nn}$ does not necessarily
entail a stable planar interface,
whereas setting $\sigma_{tt}-\sigma_{nn}=\chi_0$, 
i.e., providing an
{\em anisotropic} stress tensor, 
we will have a linearly stable planar front solution
with interface position $z=0$. This is easily seen from the fact that the
terms containing $n_x^2$ and $n_x^4$ do not contribute in a linear stability
analysis, because $n_x$ is directly proportional to the perturbation
and hence its square and fourth power have to be dropped.
Note also that the symmetry of the dynamics with respect to a replacement
of $\sigma_{tt}-\sigma_{nn}$ by its negative value
 does not hold anymore in this situation.

While this equation opens a new line of research, we will refrain here from  pursuing
 this topic any further.

\section{Numerical results}

\subsection{Validation of the model}

In order to verify that our phase-field description leads to a quantitatively correct
description of the instability, at least before cusps set in,
 we have performed a number of 
numerical tests.
Based on a simple finite-difference
scheme, the numerical implementation
is set up in a rectangular geometry. 

The bottom half of the rectangle is filled with solid, 
the top with liquid. This is realized by setting  
the phase field $\phi$ equal to a tanh-like function taking the value
 one in the bottom region and
 zero in the top region of the geometry. $\phi$ is kept at
these values one and zero exactly
at the bottom and top lines of the numerical grid, respectively. 
Periodic boundary conditions are applied at the lateral boundaries. 
 The initial interface is set by an appropriate 
modulation of the region where $\phi$ 
crosses the value $\frac12$
 and was in most cases taken to be sinusoidal or 
flat with a random perturbation.

The boundary and initial conditions for the fields 
$u_x$ and $u_z$ are chosen differently for the KM and MG models
 as will be described now.

Within %our straightforward naive approach, 
the KM model, where % which
we assume strains to vanish at equilibrium (hence $\Delta p = 0$),
we  took the $x$ {\em derivatives} of both displacement fields periodic
in the $x$ direction in our initial simulations, 
in order to obtain periodic
{\em strains}. Later, we switched to simpler {\em helical} boundary 
conditions for $u_x$, i.e., we took
$u_x(L,z) =u_x(0,z) + L  u_{xx,0}$, where $L$ is the 
length of the rectangle along the $x$
direction, and periodic boundary conditions for $u_z$.
This change in boundary conditions did not affect results
in any essential way.
All the simulations of the KM model discussed here were carried out
 with these boundary conditions
(whereas those in \cite{Kassner99} were done with periodic $x$ derivatives).
At the  bottom of the system, 
the values of the fields are fixed to values
corresponding to a homogeneously strained solid;
 at the top, $u_x$ is
fixed and the derivative $\partial_z u_z$ chosen such that  the condition  
$\nabla {\bf u} = 0$
is satisfied. $u_x$ is initialized as a linear function $u_x = x u_{xx,0}$
and the inital $u_z$ is determined
 via integration of Eq.~(\ref{urhorhorhodep}).

For simulations of the MG model (or rather its variant considered here), 
the fields were all
taken periodic in the $x$ direction,
 whereas the boundary conditions in the $z$ direction
were as in the KM model. We did not yet attempt to use spectral methods for the solution
which would require periodicity in the $z$ direction as well (achievable
by simply reflecting  the system
 at its bottom, and  including the
 image into the numerical box  \cite{Mueller99}). 
Initialization was done by setting  $u_x = 0 $ everywhere and computing
$u_z$ from (\ref{urhorhorhodep}) again.

The elastic equations were solved by successive overrelaxation,
 the time integration was
performed by a formally second-order accurate midpoint scheme. 
%(This accuracy is formal only,
%because 
Since we did not update the elastic fields at the half time step,
the formal accuracy was not attained. 
The most time-consuming
part of the simulation was the relaxation scheme 
and a way to overcome its
restrictions has been given in \cite{Mueller99} as is
discussed in appendix \ref{APPMGMAPP}. Since it requires an 
approximation to the solution of the elastic problem even at the analytic
level ($\mu/K$ has to be small), we did not implement it in our two-dimensional
simulations. We intend to compare the quality of this approximation to 
the solution of the full problem, before employing it in a  3D simulation,
where its use is essential for reasons of computational efficiency.
Most of our computations were done with
the material parameters of Helium to facilitate comparison with experiments
Therefore, whenever we do not indicate different choices, our parameters
were chosen as described in 
 \cite{footnote2}. Times and lengths  given without units
 are in seconds and centimetres, respectively. Since our nondimensional
time unit is about one second and the nondimensional length scale about
0.1~cm, this simply corresponds to using 10$\ell_1$ instead of
$\ell_1$ as the basic length scale.

One of our numerical tests
 consisted in reproducing the instability threshold to within 2\% accuracy,
another one in verifying the subcritical nature of the bifurcation, first demonstrated
analytically
by Nozi\`eres \cite{Nozieres93}. A short discussion of the last feature has been given
in \cite{Kassner99}, so we will not elaborate on it here \cite{footnote6}. We
 consider a few more tests, however. 

Figure 1 gives the dispersion relation determined for three values of 
the external stress
and compares it with the analytic result from linear stability theory.
The KM model was used here as it gave more accurate results at finite
$\epsilon$. %, a fact on which we will comment in more detail below.

To obtain the dispersion relation, we simply followed the dynamics of a
system initialized with a small-amplitude cosine profile for a number
of different wavelengths, and computed the amplitudes of the evolving
structure for a series of times. Then the amplitudes were fitted to 
an exponential function which provided the growth rate of the interface.
Taking $\sigma_{00}$ equal to the applied stress $\sigma_{0}$, we fixed
the average position of the interface. Moreover, the amplitudes were
computed in two different ways, both of which are not influenced by
the average interface position. The first method was simply to take the
square root of the spatial variance of $\zeta(x)$; as a second measure
for the amplitude we took the modulus of the Fourier component 
corresponding to the wavelength chosen. On the figure, these
two methods give essentially indistinguishable results within the size
of the symbols. System sizes used were the wavelength $\lambda_f$
of the fastest-growing mode and a number of rational multiples and
fractions thereof (ranging from $\frac14\lambda_f$ to $3\lambda_f$).
Since we kept the number of numerical grid points the same for all
the systems at $\lambda_f$, the mesh size had to be varied. 
The interface thickness $\epsilon$ was in general kept above
$\frac32$ of the mesh size, which gives a resolution of 5 points
for the region where the phase field varies between 10\% and 90\%
of its maximum value. For smaller values of $\epsilon$, locking effects
to be discussed shortly became conspicuous \cite{Boesch}.

The agreement between analytic results and numerically determined
points is satisfactory both above ($\sigma_0 = 2.8 \times 10^4$ dynes/cm$^2$)
and below  ($\sigma_0 = 2.4 \times 10^4$ dynes/cm$^2$) the instability 
threshold. Two points are worth mentioning. First at $q \approx 30$ cm$^{-1}$,
there are {\em two} symbols each for the growth rates corresponding to
the two larger  stresses. These were given to roughly indicate the
possible error in
the numerical result when %for $q$ values where 
the growth rate has a large
negative value. Points below $q\approx 20$~cm$^{-1}$ did not show a
comparable error. The two different values were obtained by fitting with
the initial and the final half of the data points, respectively.
%the former giving the better result. 
We ascribe the difference to the fact
that the amplitude of the interface becomes smaller 
than its width $\epsilon$, a
situation in which the phase-field description is no longer reliable. 
For example, the final planar interface is not located exactly at
$z=0$, about which the initial cosine was centered, albeit the deviation is 
smaller than the interface width. Second, the overall agreement is surprisingly
good in view of the fact that a phase-field model is not particularly
well-suited to the determination of a dispersion relation at all. For in order
to approach the limit of an infinitesimal perturbation of a planar 
interface one should choose very small amplitudes, but they must not
be smaller than the interface width $\epsilon$. Reduction of 
 $\epsilon$ is possible in principle but soon leads to prohibitive 
computation times. With a sharp-interface model that we investigated in
parallel \cite{Kappey2000}, it was no problem to take amplitudes
of $10^{-4}$ and to obtain nice exponential growth or decay during
long time intervals, whereas here we were restricted to starting
amplitudes on the order of 0.05 or larger.

\begin{figure}
\centerline{ \epsfig{file=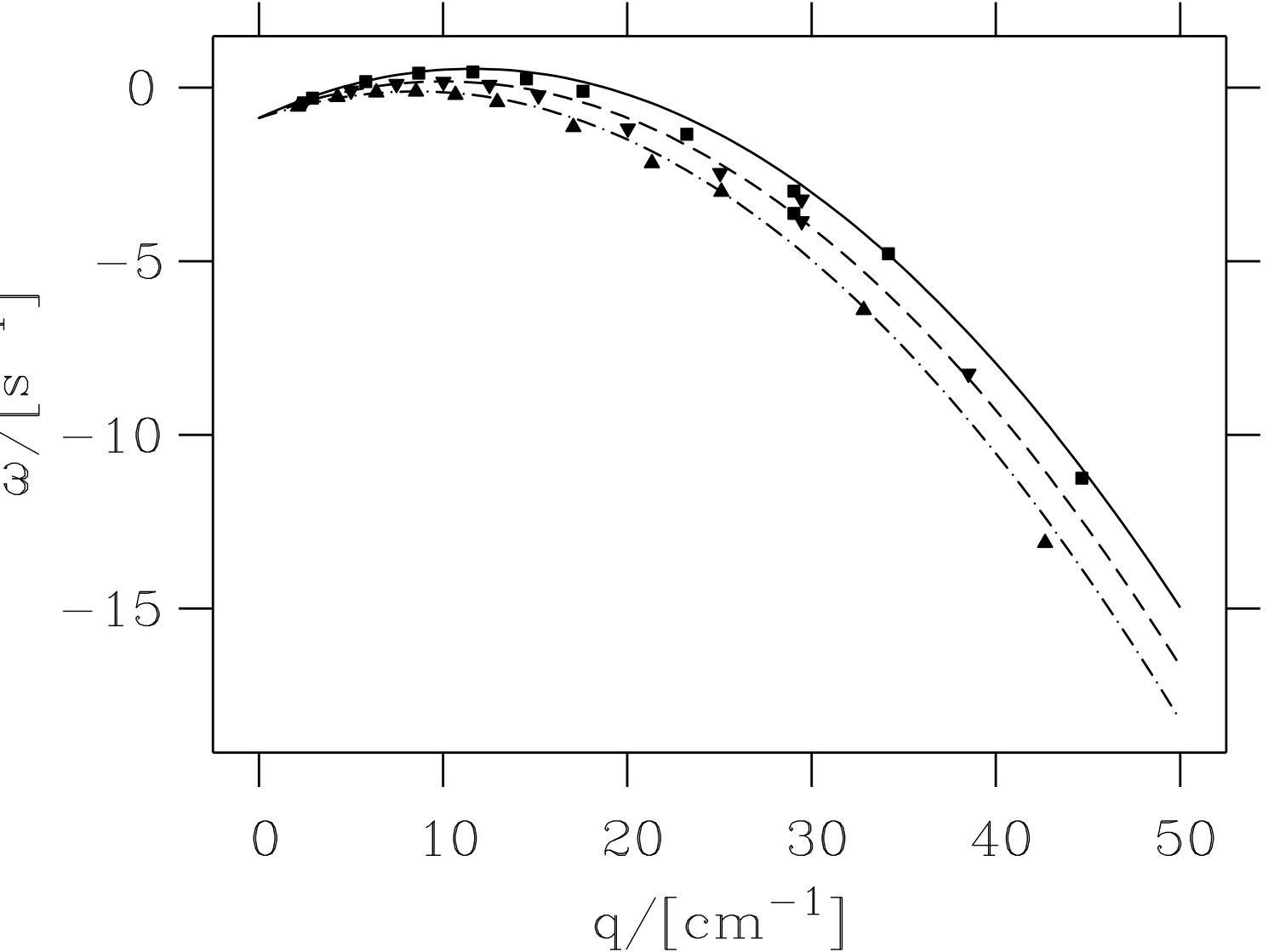,height=16cm,angle=90} }
\caption{Dispersion relation. Symbols indicate the results of numerical
simulations, lines depict the analytic theory. Material parameters are
those of Helium. 
Solid line and squares: 
$\sigma_0 = 2.8 \times 10^4$ dynes/cm$^2$, $\epsilon=0.009$; 
dashed line and inverted triangles: $\sigma_0 = 2.6 \times 10^4$ dynes/cm$^2$,
$\epsilon=0.01$;
dash-dotted line and  triangles: $\sigma_0 = 2.4 \times 10^4$ dynes/cm$^2$,
$\epsilon=0.012$. Mesh sizes: $h=0.0054$, $h=0.0063$,  $h=0.0074$, respectively.
}
\end{figure}
 
Our next test consists in investigating the dynamics of a planar
interface with both the KM and MG models.
%In Figure 2, we investigate 
%the dynamics of a planar interface with both models;
{}{}From (\ref{dmu1}, \ref{vn}) we obtain the equation of motion

%\textsl{had to take hlatt=0.002, eps =0.008, at hlatt = 0.0069 and eps = 0.011
%strong locking effects, MG goes to wrong final value}

\begin{equation}
\dot\zeta = -\frac{1}{k\rho_s} \left({(1-\nu^2)\over 2E}\sigma_0^2 + \Delta \rho g \zeta\right) \>,
\end{equation}
which is, given the initial condition $\zeta(0)=0$, solved by 
\begin{equation}
\zeta(t) = - \left( 1- e^{- \Delta \rho g t/k\rho_s} \right)
 {(1-\nu^2)\over 2\Delta \rho gE}\sigma_0^2 \>.
\label{planaranalytic}
\end{equation}
This analytic result is compared with simulations of the two models 
in Fig.~2.

%We plot $\zeta(t)+ \ell_2$ logarithmically to display  the exponential dependence on time
%and compare with the analytical slope.

What is cleared up by  the figure is that even with a well-resolved
interface width (we have $\epsilon = 4 h$) the MG model is slightly
off the analytic final position, whereas the KM one converges well
towards it. With larger values of the numerical
mesh size, convergence of the former model
gets even worse. For  $h=0.007$, $\epsilon=0.011$
the KM model still agrees reasonably well with 
the analytic curve while the MG one is off by about 10\% for
$t = 4$. Both models show deviations from exponential behavior with
this set of parameters due to metastability effects of the discrete set
of interface points.
This problem, which is particularly critical for interface
pieces parallel to one of the coordinate axes, has been discussed in
detail in \cite{Boesch}. At small interface velocities,
the sum of the energies of the discrete points of the phase field
in the double well potential may vary at successive time steps
(whereas the energy of a continuous field is degenerate under
arbitrary translations). Therefore, the interface is
slowed down, if the energy of its discretization increases due
to motion and accelerated if it decreases. For sufficiently small
driving force, the interface may stop moving at all, i.e., lock
into some favorable position.
Apparently, the MG
model is more susceptible to these effects than the KM one.

The ultimate reason for the different behavior of the two models
is that they are only asymptotically equivalent, i.e., they describe
the same system only in the limit  $\epsilon \to 0^+$. For any finite
  $\epsilon$, the equations obeyed by the phase field and the 
displacements are not the same in the two models. One can observe
this directly by comparing the different terms contributing to, e.g.,
$\partial_t \phi$. In the MG model, the term $\Delta p \, u_{ii}$
of eq.~(\ref{phasebas}) is frequently the largest interface term affecting 
$\partial_t \phi$, whereas this term is equal to zero in the 
KM model. Moreover, the sum of all terms multiplying $h'(\phi)$ is not
the same in both equations.

\begin{figure}[h]
\centerline{ \epsfig{file=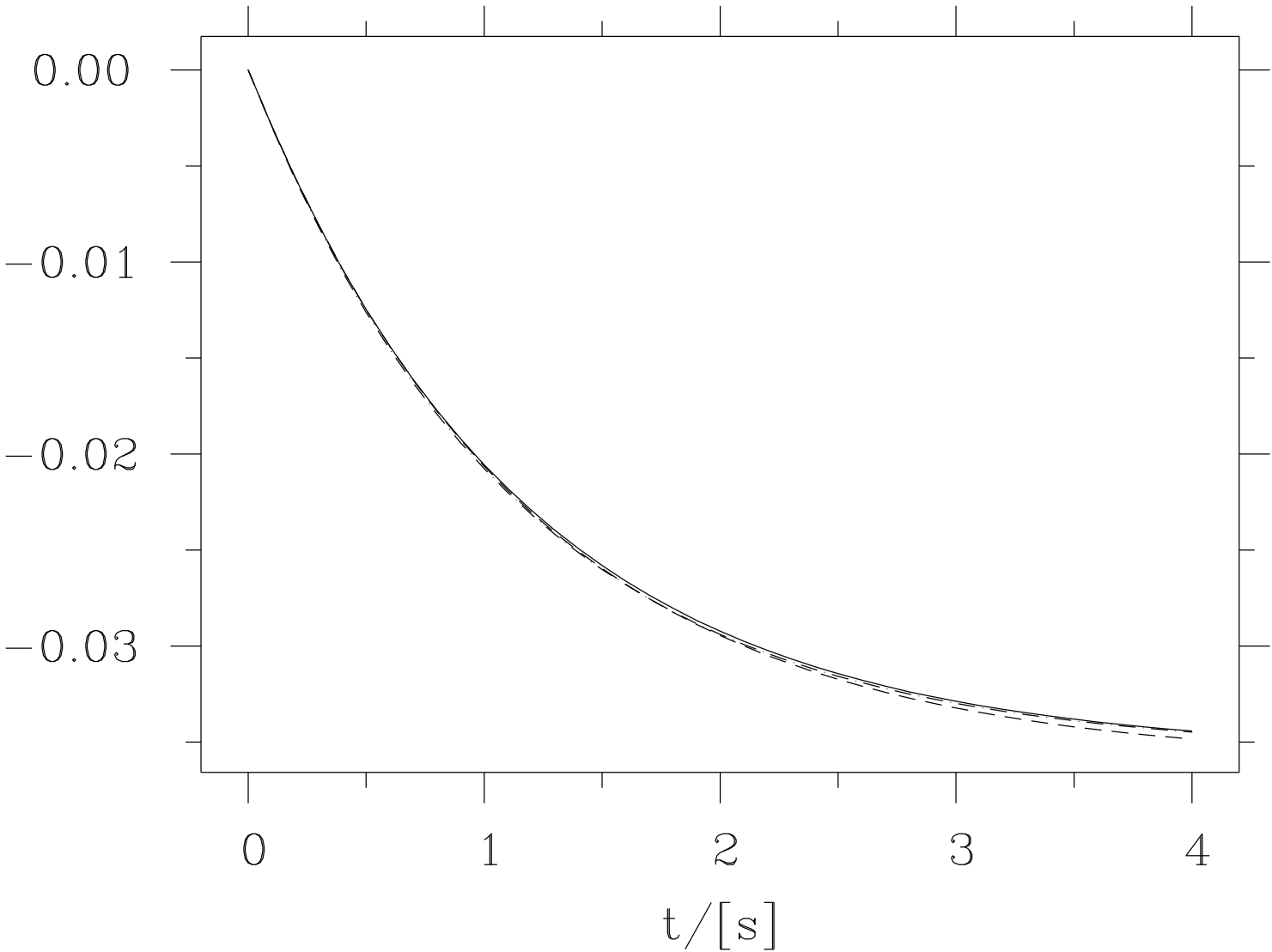,height=16cm,angle=90} }
\caption{Recession dynamics of a planar interface. Solid line: KM model;
dashed line: MG model; dash-dotted line: analytic result 
(\protect\ref{planaranalytic}). The dash-dotted line is hard to see;
it almost coincides with the solid one. To discern it, one should look at the
left part of the figure.
$\sigma_0=2\times 10^4$ dynes/cm$^2$, $\epsilon = 0.008$, $h=0.002$.}
\end{figure}

The difference can also be seen in comparing a numerical simulation of the
 the sharp-interface model (\ref{dmu1},\ref{vn})
 itself, using an integral equation approach, with the phase-field 
models.  We will report on 
details of this alternative approach 
elsewhere \cite{Kappey2000}.  Figure 3 shows the interface 
evolving in the phase-field calculation for two different values of
$\epsilon$ and compares them with the sharp-interface result starting
from the same initial condition, after the same time interval.

Again the KM model fares slightly better in the comparison for the same
value of $\epsilon$. In the groove, however, both models deviate
from the sharp-inteface result but approach it more closely for
the smaller interface thickness. The sharp-interface model produces
a more strongly pointed groove, as expected. It should be emphasized
that this simulation is not far from the limit of the temporal 
validity range of the sharp-interface model. This limit is 
signalled either by a crash of the program due to the singular
behavior of the bottom of the groove or by the appearance of a
spurious steady state, which can be achieved by overstabilization of
the interface.

Why the KM model agrees better with the sharp-interface model for a given
value of the interface width is a difficult question, to which we cannot
offer any deep answer. Also we cannot exclude that for a different choice of
the functions $h(\phi)$ and $g(\phi)$, the MG model would be superior.
It should be kept in mind that the functions employed in \cite{Mueller99}
are not the same as those used here (see App.~\ref{APPMGMAPP}).

Normally, we would thus prefer to use the KM model  for calculations to
be presented. However, since we wish to make sure that effects
of translational symmetry breaking are not due to our using a 
model in which periodicity is imperfectly implemented, we will use
the MG variant in  the following simulations.
The differences between the two models are small, after
all. Also  the MG model has the advantage to be
more easily treated using pseudospectral methods based on Fourier series
due to its periodic boundary conditions,
with a gain in accuracy that might allow to offset its apparent
disadvantage.

\begin{figure}[h]
\centerline{ \epsfig{file=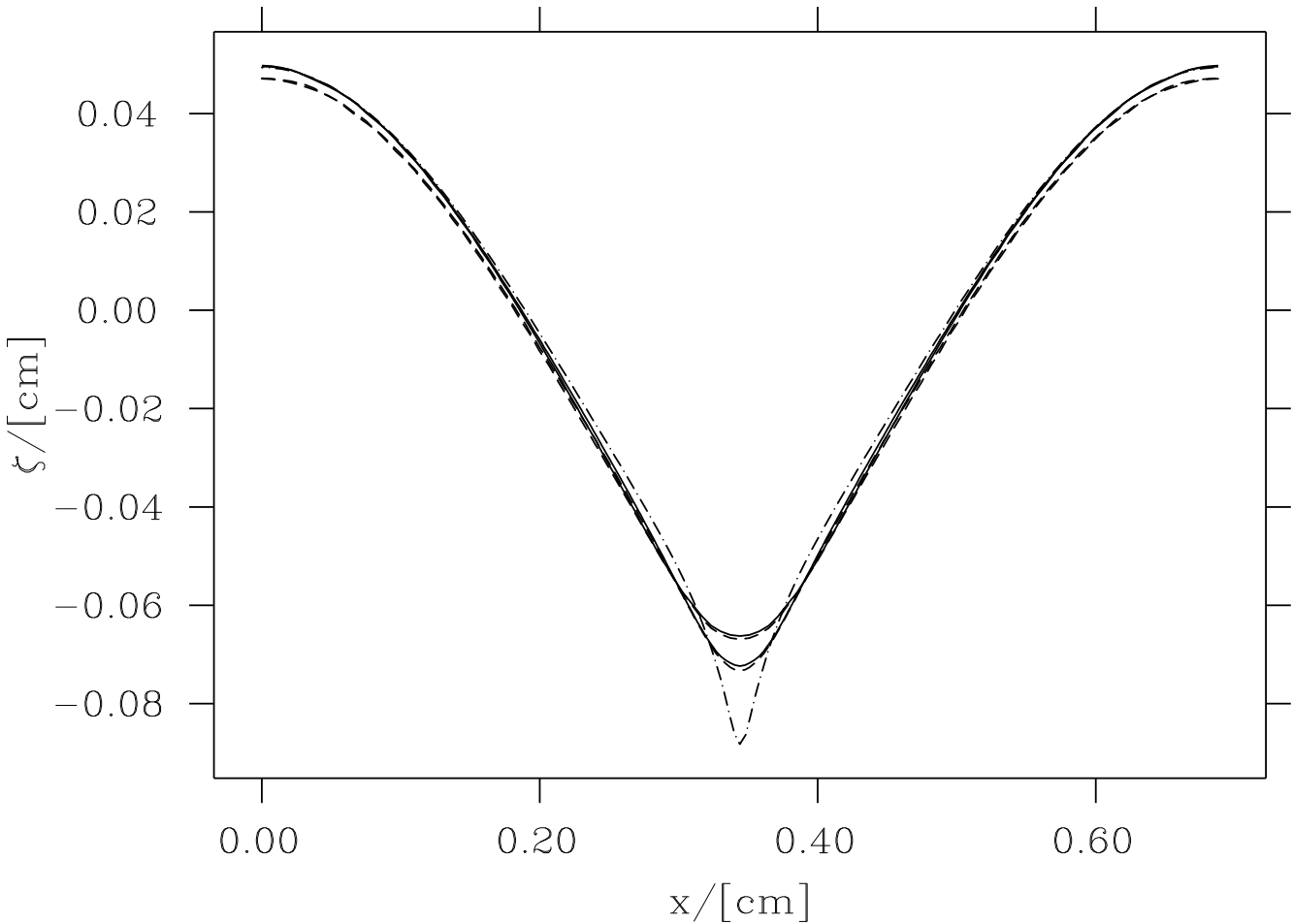,height=14cm,angle=90} }
\caption{Comparison of phase-field models for different 
interface widths
with the sharp-interface model. Solid lines: KM model; dashed lines: MG model;
dash-dotted line: sharp-interface model. The phase-field
curves with the shallower minima
correspond to $\epsilon=0.01$ (mesh size $h=0.0688$), those with 
the deeper minima to
 $\epsilon=0.005$ ($h=0.0344$). $\sigma_0 = 2.5\times 10^4$ dynes/cm$^2$,
$t=0.25$. Initial interface amplitude: 0.05.}
\end{figure}

% It must be emphasized, however, that the sharp-interface
%code crashes shortly after this time and we cannot be 100\% 
%sure that %there is indeed an acceleration of the tip of the groove
%the interface really takes a turn downward before becoming cuspy. This is
%because our sharp-interface code does not have spectral accuracy. The simulations of 
%Spencer and Meiron\cite{Spencer94}, in which spectral accuracy was achieved rather show
%morphologies, which develop an almost straight slope that extends to the point, where
%a cusp is formed. {}From the analytic calculations of Gao \cite{Chiu93}, one would suspect
%this shape to essentially be part of a self-crossing cycloid. On the other hand, the
%particular behavior may well depend on parameters, and we do observe this kind of bending
%down also in phase-field simulations at larger external stress (see below).

The conclusion from Fig.~3 is that the phase-field models
 give decent agreement with a
sharp-interface calculation in regions where the curvature is not too large.
Whereas the sharp-interface computation 
cannot be meaningfully continued by very much beyond the  time
shown in the figure ($t=0.25$), 
the phase-field models both have no problem in continuing
the simulation to times well beyond $t=1$. 

As anticipated above, we  take the point of view 
that a real solid cannot develop exact
cusps, because plastic effects such as the 
generation of dislocations  will intervene.
These will relieve stresses and thus prevent 
infinite densities of the elastic energy.
The phase-field model  does the same thing and we shall see below that
it does so by introducing a cutoff to the curvature.
 More quantitative modeling would require
to explicitly take into account models of nonlinear elasticity or plasticity, 
which is beyond the scope of this article.
Nevertheless, as we can see from Fig.~3, the behavior far from
the sharp tip of the groove is described reasonably
well by the phase-field model
for both values of $\epsilon$ and is almost independent of the interface
width. Therefore,
we believe that the phase-field approach correctly reproduces the qualitative
behavior of a situation in which plastic effects occur only in 
the minima of the grooves.

Results obtained under this hypothesis will be discussed in the next section.

\subsection{Dynamics of extended systems}

When simulating periodic structures, one realizes 
that for small supercritical stresses,
where the system takes a long time to develop deep grooves,
 one often observes symmetry breaking
and one of the grooves getting ahead of the others. 
This symmetry breaking must be triggered
by numerical noise from roundoff or truncation errors. For  high stresses,
where the system develops grooves reaching the system bottom 
within a relatively short lapse of time,
this does not happen. Figure 4 gives an example of a structure
 grown at about three times
the critical stress. The interface is plotted at constant
time increments ($\Delta t=0.05$).
 A shift in the chemical potential of the 
solid has been made to keep the
position of a planar front fixed.

\begin{figure}[h]
\centerline{ \epsfig{file=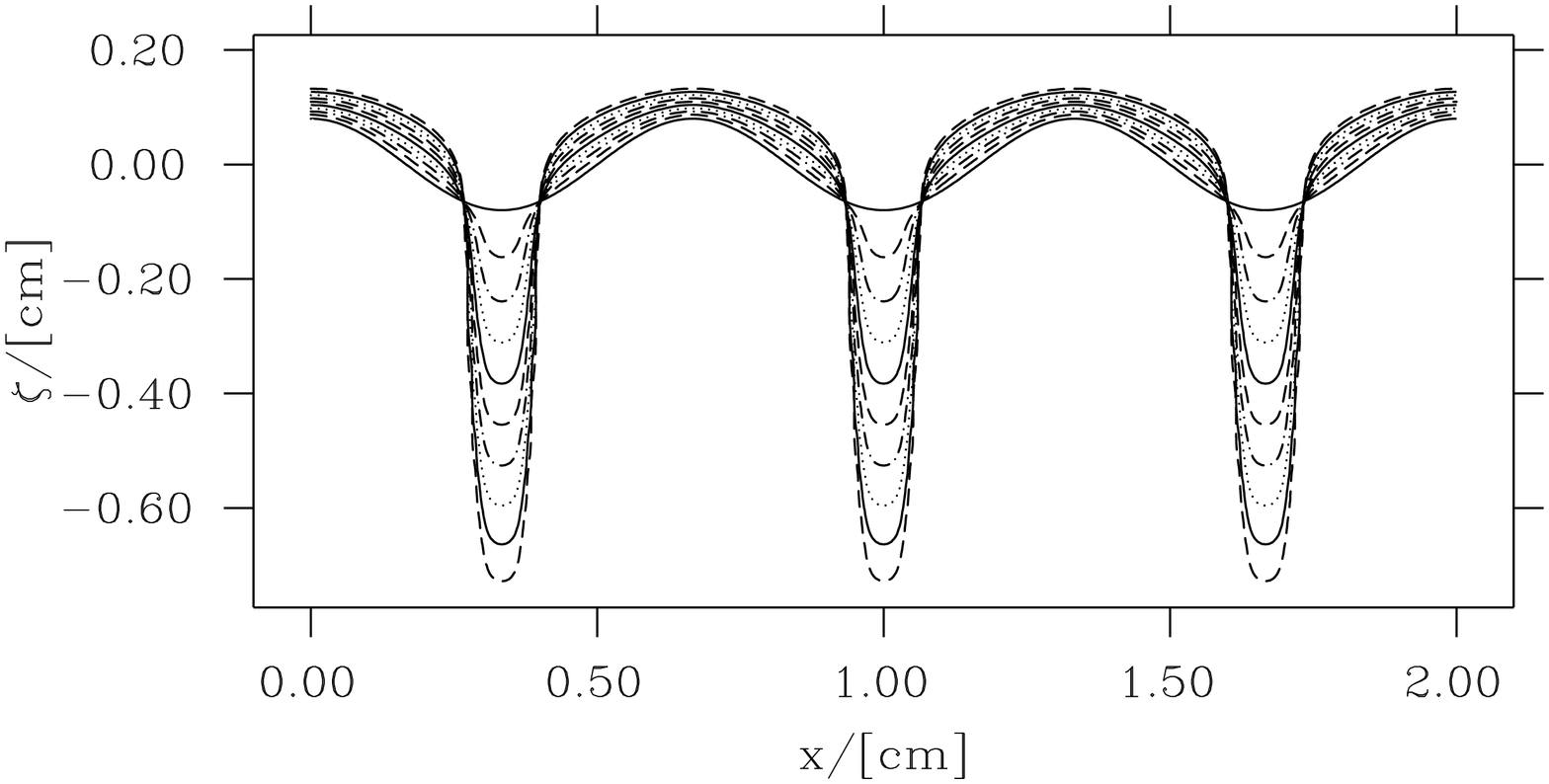,height=15.0cm,angle=90} }
\caption{Dynamics for $\sigma_0=8\times10^4$ dynes/cm$^2$,
$\epsilon=0.02$. MG model
with prestress $\sigma_{00}=8\times10^4$ dynes/cm$^2$.
The fastest-growing mode is at $\lambda_f=0.066$, the wavelength of the
pattern is $\lambda=\frac23$. Note that 
%(contrary to most other figures) 
the two axes are the same scale.}
\end{figure}

 We see that the structure remains 
 periodic
in the time interval considered and that three 
equally deep grooves evolve. Note the
peculiar shape of the cells. 
{}From  flat tips there emerge slightly curved slopes on the side
of the cells. Then there is a sharp bend downward into the deep groove. 
The appearance of this bend renders it plausible
 that the time of formation of a cusp
in the sharp-interface description has already passed and from then on
the dynamics should be governed by the curvature bound. In the final stage of 
this dynamics all grooves move at constant velocity. Figure 5 gives
the curvatures of the interfaces displayed in Fig.~4 and demonstrates 
that the radius of curvature at the bottom of the grooves remains
constant and is close to $\epsilon$, which was equal to 0.02 in this
simulation.

The curvature was calculated from the contour line defining
the interface position. Since the representation of this line ($\phi=0.5$)
was constructed by determining its intersection points with the squares
of the numerical grid, the discretization points
were unevenly spaced (two intersections with 
grid lines parallel to the $x$ and $y$ axes can be arbitrarily
close to each other, the next may be as far away as $\sqrt{2} h$).
Therefore, our curvature results are pretty noisy, even after application
of a smoothing procedure. A superior method
for their determination would  be to use the full representation
or the phase field instead of just the contour line information. 
Nevertheless, they clearly indicate the approximate 
constancy of the curvature in
the groove tips. 

\begin{figure}[h]
\centerline{ \epsfig{file=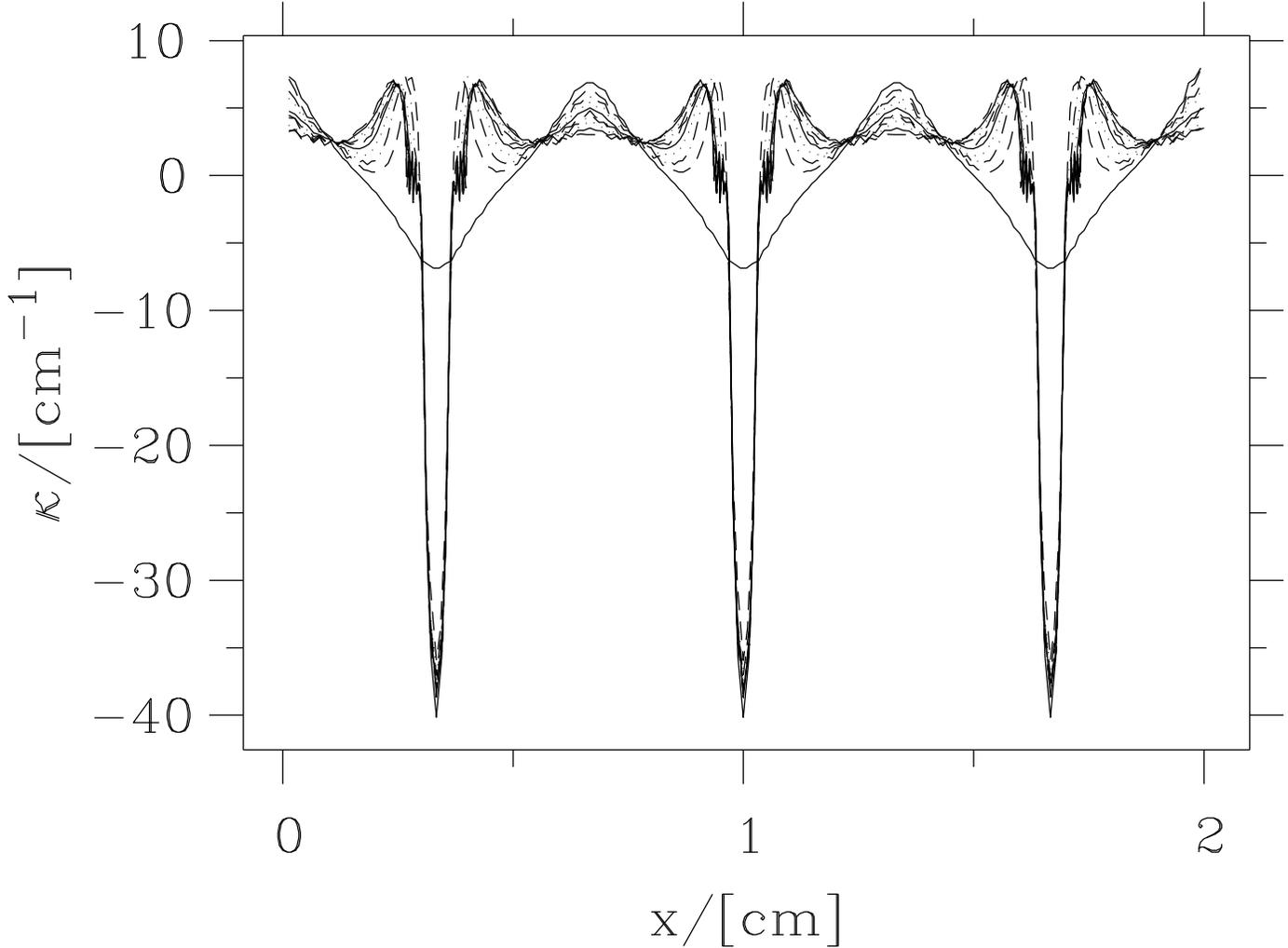,height=15.0cm,angle=90} }
\caption{Evolution of interface curvature, corresponding to the interfaces
of Fig.~4.}
\end{figure}

Since we have the stresses at our disposal, too, we can 
calculate the final velocity of the grooves. Figure 6 gives a contour
plot of the stresses $\sigma_{xx}$, corresponding to the final time
of the simulation from Fig.~4, $t=0.45$. The interface is drawn as
as solid line, the contour lines are broken lines in different styles.
What we have plotted here, is not a generalized stress tensor 
component, as defined by
(\ref{defsigtil}), but simply the stress in the solid. Therefore, the
contour lines for stresses far in the liquid  are  meaningless
(in the dynamic equations, they are multiplied by $h(\phi)\approx 0$),
although they become important when  entering the interface region. 
%or entering the solid. 
{}From the figure, we  estimate a maximum value of
$\sigma_{tt} \approx 2\times 10^5$ in the bottom of the groove
(and a similar value is obtained from the corresponding 
figure for $\sigma_{zz}$). Inserting this, the value of the curvature
and the position $\zeta$ of the groove bottom into (\ref{dmu1}),
we obtain for the velocity $v_n = -1.6$. Assuming that the interface
grew at this velocity from the outset, we obtain for its final
position the value $\zeta=-0.7$. The data show that it is actually
at $\zeta=-0.72$, which is easily explicable by the inaccuracy
of our estimate of the maximum stress. {}From the contour plot we do not
obtain more than a rough figure as stresses vary rapidly in the interface
region.

In a straight and narrow crack, the stress scales with the square root of
the distance from its tip \cite{Chiu93}. Therefore, a reduction of the
tip radius by a factor of two will increase both the stress term and
the curvature term of (\ref{dmu1}) by a factor of two as well. As long
as the gravity term in (\ref{dmu1}) is negligible (which, incidentally,
it is not in the simulation of Fig.~4, its contribution is about
as large as that of the curvature for the last curve), this means that
the velocity of the groove will roughly double when $\epsilon$ is halved.
This trend has been confirmed in the simulations, although the 
observed ratio
is slightly smaller
than the predicted one, but then our grooves do not yet really have an
extremely small width compared with their length.

\begin{figure}[h]
\centerline{ \epsfig{file=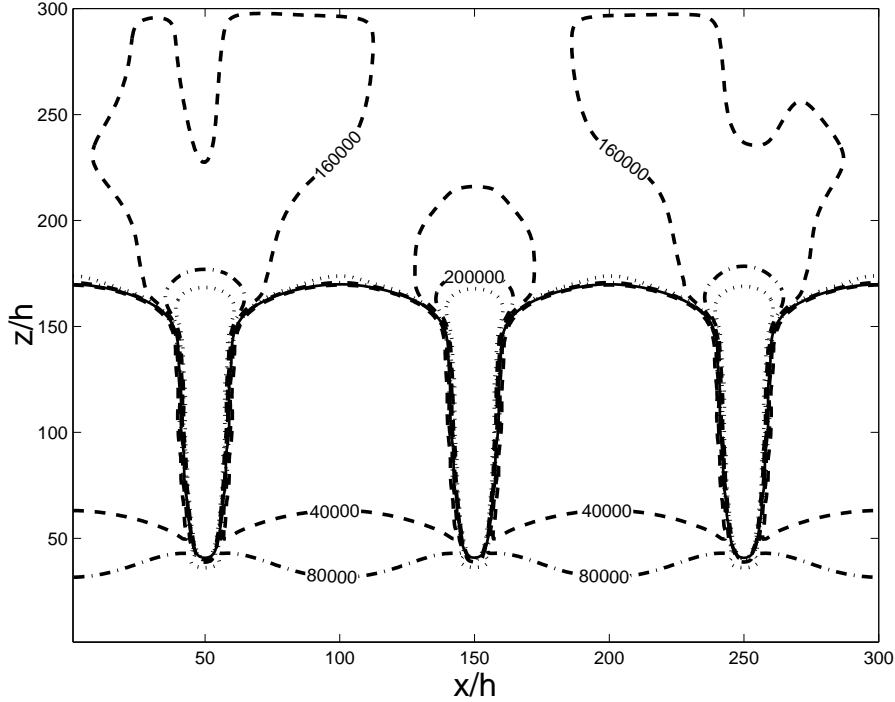,width=12.0cm,angle=0} }
\caption{Stresses  at $t=0.45$ (lowest curve in Fig.~4). 
Six contour values are displayed, the sequence of line patterns is
dashed, dash-dotted, dotted with increasing value of $\sigma_{xx}$
and an increment of $4\times10^4$ dynes/cm$^2$ between successive curves.
To distinguish lines with the same pattern from each other,
the values have been explicitly marked at those curves where there
was enough room, e.g., for $\sigma_{xx}=4\times10^4$ dynes/cm$^2$. The lower 
dash-dotted curve corresponds to the value of the applied external
stress. 
  }
\end{figure} 

The next three figures show a simulation at a stress roughly 20\% 
above the critical value.  Our numerical box contains six
wavelengths of the pattern initially. One of the grooves has 
however been made by 2\%
deeper than its neighbours. Contrary to the situation in Fig.~4,
no prestress was applied, so a planar interface would
move downward to a new equilibrium position. This kind of motion
is superimposed on the shape-changing dynamics and serves nicely
in separating the curves on the plot.

\begin{figure}[h]
\centerline{ \epsfig{file=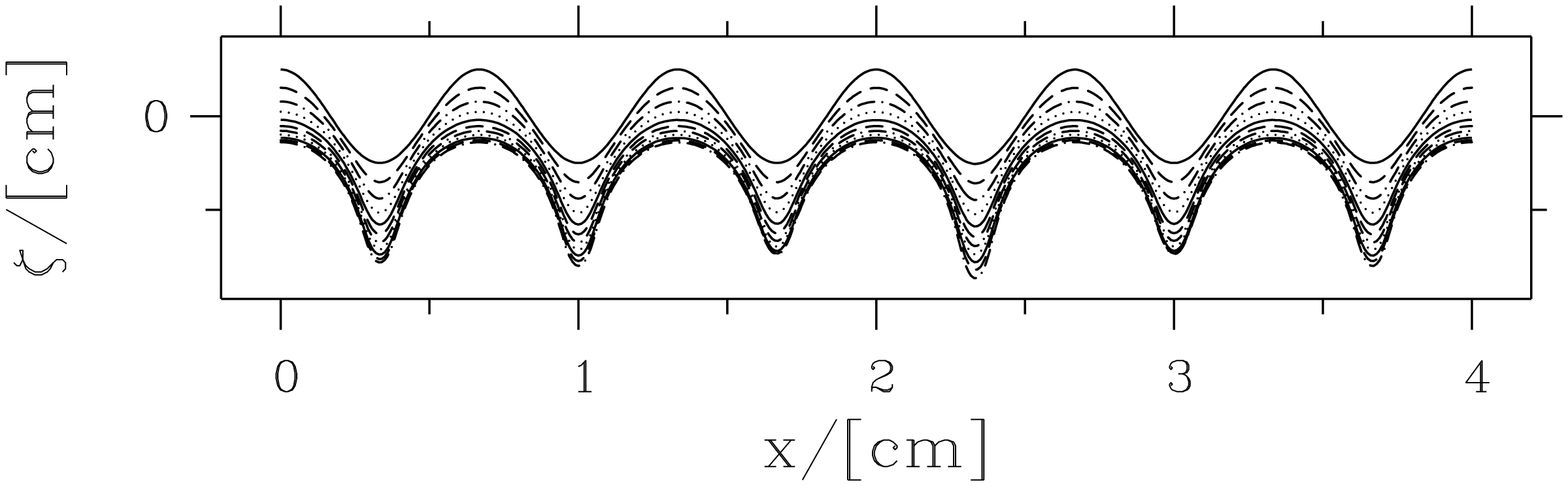,height=15.0cm,angle=90} }
\caption{Dynamical evolution of a perturbed interface at 
$\sigma_0=1.16\,\sigma_c$ ($\sigma_0=3\times10^4$ dynes/cm$^2$,
 $\epsilon=0.025$, 
$E =3.2\times10^8$ dynes/cm$^2$). MG model, no prestress.
%The fastest-growing mode has $\lambda_f = 0.5$, the wavelength of the
%initial pattern (neglecting the perturbation) is $\lambda=\frac23$.
Time interval between curves 0.25; the final time is 2.5.
After an initial phase of acceleration, the interface slows down
and approaches a cycloid-like curve.}
\end{figure}

The temporal dynamics can be divided into several stages.
At first, the sinusoidal pattern changes its shape in the 
way already discussed by Nozi\`eres \cite{Nozieres93}:
the tips become flat, the grooves pointed. After some time,
the interface becomes similar to a cycloid
%, which has been 
%identified by Chiu and Gao \cite{Chiu93} to be an almost
%steady-state structure, 
but with different depths of the grooves. Also, the dynamics
almost comes to a halt. 
Below, we shall discuss
the similarity with a cycloid in more detail (see Fig.~10).
It holds up to $t=2.5$ approximately, which is the
time of the lowest curve in 
Fig.~7. At this point the apparent periodicity of the pattern has 
doubled. (Of course, strictly speaking this periodicity has
been broken from the outset by our making one groove a little 
deeper. But this was only to avoid its being broken by numerical
noise in an uncontrolled manner, i.e., to introduce a well-defined
perturbation.)

The groove that was ahead initially, wins
the competition for the elastic field; the losing grooves fall back
and even close again. This is shown in Fig.~8, displaying
the temporal continuation of Fig.~7. 
In the initial structure of Fig.~8 (the solid line that is shallowest
in the big grooves), the smaller grooves are deeper than in the final
one (the dashed line which is deepest in the big grooves but shallowest
in the small ones). At the end of the period of time depicted
in Fig.~8, there are three clear survivors and three losers of the
competition.

\begin{figure}[h]
\centerline{ \epsfig{file=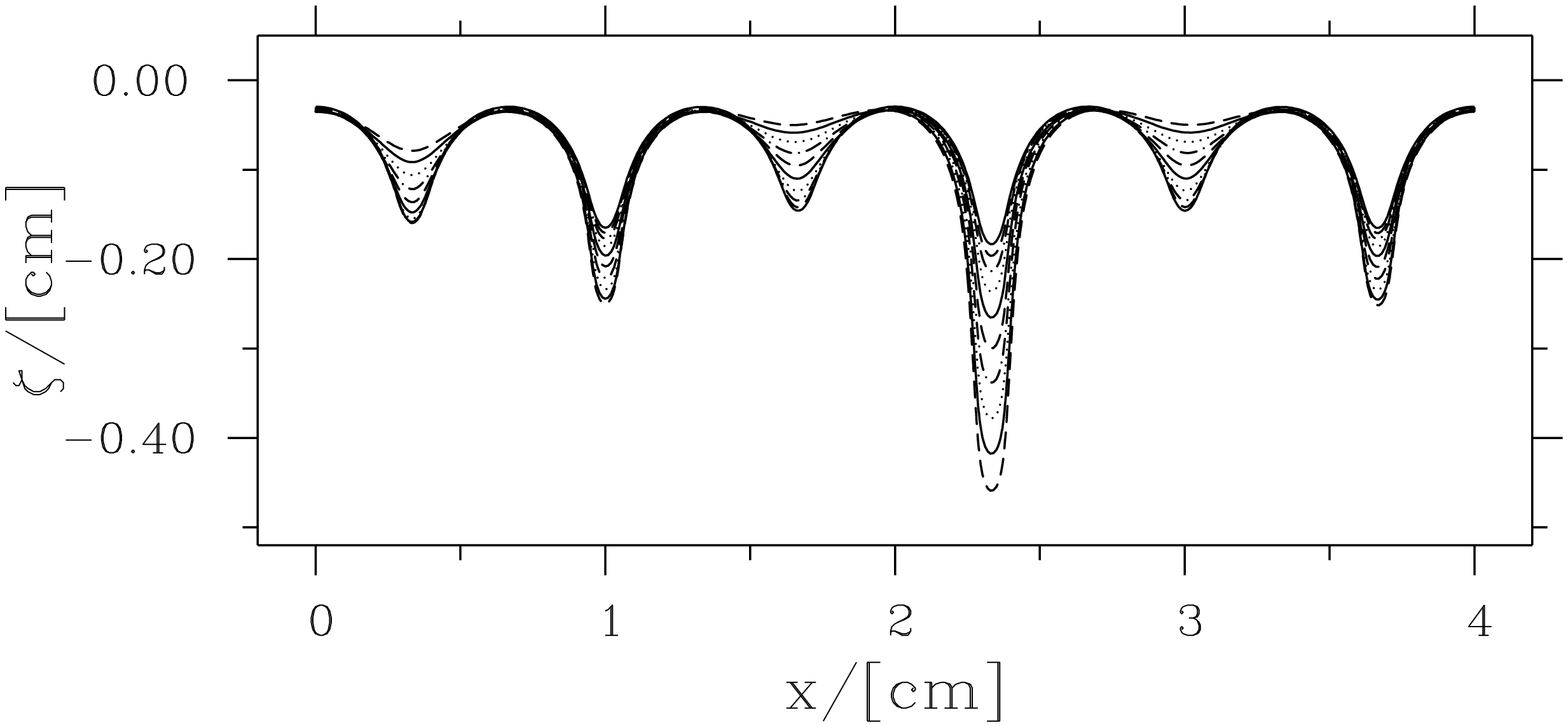,height=15.0cm,angle=90} }
\caption{Continuation of the evolution from Fig.~7. Initial time
$t=2.75$, time step $\Delta t=0.25$, final time $t=5.0$. 
The biggest  groove is still accelerating, while the other even-numbered
grooves have a roughly constant velocity and seem to decelerate
towards $t=5.0$. Odd-numbered grooves retract. They are deepest at
$t=2.75$ and have almost closed at $t=5.0$.
}
\end{figure}

 Finally, as shown in Fig.~9, only one groove survives.
Its velocity is almost constant over a range of times. Eventually it
slows down and grows sideways towards the end, which may
have to do with the fact that it gets too close to the bottom
of the numerical box (which is at $\zeta=-1$). Also gravity has a decisive 
decelerating effect here.
%it outgrows its
% neighbours and accelerates while these
%even start to decelerate and disappear eventually.

What we observe, then, is a coarsening process that seems to 
proceed via imperfect period doubling transitions. Because our system has
only six grooves, we cannot explicitly see more than the first period
doubling
here. These transitions are local in the following sense. Not all
grooves surviving the first period doubling get ahead of the others
simultaneously. Rather what  happens is that
 first the winning groove gets ahead of its 
nearest neighbours, screening each of them off the 
stress field on one side a little.
This causes these neighbours to grow more slowly,
 making them screen off {\em their}
next neighbours on the other side {\em less}. 
So these get ahead of their 
neighbour grooves, and so on. The perturbation 
made by one groove moves through the
array in an alternating fashion. 
In an infinite system, one could imagine a series
of ``near'' period doublings propagating through the array. 
These morphology changes are not exact period doubling transitions,
because there is no restabilization of a structure with doubled 
periodicity. The system remains dynamic (but see the discussion on
gravity below) which means that the foremost groove does not get
slower than its competitors, which would be necessary for length
adjustment.

\begin{figure}[h]
\centerline{ \epsfig{file=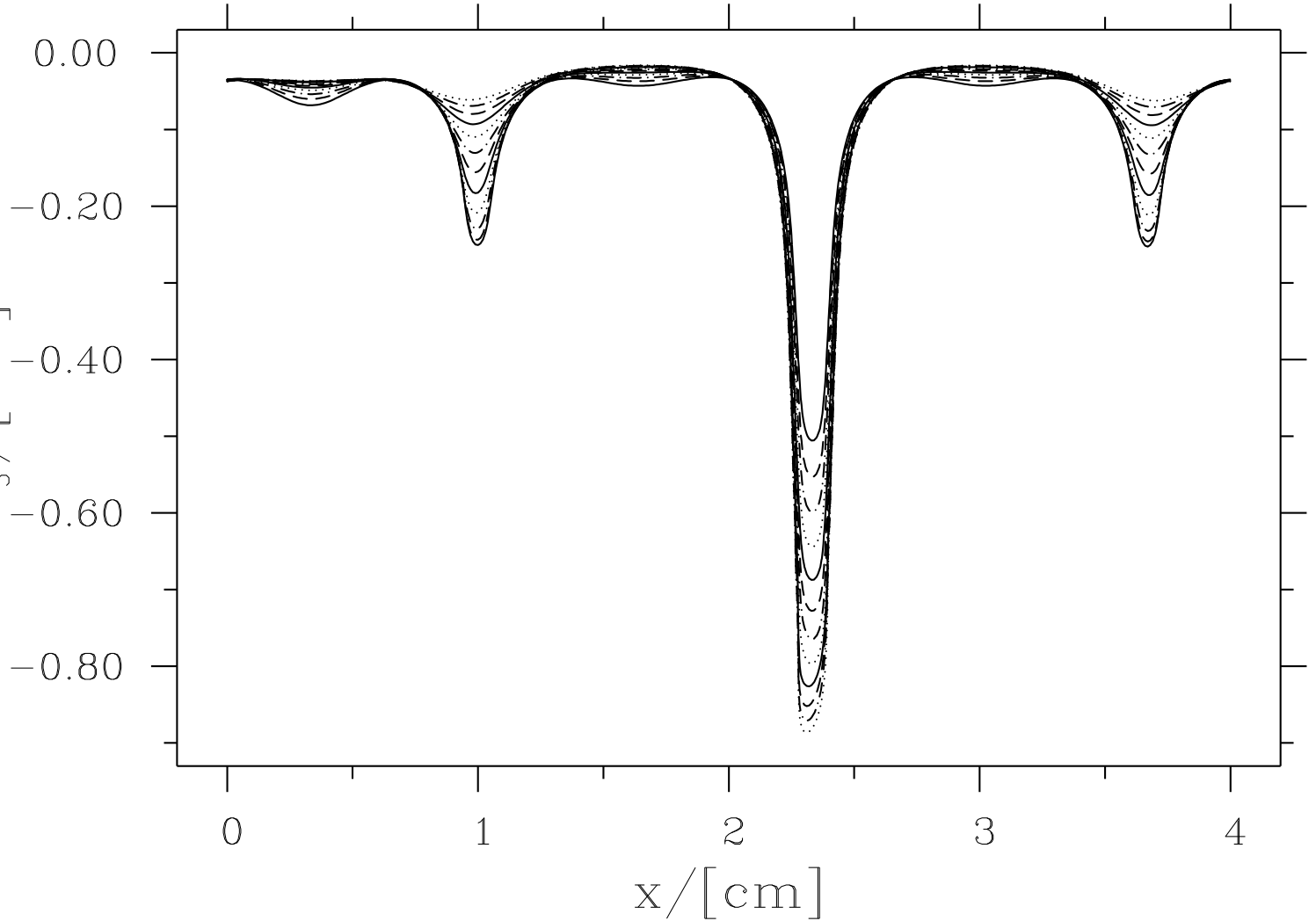,height=15.0cm,angle=90} }
\caption{Continuation of the evolution from Fig.~8. Initial time
$t=5.25$, time step $\Delta t=0.25$, final time $t=8.0$. The winning
groove has a constant velocity most of the time.}
\end{figure} 

The first of these period doublings may be discussed  analytically
in some detail. Consider the shape of the interface close to the 
last time of Fig.~7. It can be modeled approximately by a curve that
we would like to call a ``double cycloid''. A parameter representation
of this curve is given by
\begin{eqnarray}
x &=& \xi - A \sin k \xi - B \sin 2k \xi \\
z &=& \phantom{\xi} - A\cos k \xi - B \cos 2k \xi
\label{doubcyc1}
\end{eqnarray}

Figure 10 compares a double cycloid with the interface at $t=2$.
The wavenumber $2k$ (= 9.425) is given by the basic periodicity of the 
initial interface (before it is perturbed),
the amplitudes $A$ and $B$ have been
fitted ``by eye'' and the double cycloid has been shifted using translational
invariance in the $x$ direction. (Its position in the $z$ direction
can also be adjusted, which corresponds to a particular choice of the
initial chemical potential of the solid.) 

\begin{figure}[h]
\centerline{ \epsfig{file=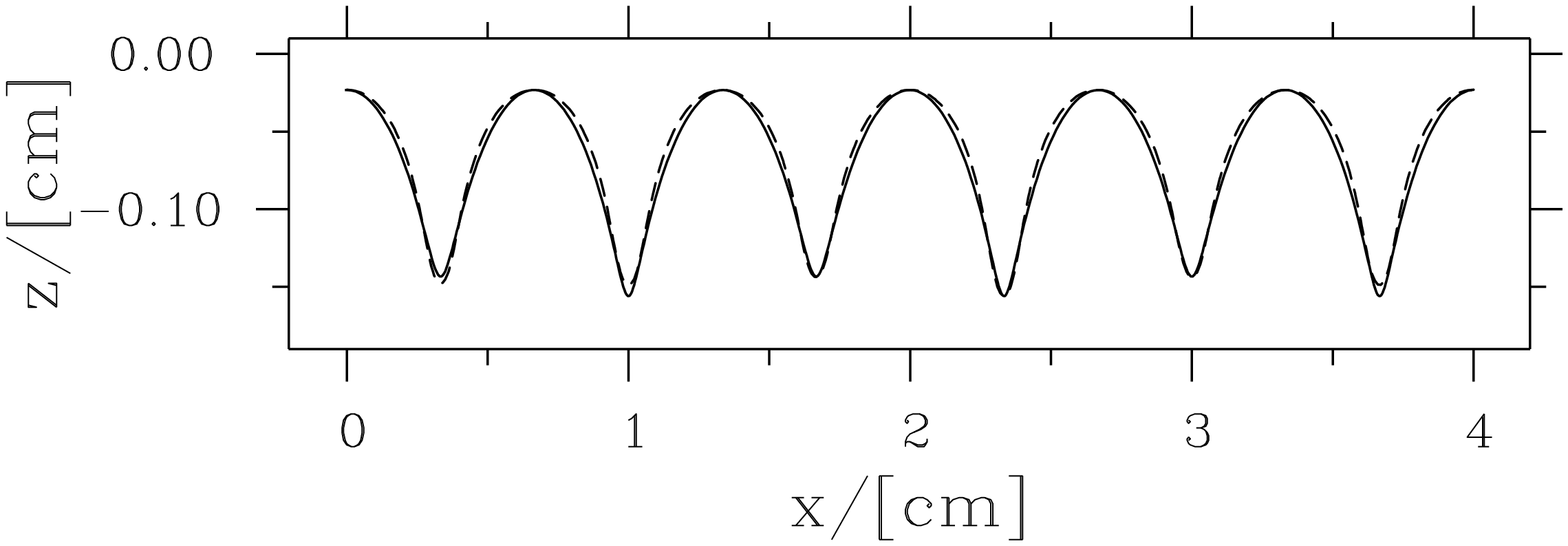,height=15cm,angle=90} }
\caption{Comparison of the interface at $t=2$ from Fig.~7 (dashed line)
with a double cycloid (solid line).}
\end{figure}
 
Since we made only one of
the grooves deeper than the others, the agreement of the groove
minima is not quite
perfect, as we can adjust only the depths of this groove and its
nearest neighbours by an appropriate choice of the two constants
$A$ and $B$. Had we taken an initial perturbation of periodicity
length $2\pi/k$ instead of a local one in the simulation, 
a much better agreement would have been obtained.
The purpose of this comparison, however, is not to claim
that the interface shape goes precisely to a double cycloid
but only to show that it may be well approximated by such a curve,
which can be considered a cycloid (with amplitude $B$)
modified by a small perturbation of twice its wavelength. In our
fit shown in Fig.~10, we have $A\approx B/10$.

Our key observation is then that we can solve the 
sharp-interface elastic problem for
a double cycloid {\em exactly} in an extension of the work 
of Gao {\em et al.} \cite{Chiu93}, using a conformal mapping technique.
This solution is given in appendix \ref{APPCONFMAPP}.
 %Let us first rewrite the velocity equation (\ref{dmu1},\ref{vn})
%for nondimensional variables. Setting $v_n = \tilde{v}_n \ell_1/\tau$,
%$\kappa=\tilde\kappa/\ell_1$, and $\zeta=\tilde\zeta \ell_1$,
%we have 
%\begin{equation}
%\tilde{v}_n = -\left\{ \frac12 \frac{(\sigma_{tt}-\sigma_{nn})^2}{\sigma_0^2}
%+\tilde\kappa + \frac{\ell_1}{2\ell_2} (\tilde\zeta-\tilde\zeta_0) \right\} \>.
%\label{vntil}
%\end{equation}
In what follows, we will neglect the gravity term, a procedure that we
justify later.
The evaluation of the nondimensional
velocity via Eq.~(\ref{vntil}) for the double cycloid yields, 
in the bottoms of the grooves
[see appendix, %\ref{APPCONFMAPP}, 
Eq.~(\ref{vntilcycl})]
\begin{equation}
\tilde{v}_n = - \frac{1}{2\left[1-k\left( A (-1)^m + 2B\right)\right]^2}
\left\{\left(1+2B k +\frac{1+Bk}{1-Bk} Ak (-1)^m\right)^2 
-\alpha k \left[ A (-1)^m + 4B\right]\right\} \>,
\label{vntilcyc}
\end{equation}
where $\alpha = 2k/q_f$ is the ratio of the actual wavenumber of the basic
cycloid and the wavenumber of the fastest-growing mode. The formula
with odd $m$ holds for the minima with depth $-2Bk+Ak$, that with
even $m$ for those with depth $-2Bk-Ak$. A condition for the solution
to hold is that there are no self-crossings of the curve, therefore
we must require $Ak+2Bk < 1$. Let us now assume that $A \ll B$, i.e.
that the pattern actually is a slightly perturbed cycloid (where the
perturbation has twice the basic wavelength $\pi/k$). Then the denominator
in Eq.~(\ref{vntilcyc}) in front of the braces  goes to zero for even $m$ 
as $2Bk$ approaches the value 1. This is the finite-time singularity,
already identified by Gao {\em al.} \cite{Chiu93}. The velocity goes to
$-\infty$, if the braces remain positive, which they do for small
enough $\alpha$, i.e., when the wavelength is large enough. For small
$A$, we can expand (\ref{vntilcyc}). This gives
\begin{eqnarray}
\tilde{v}_n & = & - \frac{1}{2\left[1-k\left( A(-1)^m + 2B\right)\right]^2}
\Biggl\{(1-2Bk)^2 + 4(1-\alpha) Bk \nonumber \\
&&\quad \quad \quad \quad \mbox{}
 + Ak (-1)^m\left[2\frac{(1+Bk)(1+2Bk)}{1-Bk}
-\alpha\right]\Biggr\} \>,
\label{vntilsmallA}
\end{eqnarray}
a formula that shows that the marginal
value of $\alpha$ is 1. Thus, for wavelengths larger than that corresponding
to the fastest-growing mode ($\alpha\approx 1$), the velocity will diverge in the deepest
minima, leading to cusps in the sharp-interface limit.
We could leave the gravity term out of this consideration, because it
never diverges for finite $\zeta$.

Now assuming we are at or slightly above the wavelength of the
fastest-growing mode, we can see
 from (\ref{vntilsmallA}) that for $(1-2Bk)^2 < Ak$ the velocity
is {\em positive} in the secondary minima corresponding to
 odd $m$ \cite{footnote7}.
This means {\em resolidification} and closure of the corresponding grooves.

Suppose for a moment that $A=0$. Then the system with a 
sharp interface will evolve towards a cusped cycloid, i.e., 
$2Bk$ will increase towards 1. But this means that eventually
a point will be reached where $1-2Bk$ is small enough that {\em any}
perturbation will be larger than $(1-2Bk)^2$. In this case, our 
equations state that (for $1-\alpha\ll 1$) the tip perturbed 
in this way in the right
direction (i.e., the perturbation must reduce the depth of the groove)
will recede again, its velocity will become positive. A groove tip
that is perturbed in the other direction will approach
the cusp singularity  even faster and reduce the speed of its neighbours. 
Of course, not all perturbations are periodic; what happens when
only a local perturbation is applied, can be seen from the simulation.

What the analytic calculation shows, then, is that the {\em first} 
period-doubling bifurcation happens  before the cusp singularity
is reached, if the periodicity of the system is {\em equal} to  the
wavelength of the fastest-growing mode. Whereas the bifurcation to a set of
alternatingly receding and advancing grooves may happen for any
wavelength {\em larger} than this one,  whether it happens before or
after the predictable time of cusp formation %is predicted 
%(within the linear-elasticity sharp-interface model) 
will in general depend on the strength of the
perturbations present in the system.
In the simulation of Figs.~7-9,
the periodicity of the unperturbed system is $\frac23$, the wavelength
of the fastest-growing mode is 0.5.

Ordinarily, the time when the finite-time singularity appears in
the sharp-interface system will be too short for the losing grooves
to have appreciably retracted. In our phase-field model, there are
no finite-time singularities, so the evolution can continue. It is then
highly plausible that further period doublings occur, even though
we have no analytic model for these. But on general grounds, we
expect screening of neighbouring grooves to become more effective
as all grooves get deeper. Hence the process should repeat, even at
wavelengths far from, but above, $\lambda_f$.

The difference between the cases of a wavelength close to that
of the fastest-growing mode and one far above it is that in the former
case, the first period doubling will happen {\em before} the
time $t^\ast$, at which cusps form in the sharp-interface limit, whereas
in the latter case, it will happen afterwards. This case is, in fact,
realized in Fig.~4, where the wavelength of the fastest-growing mode
is about one tenth of the periodicity length. {}From Fig.~6, we can
infer that the translational symmetry with respect to the basic
wavelength $\lambda=\frac23$ has already been broken by numerical noise
(the stress pattern does not show this periodicity in the upper half of
the picture, this symmetry breaking will slowly propagate into the 
lower half where everything still appears periodic).

Another interesting conclusion from formula (\ref{vntilcyc}) is that
for $\alpha > 1$, i.e., for systems with small enough wavelength,
stable steady states may be possible, because then surface tension 
may succeed in overwhelming the effects of stress.
For  $\alpha > 1$ and $A=0$, the formula predicts that a cycloid
becomes stationary in its minima before the appearance of cusps.
 We hope to report on this aspect in the future.

Finally, let us have a look at a system with a random initial condition.
Figure 11 shows the evolution starting from an interface
 resulting from uniformly random 
perturbations of a planar front. We see that first some 
10 waves develop, which is already a coarsened structure,
as the wavenumber of the fastest-growing linear mode 
would correspond to about 24 waves fitting into the system.
However, the initial amplitude is too small for this wavelength
to become clearly visible.
Some time later, there are much fewer
features and eventually, only two grooves remain. 

\begin{figure}[h]
\centerline{ \epsfig{file=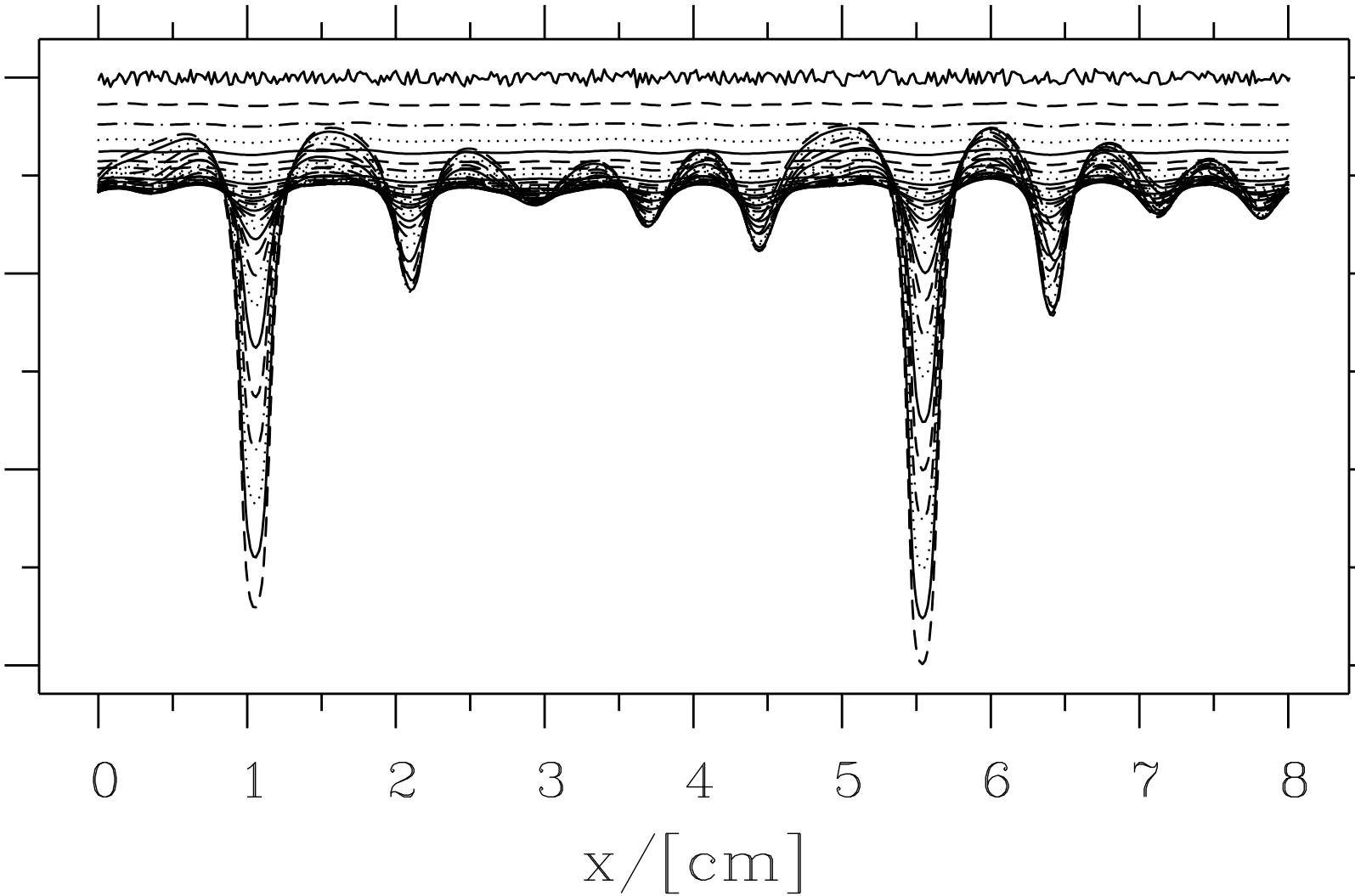,height=15.0cm,angle=90} }
\caption{Evolution of a random interface.
 $\sigma_0=3.5\times10^4$ dynes/cm$^2$, 
$\epsilon=0.05$.
Periodic boundary conditions, i.e., MG model.
 $0 \le t \le 11.6$, $\Delta t = 0.4$.}
\end{figure}
Whether
 one of the two will die off in the end is not clear, since this is a 
simulation with gravity. Hence the largest groove is bound to 
stop at some time, because the stress and curvature terms remain
constant once all other grooves are sufficiently small, but
 the gravity term continues to increase. If the second-largest
groove still has a positive velocity when the first stops, it
will not reverse its growth direction, but only grow to a point
where its velocity becomes zero. 

An example, where the final state actually consists of two grooves, is
shown in Fig.~12. Here the applied stress is smaller than in Fig.~11,
so the pattern actually does come to a halt within the numerical
box, after a long time ($t\approx 60$). Note that during most of the
period where two grooves are dominant, one of them is ahead. Once it
stops, the second approaches and in the end it has the same length as
the first, to numerical accuracy. In the case of two periodically
repeated grooves, this
is to be expected for symmetry reasons. With three or more grooves, it
is also conceivable that not all of them are the same length in
the steady state.

\begin{figure}[h] 
\centerline{ \epsfig{file=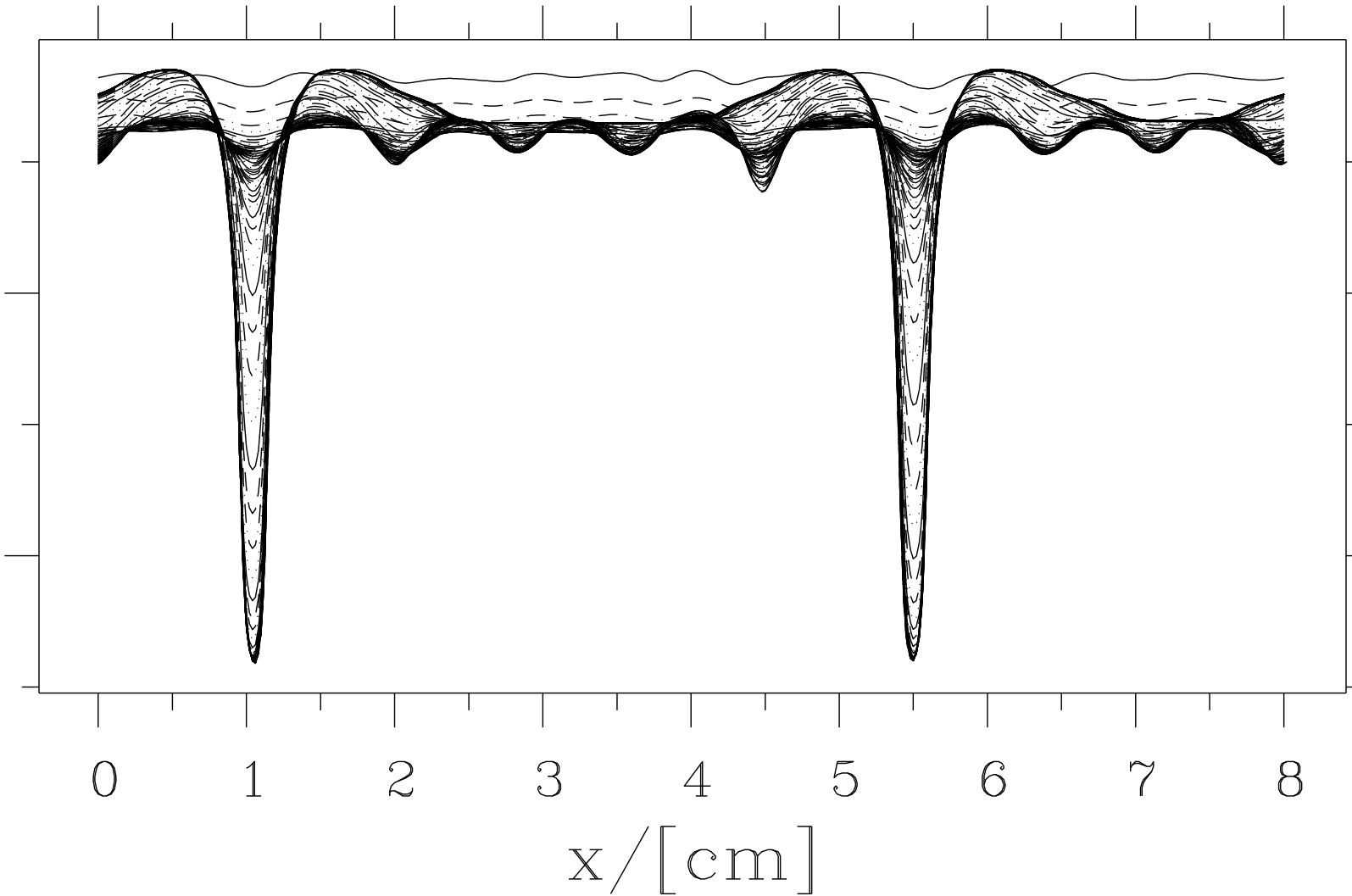,height=15.0cm,angle=90} }
\caption{Evolution of a random interface
with periodic boundary conditions. $\sigma_0=2.8\times10^4$ dynes/cm$^2$, 
$\epsilon=0.035$,  
$0 < t \le 80.0$, $\Delta t = 0.8$.
The final interface has only two grooves and no further
substructure. It is symmetric with respect to two symmetry axes at 
the two possible central positions between the two grooves.}
\end{figure}

We think that in the absence of gravity, the situation in this strongly 
nonlinear region is very similar to
the evolution of a Saffmann-Taylor finger in a Laplacian field. 
The Lam\'e equations determining the displacement field are scale invariant
just as the Laplace equation (and in fact, Eq.~(\ref{vntil}) is scale
invariant for $\ell_2=\infty$).
Once a strongly nonlinear state has been reached,
none of the length scales discussed in section II can play 
a  role anymore, since
they only govern the local behavior of the growth pattern.
 The long-range elastic
field will determine the factor $\sigma_{tt}-\sigma_{nn}$ of the destabilizing 
term in (\ref{vntil}) and this factor will be the larger, the fewer
 competitors of
a groove have grown to  the same  depth. This will lead to smaller grooves not
growing anymore. This situation bears strong similarities
 to the growth of thermal 
cracks described in \cite{Bahr92}. The main difference is that there a loser in
the competition will simply stop growing. 
In our case, it will even shrink again,
for the crystal can not only melt but also freeze again, and whether it will do
so is simply determined by the chemical potential difference (\ref{dmu1}).

An analogous behavior is found in the side branching activity of 
a dendrite in the region about 20 to 50 tip radii behind the tip 
\cite{Li98,Corrigan99}. 
There coarsening is observed, too, which also proceeds via
imperfect period doubling. If this dynamics can be described in terms
of a series of nonequilibrium phase transitions at all, 
these  would have to be considered first-order transitions because of
the discussed locality aspect. There is no diverging length scale
in a single transition.

We expect that the dynamics of large systems can be described by
scaling laws similar to those given previously for the 
growth of needles in a Laplacian field \cite{Krug93}. The fact that
``needles'' can shrink again in the elastic problem should modify the
long-time behavior of the needle density, which must pass through a
maximum and then go to zero as a function of time, for any needle length.

To some extent, this expectation is
supported by the coarsening scenario described in \cite{Mueller99}
in which extended systems without gravity are studied using random initial 
conditions. They measured the Fourier transform of the height-height 
correlation function $S(q,t)$ and observed dynamical scaling. For early times,
they observed a strong similarity between this behavior and early-stage
soinodal decomposition in long-range systems. For later times, when the
linear theory no longer describes the data, coarsening is evident:
The location of the peak $q_{\rm max}$ of $S(q,t)$ moves
 to smaller wavenumbers, 
as the peak height increases and sharpens. 
The peak height follows $S(q_{\rm max}, t) \sim t^{\alpha+1}$, where  
$\alpha \approx 2$, while the peak width sharpens with time as 
$w \sim t^{-\gamma}$, where $\gamma \approx 0.5$.
The former dependence is due to the 
total interface length increasing linearly with time for any unstable
wavenumber.  The latter dependence is due to competitive 
ordering between different wavenumbers, analogous to phase
ordering.  Within the accuracy of their study, 
they find that the structure factor shows scale invariance:
$S(q,t)/S(q_{\rm max}, t) = S^* (q^*)$,
where the scaled wave number $q^* = (q-q_{\rm max})/ w$.
  Fitting to $S^* \sim (q^*)^\delta$
and $S^* \sim (1/q^*)^\psi$, for small and large $q^*$ respectively,
gives $\delta \sim 1-2$, and $\psi \sim 5 - 6$.

It is however difficult to assess to which time regimes these results
correspond when compared with the present simulations, because 
the freedom to rescale parameters has been used extensively in
\cite{Mueller99}. 
Since the vanishing of grooves does  not seem to be 
a dominant mechanism of coarsening in their simulations,
it is likely that the time windows considered in \cite{Mueller99}
and here have little overlap and that the stage of needle-like
growth of the grooves is
never reached in \cite{Mueller99}. Of course, the concept of short and
long times is ambiguous in the absence of gravity due to the scale 
invariance of (\ref{vntil}) for $\ell_2=\infty$. However, one can compare
the depths of grooves with their lateral distances to decide whether
growth is best described as
competition of wavenumbers (a Fourier space concept)
or as competition of needles (a real-space concept).
In this context, it is also important to realize that the validity
of the simulations in \cite{Mueller99} is restricted to
small  values of $\mu/K$ with
the parameter $\eta_0$ (see app.~\ref{APPMGMAPP}) being $O(1)$,
hence these simulations are quantitative only for external stresses
$\sigma_0=O(\mu)$. For these stresses, the state
describable as a forest of needles is only obtained after a long 
time [of order $(K/\mu)^4$]. Therefore, the scaling exponents 
obtained in \cite{Mueller99}
 might not be relevant to the scenario discussed here,
which makes an analytic treatment along the lines of \cite{Krug93}
even more desirable, because it could
provide these exponents.

In an infinite system what we have depicted here is probably just the
continuation of the coarsening scenario described in \cite{Mueller99}.

It should also be pointed out that for sufficiently wide systems,
i.e., in particular for infinite ones, this dynamics may
be an intermediate state only. Once a groove becomes sufficiently long, 
stresses along its side may become large enough to provoke a 
Grinfeld instability of the ``side walls'' of this crack-like structure,
as has been shown by Brener and Marchenko \cite{Brener98}. Whether or
not this happens, depends on how efficiently the stresses are relaxed along 
the grooves, on the speed of the grooves, on the perturbation amplitude, etc. 
This secondary instability
might completely change the scaling behavior, possibly leading to 
tip splitting of the grooves and tree-like structures. So far, we have not
seen anything of this kind in our simulations.
%therefore we will refrain from speculating further on this issue.

%Because in their work a random interface was considered and the effective
%strain was very large, %(gravity being not taken into account, hence the
%%system is always ``far'' above threshold; without gravity
%%all systems can be mapped onto each other via a rescaling of 
%%times and lengths),
%the coarsening mechanism via vanishing of grooves was not very conspicuous.
In a system of finite lateral extent (or periodicity) we think coarsening
 will in the absence of gravity generally lead to the disappearance of
all grooves  except one which will grow
at constant velocity. If gravity is present, several grooves can survive
and in a sufficiently deep system they will  stop once they reach
a depth where the gravity term compensates the stress one. 

\section{Conclusions}

In this article, we have constructed a class of phase-field models
from a free-energy functional including the elastic energy density.
A salient feature of the model is that the liquid is treated
as a shear-free solid, which is to be contrasted with phase-field models
taking into account hydrodynamic effects in solidification,
where the solid is usually treated as a liquid of infinite viscosity
\cite{Toenhardt98}. Our approach implies the artificial introduction of
coherence conditions at the interface which is however counterbalanced
by the fact that the only relevant elastic variable in the liquid
is $\nabla {\bf u}$. A whole class of models is obtained instead of
a single one as a consequence of the freedom of choice for the state of
reference used in measuring displacements. We compared the two most
natural choices and found them to yield slightly different numerical results
despite their asymptotic equivalence.

Investigating a large number of laterally small and extended systems, we 
believe to be able now to describe the generic dynamic behavior.
For systems  smaller than the wavelength of the fastest-growing
mode of linear stability theory but larger
than that of the marginal mode (where surface tension
stabilizes the planar interface), stable steady-state strucures are
possible even in the sharp-interface limit. This is similar to the
findings by Spencer and Meiron \cite{Spencer94} for the case of transport
via surface diffusion, even though we think the whole picture is
more complex than what they described \cite{Kohlert}. Here, we did not
show detailed results on small systems but  focused on extended systems.
% (the more frequent case).

The case {\em without gravity\/} is particularly simple, as the equation of 
motion can then be made parameter free [Eq.~(\ref{vntil})
without the last term]. 
%There is essentially only one type of long-time dynamics.
 Initially,
an interface may grow periodically but as soon as perturbations break
the periodicity, coarsening will proceed via approximate
period doubling transitions. Supposing randomized perturbations,
the interface will, after a sufficient lapse of time, not look much
different from one started with random initial conditions (compare Figs.~9
and 11). If the system is of finite lateral extent (but infinitely deep),
only a single groove will survive growing at constant velocity,
determined by the final constant stress and constant 
surface tension terms near its tip. The final velocity will scale with 
the system width $L$ and the radius of curvature $\epsilon$ as
$v_n\sim L/\epsilon$. All the other grooves will eventually retract, i.e.,
they will not even survive keeping a finite depth, which is different from
the behavior of cracks \cite{Bahr92}. 
For wide systems, stresses near the groove tips may become large enough
to trigger a secondary instability \cite{Brener98}
which would considerably modify
the system behavior and allow the appearance of complex 
crack morphologies. It is however possible that this will arise
only in the case of finite perturbations, as grooves may grow too
fast in this situation for the instability to develop before it
is ``advected'' (relative to the groove tip)
into a region of very small stresses along the groove.
 If the system is laterally infinite,
the system state  will first
follow dynamical scaling as studied in \cite{Mueller99}
and should then cross over to the scaling dynamics 
described here with the number of grooves
continually decreasing according to a power law, 
possibly with logarithmic corrections. Alternatively, the
coarsening scenarios observed in  \cite{Mueller99} and in this article
might be governed by the {\em same\/} scaling laws, 
with their difference being only apparent.
The emphasis of \cite{Mueller99} was on the scaling laws governing
coarsening, that of the present study is on the mechanism of coarsening.
Obviously, this situation
calls for large-scale simulations in order to determine
the scaling exponents in cases where the mechanism presented
here is definitely at work already.
% This state should be tractable along the lines
%given in \cite{Krug93}. 

 If the mentioned secondary instability 
\cite{Brener98}  becomes important, the identified
state of competing grooves will
be only of intermediate nature.  
Of course, all our considerations hold only as long as 
linear elasticity remains valid in the bulk, nonlinear elastic effects
may alter the scenario.
%will change the picture. 
%Moreover, it is possible that secondary 
%instabilities such as the one discussed in \cite{Brener98} will
%alter the final scenario.

With gravity included (which was not considered in \cite{Mueller99}), 
there are some modifications. First, it is now
possible for a planar interface to be stable (apart from a vertical
 translation).
Once the threshold of the instability has been exceeded, the behavior
will be similar to the case without gravity. However, we  predict
that it is possible for several grooves to survive in a finite system
and that they will eventually stop growing, because the stress does
not increase beyond a certain magnitude due to the lateral system width,
whereas the gravity term increases as long as a groove gets deeper.
That several grooves may survive has to do with the fact that now
we have length scales in Eq.~(\ref{vntil}). More simple-mindedly we
can immediately see that once the biggest groove stops, the second-largest
will not retract, if it still has a downward velocity at that moment.
Once the second stops, we can repeat the argument for the third, and so on.
The final state will consist of 
 a number of grooves, probably of different lengths
and disordered. (For large-width systems, the aforementioned secondary
instability may again complicate the picture.)
 In laterally infinite systems, it seems likely that
a scaling state will prevail, possibly with a modified scaling exponent.
Because now both stresses at the tips of the largest grooves
and the gravity terms continue to grow, but they will both grow linearly
with the length of the grooves. If initially the stress was large enough
to overcompensate the gravity term, it will presumably stay like that.
It is, however, not excluded that starting from specific initial conditions
a system can be stabilized by gravity in the end.

\vspace*{0.5cm}
{\bf Acknowledgments}\\

This work was supported by the German Research Society
({\em Deutsche Forschungsgemeinschaft}) under grant Ka 672/4-2,
which is gratefully acknowledged.
Moreover, we acknowledge a PROCOPE grant for travel exchanges
by the DAAD (German academic exchange service),
grant no.~9619897, and the A.P.A.P.E.
(corresponding French organisation), grant no.~97176, as well as 
financial support by the TMR network ``Pattern formation, noise and 
spatio-temporal chaos in complex systems''.  
 
\appendix

\section{Derivation of the sharp-interface limit \label{APPSHARPINT}}

In deriving the sharp-interface limit, we will restrict ourselves to the
two-dimensional case as we did in discussing the sharp-interface equations (where we 
used only two stress components and only one curvature). The generalization
to three dimensions is, however, straightforward. 
We may use (\ref{phasebas}) and (\ref{stressbas})
as {\em outer} equations, %valid
to be used in the regions where the
gradient of the phase field is small. 
For convenience, we set $z_0=0$.

To obtain the {\em inner} equations, we transform to a local system
of (orthogonal) curvilinear coordinates comoving with the interface,
with one coordinate axis parallel to  $\nabla \phi$; the 
corresponding coordinate will be called $r$, while the
second  will be conveniently expressed by the arclength
$s$ along the interface \cite{footnote3}.
We introduce a
stretched variable setting $r= \tilde\epsilon \rho$.
It is then easy to see that a distinguished limit of (\ref{phasebas}),
leading to a nontrivial inner equation that allows to satisfy the
boundary conditions,
is obtained by setting $\tilde\epsilon = \epsilon$.
In saying this, we have assumed that the stresses and strains
behave properly under rescaling, i.e., do not diverge.
Designating by capital letters the values of the
fields in the inner domain (where the gradient of the 
phase field is large), we then have the inner equations
\begin{eqnarray}
-\frac{v}{\epsilon}\> \dbydrho \Phi &=& 
\frac{\gamma}{\ktil}\>\Biggl\{\frac1{\epsilon^2} \dbydrhot\Phi 
+\frac{\kappa}{\epsilon}\>  \dbydrho \Phi + \dbydst \Phi
- \frac1{\epsilon^2}\Biggl[ 2 g'(\Phi)  \nonumber \\
&& \mbox{} + \frac{\epsilon}{3\gamma} h'(\Phi) 
\left( \mu \, U_{ij} U_{ij}
 + \frac{\lambda-\tilde\lambda}{2}  U_{ii}^2 
+ \Delta p \, U_{ii} + \Delta W +\Delta\rho g z \right)\Biggr] \Biggr\}\>,
\label{innerphi1} \\
0&=& \frac1{\epsilon}\>\dbydrho\tilde\Sigma_{\rho\rho}
     +\dbyds \tilde\Sigma_{\rho s}  
     + \kappa \left(\tilde\Sigma_{\rho\rho}-\tilde\Sigma_{ss}\right)\>,
     \label{innersiga} \\
0&=& \frac1{\epsilon}\>\dbydrho\tilde\Sigma_{s\rho}
     +\dbyds \tilde\Sigma_{s s}  
     +\kappa \left(\tilde\Sigma_{\rho s}+\tilde\Sigma_{s\rho}\right)\>, 
     \label{innersigb} 
\end{eqnarray}
where 
\begin{equation}
\tilde\Sigma_{\alpha\beta} = h(\Phi) \Sigma_{\alpha\beta} 
                           - (1- h(\Phi)) P \delta_{\alpha\beta} 
\label{defsigtil}
\end{equation}                           
is the generalized stress tensor of the two-phase system.
In order to obtain (\ref{innerphi1}--\ref{innersigb}),
we have used
\begin{eqnarray}
\nabla &=& \frac1{\epsilon}\> \nvec \>\dbydrho 
                + \tvec \>\dbyds \>, \label{defnabla} \\
\nabla^2 &=& \frac1{\epsilon^2} \> \dbydrhot
            + \frac{\kappa}{\epsilon}\> \dbydrho
            + \dbydst \>. \label{deflaplace} 
\end{eqnarray}            
Derivatives such as $\partial_x u_x$ can then be expressed
in invariant form as ${\bf e}_x ({\bf e}_x \nabla) {\bf u}$,
which leads to the following relations for the strain tensor components
in the new coordinates:
\begin{eqnarray}
U_{\rho\rho} &=& \frac1{\epsilon}\>\dbydrho U_\rho \>, \label{Urhorho}\\
U_{s s} &=& \dbyds U_s + \kappa U_\rho \>, \label{Uss} \\
U_{\rho s} &=& U_{s\rho} 
\mbox{}= \mbox{} \frac12 \left(\dbyds U_\rho 
                      +\frac1{\epsilon}\>\dbydrho U_s 
                      - \kappa U_s\right)\>.
                      \label{Urhos}
\end{eqnarray}                      
%{}From these relations, the corresponding equations for the 
%stress tensor are obtained via Hooke's law which finally leads
%to (\ref{innersiga}) and (\ref{innersigb}). 

The next step consists in solving the 
outer and inner equations via an asymptotic
analysis that leads to a globally valid
approximation for small interface thickness $\epsilon$ which approaches
the sharp-interface equations as  $\epsilon\to 0^+$. To this end we expand both outer
and inner fields in powers of $\epsilon$:
\begin{eqnarray}
\phi(x,z,t) &=& \phi_0(x,z,t) + \epsilon \phi_1(x,z,t) + ... \>, 
           \label{phioutexp} \\
u_{ij}(x,z,t) &=& u_{ij}^{(0)}(x,z,t) + 
           \epsilon u_{ij}^{(1)}(x,z,t) + ... \>, 
           \label{uijoutexp} 
\end{eqnarray}
and
\begin{eqnarray}  
\phi(x,z,t) &=& \Phi(\rho,s,t) = \Phi_0(\rho,s,t) + \epsilon \Phi_1(\rho,s,t)
 + ... \>, \label{phiinnexp} \\
u_{ij}(x,z,t) &=& U_{ij}(\rho,s,t) = U_{ij}^{(0)}(\rho,s,t)  + 
           \epsilon U_{ij}^{(1)}(\rho,s,t) + ... \>,
           \label{uijinnexp} 
\end{eqnarray}
where, due to the transformation properties of tensors, we can think
of the subscripts $i$, $j$ as running either over the values ($x$,$z$) or
($r$,$s$) and ($\rho$,$s$), respectively.
Our basic field equations are, however, equations not for the strains
but for the displacement fields. Thus, the expansion of the 
$U_{ij}(\rho,s,t)$ induces one for the displacement components:
\begin{eqnarray}  
u_r &=& U_\rho(\rho,s,t) = U_\rho^{(0)}(\rho,s,t) 
                    + \epsilon U_\rho^{(1)}(\rho,s,t) + ... \>,
          \label{urhoinnexp} \\
u_s &=& U_s(\rho,s,t) = U_s^{(0)}(\rho,s,t) 
                    + \epsilon U_s^{(1)}(\rho,s,t) + ... \>.
          \label{usinnexp}
\end{eqnarray}  
Now the physical requirement that both $u_r$ and $u_s$ remain finite
in the limit $\epsilon\to 0^+$ allows us to conclude from (\ref{Urhorho})
and (\ref{Urhos}) that neither $U_\rho^{(0)}$ nor $U_s^{(0)}$ can depend
on $\rho$, hence
\begin{eqnarray}  
U_\rho^{(0)} &=& U_\rho^{(0)}(s,t) \label{urho0} \>, \label{urho0norhodep} \\
 U_s^{(0)} &=&  U_s^{(0)}(s,t) \label{us0} \>. \label{us0norhodep}
\end{eqnarray}

Furthermore,  we
have matching conditions for $1\ll \rho \ll \epsilon^{-1}$ that can be obtained
from the inner and outer expansions by equating equal powers of 
$\epsilon$ (and taking into account that the variable $r$ is itself
$\epsilon$ dependent):
\begin{eqnarray} 
\Phi_0(\rho,s,t) &\sim& \phi_0(r,s,t) \vert_{r=\pm 0}\>, 
                     \quad\quad \rho\to\pm\infty\>, \label{matchphi0} \\
\Phi_1(\rho,s,t) &\sim& [\phi_1(r,s,t) 
       + \rho \>\partial_r\phi_0(r,s,t)] \vert_{r=\pm 0}\>,
       \quad\quad \rho\to\pm\infty\>, \label{matchphi1} \\ 
U_{\alpha\beta}^{(0)}(\rho,s,t) &\sim& u_{\alpha\beta}^{(0)}(r,s,t)\vert_{r=\pm 
0}\>, 
                     \quad\quad \rho\to\pm\infty\>, \label{matchuij0} 
\end{eqnarray}
where we use the $\sim$ symbol in the sense of asymptotic equality,
i.e., $f(x) \sim g(x)$, $x\to x_0$ is equivalent to 
$\lim_{x\to x_0} f(x)/g(x) = 1$, and for two series in $x-x_0$ we require 
this relation for each corresponding pair of terms. 

The relations induced by (\ref{matchuij0}) for the displacements are
more complicated. We just give two examples. Because each derivative
with respect to $r$ comes with a factor $1/\epsilon$ when transformed
into a derivative w.r.t. $\rho$,
we have $U_{\rho\rho}^{(0)}= \partial_\rho U_\rho^{(1)}$ and hence
\begin{equation}
\lim_{\rho\to\pm\infty} \partial_\rho U_\rho^{(1)}(\rho,s,t)
 = \partial_r u_r^{(0)}(r,s,t) \vert_{r=\pm 0}\>.
\end{equation}
Our second example is even more instructive. We write
\begin{eqnarray}
\lim_{\rho\to\pm\infty} U_{ss}^{(0)}(\rho,s,t) 
 &=&  \lim_{\rho\to\pm\infty} \left(\partial_s   U_s^{(0)}(s,t) 
       + \kappa  U_\rho^{(0)}(s,t)\right) \nonumber \\
 &=& \partial_s   U_s^{(0)}(s,t) + \kappa  U_\rho^{(0)}(s,t) \nonumber \\
 &=& [\partial_s   u_s^{(0)}(r,s,t) + \kappa  u_\rho^{(0)}(r,s,t)] \vert_{r=\pm 
0}\>, \label{ussfuncofsonly}
\end{eqnarray}
which shows that the linear combination $ u_s^{(0)} + \kappa  u_\rho^{(0)}$
must be continuous across the interface.

Finally, we need the expansions of all functions of $\phi$ in powers of 
$\epsilon$, e.g.:
\begin{equation}
h(\phi) = h(\phi_0) + \epsilon h'(\phi_0) \phi_1
          + \epsilon^2 \left(h'(\phi_0) \phi_2 +
 \frac12 h''(\phi_0) \phi_1^2\right) + ...
\>, \label{hexpans}
\end{equation}
and we will use the obvious abbreviations $h_0$, $h_0'$, etc., for functions
of $\phi_0$. Let us note a few useful relations in passing:
\begin{eqnarray}
%h(\phi) &=& \phi^2 (3-2\phi) \>, \label{useful1} \\
h'(\phi) &=& 6 \phi (1-\phi)  \>, \label{useful2}\\
%h''(\phi)  &=& 6 (1-2\phi) \>, \label{useful3}\\
g(\phi) &=& \phi^2 (1-\phi)^2 = \left(\frac16 h'(\phi)\right)^2  
  \>, \label{useful4}\\
g'(\phi) &=& 2\phi(1-\phi)(1-2\phi) = \frac1{18}h'(\phi) h''(\phi)\>, 
\label{useful5}
%g''(\phi) &=& 2(1-6\phi(1-\phi)) = 2 (1-h'(\phi)) \>. \label{useful6}
\end{eqnarray}

We have  now collected all the prerequisites to perform the 
asymptotic analysis providing the sharp-interface limit. 

First, we note that the outer solution to lowest order [order $\epsilon^{-2}$
of (\ref{phasebas})] is 
simply given by $g'(\phi_0)=0$, which yields the solutions $\phi_0=0$,
$\phi_0=1$, and $\phi_0=\frac12$ [see (\ref{useful5})]. 
The last of these is unstable and 
also not compatible with the boundary conditions in typical numerical setups.
We assume $\phi_0=0$ for $r>0$, corresponding to the liquid phase,
and $\phi_0=1$ for $r<0$,  corresponding to the solid phase.
Equation (\ref{useful2}) tells us that $h'(\phi_0)=0$, and hence these
solutions are valid at {\em all} orders of $\epsilon$. 
Using $h(\phi_0) = 0$ in the liquid and $h(\phi_0) = 1$ in the solid,
we immediately see that (\ref{stressbas}) turns into the mechanical 
equilibrium condition for  the liquid and solid, respectively:
\begin{eqnarray}
\partial_i p &=& 0 \quad \quad \mbox{(liquid)}\>,\\
\partial_j \sigma_{ij}  &=& 0 \quad \quad \mbox{(solid)}\>.
\end{eqnarray} 
This is again true at all orders of $\epsilon$, and we can write the
zeroth-order piece of the result in the form:
\begin{eqnarray}
p^{(0)} &=& p_{0\ell}-\lamtil u_{kk}^{(0)} = p_0 = {\rm const.}\>, \label{p0} \\
\sigma_{ij}^{(0)} &=& - p_{0s} \delta_{ij} + 2\mu u_{ij}^{(0)} 
+ \lambda u_{kk}^{(0)} \delta_{ij} \>. \label{sig0}
\end{eqnarray}
Later, we will  look at two  reference states in particular. One is
the ``natural'' choice $p_{0s}=p_{0\ell}$, i.e., the unstrained
state is hydrostatic and corresponds to the same pressure in the 
liquid and in the solid. If moreover, this pressure is chosen equal
to the equilibrium pressure $p_0$, then we have
$u_{kk}^{(0)} \equiv 0$ in the liquid at equilibrium. 
The second choice corresponds to 
assuming a finite difference $\Delta p = p_{0\ell}-p_{0s}$
while keeping  $p_{0\ell}=p_{0}$.
This means that zero strain corresponds to a prestressed solid, with a
stress tensor $\sigma_{ij}=-p_0\delta_{ij} + \Delta p \,\delta_{ij}$, 
i.e., the deviation
from equilibrium is the isotropic tensor $ \Delta p \,\delta_{ij}$.
%equal to a prestress of the 
%solid and to have zero displacements in the solid when the latter
%has an  isotropic stress tensor corresponding to that value.
Both approaches can be exploited numerically.

We now consider the inner solution. The lowest order of (\ref{innerphi1})
gives 
\begin{equation}
\partial_\rho^2 \Phi_0 - 2 g'(\Phi_0) = 0 \>. \label{phidiffeq}
\end{equation}
This equation can be solved by standard methods. Multiplying by
$\partial_\rho \Phi_0$, we immediately obtain a first integral, 
written down here for further reference,
\begin{equation}
\partial_\rho \Phi_0 = - 2 \Phi_0 (1-\Phi_0)= -\frac13 h_0' \>, 
\label{firstintphi}
\end{equation}
which is solved by $\Phi_0 = \frac12 (1-\tanh \rho)$, and this solution
satisfies the matching conditions (\ref{matchphi0}).

{}From the  strain equations (\ref{innersiga}), (\ref{innersigb}) we obtain
to lowest order
\begin{eqnarray}
\partial_\rho \tilde\Sigma^{(0)}_{\rho\rho} &=& 0 \>, \label{lowordsignn} \\
\partial_\rho \tilde\Sigma^{(0)}_{s \rho }  &=& 0 \>, \label{lowordsigtn} 
\end{eqnarray}
which on integration from $\rho=-\infty$ to $\rho=\infty$ together
with the matching conditions (\ref{matchuij0}) yields
\begin{eqnarray}
\sigma^{(0)}_{rr} \vert_{r=0^-} &=& \tilde\Sigma^{(0)}_{\rho\rho} (-\infty) =
\tilde\Sigma^{(0)}_{\rho\rho} (\infty) = -P = -p_0 \>, \label{mecheq1} \\
\sigma^{(0)}_{sr} \vert_{r=0^-} &=& \tilde\Sigma^{(0)}_{s\rho} (-\infty) =
\tilde\Sigma^{(0)}_{s\rho} (\infty) = 0 \>. \label{mecheq2}
\end{eqnarray}
The limiting values of $\tilde\Sigma^{(0)}_{\rho\rho}$ and $\tilde\Sigma^{(0)}_{s\rho}$
can be gathered from (\ref{defsigtil}). Obviously, these two equations
constitute the condition of mechanical equilibrium at the interface, as
$\sigma^{(0)}_{rr}$ and $\sigma^{(0)}_{sr}$ are the normal and shear components of 
the stress tensor of the outer solution there.

However, the strain equations provide more information than just mechanical
equilibrium on the outer scale. 
We write  (\ref{lowordsignn}) explicitly in terms of the strains
and integrate indefinitely with respect to $\rho$, which yields
\begin{equation}
h_0 \left[(2\mu+\lambda-\lamtil) U_{\rho\rho}^{(0)}
 + (\lambda-\lamtil) U_{ss}^{(0)} + \Delta p \right]
 + \lamtil (U_{\rho\rho}^{(0)}+ U_{ss}^{(0)}) = f(s,t) \>, \label{indefintu}
\end{equation}
where %we have set $\Delta p = p_{0\ell} - p_{0s}$ and 
$f(s,t)$ is a
function of integration, to be determined from the matching
conditions. This is straightforward and yields 
$f(s,t)=p_{0\ell} - p_0$. 
Moreover, we know that the only spatial dependence of $ U_{ss}^{(0)}$ is that
on $s$  [see (\ref{ussfuncofsonly}) 
and (\ref{us0norhodep})], which suggests to solve
for $ U_{\rho\rho}^{(0)}$ in terms of $ U_{ss}^{(0)}$. The result is
\begin{eqnarray}
U_{\rho\rho}^{(0)} &=& \frac{-1}{(2\mu+\lambda-\lamtil)h_0+\lamtil}
\biggl\{\Delta p \> h_0  + p_0 - p_{0\ell}
 + [\lamtil+(\lambda-\lamtil)h_0] U_{ss}^{(0)}\biggr\} \>. 
\label{urhorhorhodep}
\end{eqnarray}
The advantage of this equation is that it provides us with the full
$\rho$ dependence of $U_{\rho\rho}^{(0)}$, allowing the explicit evaluation
of integrals on $\rho$ containing the strains. An analogous procedure
for the second strain equation determines $U_{s\rho}^{(0)}$ to be equal to
zero.

Now we proceed to the next-order equation for $\Phi$. Written out explicitly,
it reads
\begin{eqnarray}
-v \partial_\rho \Phi_0 &=& \frac{\gamma}{\ktil}\biggl\{ \partial_\rho^2 \Phi_1
 +\kappa \partial_\rho\Phi_0 - 2 g''_0 \Phi_1
\nonumber \\
&&\mbox{} - \frac{h'_0}{3\gamma} \Bigl(
\mu \bigl[{U_{\rho\rho}^{(0)}}^2 + {U_{ss}^{(0)}}^2 
\bigr]+\frac{\lambda-\lamtil}{2}\bigl(U_{\rho\rho}^{(0)}+U_{ss}^{(0)}\bigr)^2 
\nonumber \\
&&\mbox{} + \Delta p \bigl(U_{\rho\rho}^{(0)}+U_{ss}^{(0)}\bigr) + \Delta W + 
\Delta\rho g z(s)\Bigr)\biggr\}  
\end{eqnarray}
where we have used $U_{s\rho}^{(0)}=0$. With the help of (\ref{useful2})
 and (\ref{firstintphi}), we can arrange this as
\begin{eqnarray}
L \Phi_1 &=& \frac{h'_0}{3\gamma} \biggl\{ \ktil v + \gamma \kappa
+\mu \bigl[{U_{\rho\rho}^{(0)}}^2 + {U_{ss}^{(0)}}^2 \bigr]
+\frac{\lambda-\lamtil}{2}\bigl(U_{\rho\rho}^{(0)}+U_{ss}^{(0)}\bigr)^2
 \nonumber \\
&&\mbox{} + \Delta p \bigl(U_{\rho\rho}^{(0)}+U_{ss}^{(0)}\bigr) + \Delta W + 
\Delta\rho g z(s)\biggr\} \>, \label{inhomlineq}
\end{eqnarray} 
with $L\equiv \partial_\rho^2 - 2 g''_0 $ 
being a self-adjoint linear operator. The solvability condition for this
inhomogeneous linear equation is that the right-hand side be orthogonal
to the left-sided null eigenspace of $L$. Since
$L$ is hermitean, 
we know that the translational mode $\partial_\rho \Phi_0$ is an appropriate
eigenvector:
\begin{equation}
\partial_\rho \Phi_0 L = L \partial_\rho \Phi_0 = 0 \>. 
\end{equation}
Multiplying (\ref{inhomlineq}) by $3\gamma \partial_\rho \Phi_0$ from the
left and integrating on $\rho$, we find
\begin{eqnarray}
0 &=& \int_{-\infty}^{\infty} \,d\rho\> \biggl\{ \ktil v + \gamma \kappa
+\mu \bigl[{U_{\rho\rho}^{(0)}}^2 + {U_{ss}^{(0)}}^2 \bigr]
+\frac{\lambda-\lamtil}{2}\bigl(U_{\rho\rho}^{(0)}+U_{ss}^{(0)}\bigr)^2
 \nonumber \\
&&\mbox{} + \Delta p \bigl(U_{\rho\rho}^{(0)}+U_{ss}^{(0)}\bigr) + \Delta W + 
\Delta\rho g z(s)\biggr\} h'_0  \partial_\rho \Phi_0 \>. \label{fredholm1}
\end{eqnarray}  

Now we can exploit (\ref{urhorhorhodep}), telling us that the 
$\rho$ dependence of the 
braces in (\ref{fredholm1}) is fully 
contained in their dependence on $h_0$. 
All the integrals
can be done analytically, using
\begin{equation}
I\equiv -\int_{-\infty}^{\infty}\,d\rho\> f(h_0) h'_0  \partial_\rho \Phi_0
= -\int_1^0 \,d\Phi_0 \>  f(h(\Phi_0)) h'(\Phi_0) = \int_0^1 \,dh\> f(h)
\>.
\end{equation}
Integrals that appear in (\ref{fredholm1}) are
\begin{eqnarray}
I_1 &=&  -\int_{-\infty}^{\infty}\,d\rho\>  h'_0  \partial_\rho \Phi_0 = 1 \>,
\label{I1} \\
I_2 &=& -\int_{-\infty}^{\infty}\,d\rho\> U_{\rho\rho}^{(0)}  h'_0  
\partial_\rho \Phi_0 \>, \label{I2} \\
I_3 &=& -\int_{-\infty}^{\infty}\,d\rho\> {U_{\rho\rho}^{(0)}}^2  h'_0  
\partial_\rho \Phi_0 \>. \label{I3}
\end{eqnarray}
The evaluation of the latter two integrals is as straightforward as 
that of the first, although a little more tedious. We just give the final
result for the solvability condition, taking  $p_{0\ell}=p_0$
for simplicity:
%Assuming that $p_{0\ell}=p^{(0)}$
%(which defines the reference state in the liquid)
% It reads
\begin{equation}
-\ktil v = \gamma \kappa + \Delta\rho g z
+\frac{\mu}{2(\mu+\lambda)(2\mu+\lambda)} \left[2(\mu+\lambda)  U_{ss}^{(0)} + \Delta p\right]^2
- \frac{2\mu+\lambda}{8 \mu (\mu+\lambda)} \sigma_{00}^2 
\>,
\label{finres1}
\end{equation}

At this point,  we may specify our choice of reference state 
for the solid. 
First, let us assume that the unstrained state corresponds to 
a state of equal hydrodynamic pressure in the two phases, i.e. $p_{0s}=p_{0\ell}$,
or $\Delta p = 0$. This is the KM choice.
Then taking the limit $\rho\to -\infty$ of 
(\ref{urhorhorhodep}) we get ($U_{ss}^{(0)} = u_{ss}^{(0)}$)
\begin{equation}
u_{rr}^{(0)} = \frac{-\lambda u_{ss}^{(0)}}{2\mu+\lambda} \>,
\end{equation}
implying $\sigma_{ss}^{(0)} -\sigma_{rr}^{(0)} = 2\mu (u_{ss}^{(0)} - 
u_{rr}^{(0)}) = 4\mu(\mu+\lambda) u_{ss}^{(0)} /(2\mu+\lambda)$, from which we 
obtain
\begin{equation}
u_{ss}^{(0)} = \frac{2\mu+\lambda}{4\mu(\mu+\lambda)} (\sigma_{tt}^{(0)}
- \sigma_{nn}^{(0)}) \>, \label{sigmadif1}
\end{equation}
where we have now switched to  the conventional notation for the principal components
 of the stress tensor in the  normal and tangential directions
($\sigma_{rr}=\sigma_{nn}$, $\sigma_{ss}=\sigma_{tt}$).
Finally, expressing the Lam\'e constants by Young's modulus and the
Poisson ratio, we arrive at
\begin{equation}
v = - \frac1{\ktil}\biggl\{ \frac{1-\nu^2}{2E}
\left[(\sigma_{tt}^{(0)}-\sigma_{nn}^{(0)})^2-\sigma_{00}^2\right] +  
\gamma \kappa + \Delta\rho g z  \biggr\} \>, \label{finres2} 
\end{equation}
which is the desired sharp-interface limit. [In eqs.~(\ref{dmu1}, \ref{vn}), $\sigma_{00}=0$.]

A remark is in order here. The phase-field equations imply the continuity
of $u_{ss}^{(0)}$ across the interface. As this quantity is obviously
nonzero whenever the solid is strained, this means that we will {\em not} 
have $u_{ss}^{(0)}=0$ in the liquid.
% i.e., $u_{ss}^{(0)}$ does
%not faithfully represent a true strain component in the liquid.
However, we will still have $u_{ss}^{(0)} + u_{rr}^{(0)} = 0$, i.e., the
divergence of the displacement vector  vanishes in the liquid.
But this is all that matters, because it is only this quantity that
enters the description of the liquid. 

The reason for $u_{ss}^{(0)}\ne 0$ in the liquid is that the phase-field description
imposes {\em coherence} of the strain across the interface, which ultimately goes
back to our viewing the liquid as a (shear-free, but nonetheless)
solid.  For a true liquid in contact with a solid there is no such coherence condition
as it is free to slip on the solid surface. Therefore, it could
 always keep its 
strain tensor isotropic (if such a notion made much sense at all for a liquid). 
The reason why we can nevertheless {\em model } the liquid
as a solid is the additional
 degree of freedom that arises in the description of
a liquid by having two fields $u_x$ and $u_z$ at our disposal even though only their
combination $\nabla {\bf u}$ enters the free-energy expression. Therefore, we can
compensate, so to speak, for imposing (nonphysical) coherence by allowing (equally
nonphysical) anisotropic strain in the liquid. 

At this point we may also note that
 if we were to model the elastic properties of two
real solids by the current phase-field approach, we would
necessarily impose coherence at the interface. The treatment of noncoherent solid-solid
interfaces via phase fields would require some rethinking of the method.

Concerning computational purposes, one disadvantage of the chosen reference
state is that it is not very well-suited for the use of 
periodic boundary conditions, meaning periodically varying stresses and strains.
 The field equations are
 set up in terms of the displacements which acquire linearly increasing or decreasing
components in  directions where the strain has a nonzero average. 
This observation motivates the consideration
of a different reference state in the solid, in which the average
strain due to the external stress is subtracted. This is the MG choice (see
App.~\ref{APPMGMAPP}).
%\cite{footnote4}.
If we impose a constant stress $\sigma_0$ in the $x$ direction, the stress tensor
in the solid is $\sigma_{ij} = -p_0 \delta_{ij} + \sigma_0 \delta_{i1} \delta_{1j}$
and requiring $u_{xx}=0$, we find that this is achieved by setting 
\begin{equation}
\Delta p = \frac{2\mu+\lambda}{2\mu} \sigma_0 \>. \label{delpsig}
\end{equation}
The corresponding homogeneous strain tensor is given by
$u_{xx}=0$, $u_{xz}=0$, and $u_{zz}=-\Delta p/(2\mu+\lambda)$.
We then obtain
% Instead of modeling displacements
%faithfully in the solid and foregoing their exact representation in the 
%liquid we may also do the reverse, requiring $u_{ss}^{(0)}= u_{rr}^{(0)} = 0$
%in the liquid. This immediately implies  $u_{ss}^{(0)}=0$ in the solid.
%Hence, the reference pressure $p_{0s}$ must be chosen differently from
%$p_{0\ell}$ in order to describe a nonhydrostatically strained solid.
taking the limit $\rho\to -\infty$ of (\ref{urhorhorhodep})
\begin{equation}
u_{rr}^{(0)} = \frac{-\Delta p-\lambda u_{ss}^{(0)} }{2\mu+\lambda} \>.
\end{equation}
This can be used to express
\begin{equation}
\sigma_{ss}^{(0)} -\sigma_{rr}^{(0)} = 2\mu (u_{ss}^{(0)} - u_{rr}^{(0)})
= \frac{2\mu}{2\mu+\lambda} \left[2(\mu+\lambda)  u_{ss}^{(0)} + \Delta p\right]\>,
\end{equation}
wherefrom we obtain  $2(\mu+\lambda)  u_{ss}^{(0)} + \Delta p = (2\mu+\lambda)
(\sigma_{tt}^{(0)}- \sigma_{nn}^{(0)})/2\mu$, which on insertion in (\ref{finres1})
leads back to (\ref{finres2}).

%With this choice of reference state, a homogeneous isotropic strain   
%has effectively been subtracted from the strains in the solid, so that
%the initial state of linear homogeneous displacement in the solid now
%corresponds to effective displacement zero, which allows the simple
%implementation of periodic boundary conditions.

Note that even here we cannot require $u_{ss}^{(0)}= u_{rr}^{(0)} = 0$
in the liquid, which would imply $u_{ss}^{(0)}=0$ at the interface and 
thus, according to 
(\ref{urhorhorhodep}), $ u_{rr}^{(0)} = -\Delta p/(2(\mu+\lambda))$, i.e.,  $ u_{rr}^{(0)}$
would be constant along the interface. Then also $\sigma_{tt}^{(0)}- \sigma_{nn}^{(0)}$
would have to be constant, which would lead to a dynamics entirely different from
that of the Grinfeld instability [where $(\sigma_{tt}^{(0)}- \sigma_{nn}^{(0)})^2$
increases in the grooves and diminishes on the peaks]. Hence, once again we are obliged to
make use of the additional degree of freedom of the fields inside the liquid, even though
now we can impose $u_{xx}^{(0)}= u_{zz}^{(0)} = 0$ in the liquid as an initial condition
(and as a far-field boundary condition),
because this satisfies the periodicity requirement.

%%%%%%%%%%%%%%%%%%%%%%%%
\section{Mapping of the M\"uller-Grant model to the present formulation \label{APPMGMAPP}}
\newcommand{\gtil}{\tilde{g}}
\newcommand{\htil}{\tilde{h}}
\newcommand{\mutil}{\tilde{\mu}}
\newcommand{\Gamtil}{\tilde{\Gamma}}

The main 
difference between the form of the MG model given in \cite{Mueller99}
and the one given here is a different choice of the functions $g(\phi)$ and 
$h(\phi)$. To make this conspicuous, we will rename their original 
functions to $\gtil(\phi)$ and  $\htil(\phi)$ (since the second function was
called  $g(\phi)$ in \cite{Mueller99},  our renaming is also useful
to avoid unnecessary confusion here).
 We shall leave the gravity and shift terms,
$f_{{\rm grav}}$ and $f_{{\rm c}}$ out of the consideration, since they
were not used by MG.

Their double well potential is defined as
\begin{equation}
f_{{\rm dw}}(\phi) = \frac1a \gtil(\phi) \>, \label{dwMG}
\end{equation}
with $\gtil(\phi) = \phi^2 (1-\phi^2)^2$, which is a sixth-order polynomial
and actually has a third minimum at $\phi=-1$. The latter does not, however,
play any role in the dynamics, provided no negative $\phi$ values are
given in the initial condition. $a$ is a constant to be identified
via the sharp-interface limit.  

Second, there is an elastic contribution to the free energy which they give
as
%\begin{equation}
%f_{{\rm el}}(\phi,\{u_{ij}\}) = \frac{1}{2} K (u_{ii}-u_{ii}^{eq})^2 
% + \mutil \sum_{ij} \left( u_{ij} - u_{ij}^{eq} - \frac{\delta_{ij}}{d} (u_{ii}
%-u_{ii}^{eq})\right)^2\>,  \label{elMG}
%\end{equation}
\begin{equation}
f_{{\rm el}}(\phi,\{u_{ij}\}) = \frac{1}{2} K  (\nabla \cdot {\bf u})^2 
 + \mutil \sum_{ij} \left( u_{ij}
 - \frac{\delta_{ij}}{d}  \nabla \cdot {\bf u}\right)^2\>,  \label{elMG}
\end{equation}
where $K$ is the bulk modulus and $\mutil$ the shear modulus which is
$\phi$ dependent:
\begin{equation}
\mutil = \mu_1 \htil(\phi).
\end{equation}
The convenient choice 
\begin{equation}
\htil(\phi) = \frac{1}{2}\phi^2 - \frac{1}{4}\phi^4,
\end{equation}
guarantees that both bulk phases keep
their equilibrium values at $\phi = 0$ (liquid) and $\phi =  1$ (solid).
This is due to the fact that $\htil'(0)=\htil'(1)=0$, a property $\htil(\phi)$
shares with $h(\phi)$ from the KM model (see Sec.~\ref{secphase}). 
Obviously, the true shear modulus
of the {\em solid} is $\mu=\mu_1\htil(1)=\mu_1/4$.

For simplicity and since it does not change the behavior qualitatively, 
the bulk modulus is assumed to be the same in both phases.
However, this restriction can be easily dropped by replacing $K$ with
\begin{equation}
K = K_0 + K_1 \htil(\phi).
\end{equation}
As reference state they chose a prestressed state of the solid with 
$\sigma_{xx} = \sigma_0$, in which the
strains $u_{xx}$ and $u_{xz}$ for a flat surface vanish.
This entails that the state in which {\em all} strains vanish is
a hydrostatic state with a different stress value. It is described
by MG using a parameter $\eta_0$ and [as may be verified easily 
from Eq.~(\ref{stressMG}) below], in this state we have  
$\sigma_{xx}=\sigma_{zz}=\eta_0 \htil(1) = \eta_0/4$. As a result,
there is a relation between the new parameter $\eta_0$ and the
external stress $\sigma_0$ in the uniaxially stressed reference state
\begin{equation}
\eta_0 = \frac{8 (K+\mu_1/4)}{\mu_1}\, \sigma_0\>. \label{etaexpr}
\end{equation} 
    
The free energy density is then given as the
sum of $f_{{\rm dw}}(\phi)$ and $f_{{\rm el}}(\phi,\{u_{ij}\})$ and
 two additional terms $(\eta_0^2/2K)\htil(\phi)^2$ 
and $\eta_0 \htil(\phi) \nabla \cdot {\bf u}$. The first of these terms
describes an energy shift, the second an additional coupling between
the phase field and the elastic field (beyond that already implied by
the $\phi$ dependence of the shear modulus and, possibly, the 
bulk modulus).
A nice feature of the approach given in the present paper is that these terms
are {\em automatically generated} by  accounting for the 
fact that
the equilibrium state does not have vanishing strain 
when the MG reference state is used: these are the terms containing
$p_0-p_{0s}$ in (\ref{hooke3}) and  the corresponding terms $\Delta p\, u_{ii}$
and $\Delta W$ in (\ref{phasebas}).
 
The free energy density is then given by:
\begin{equation}
f(\phi,u_{ij}) = \frac1a \gtil(\phi) + \frac{\eta_0^2}{2 K} \htil(\phi)^2
                + \eta_0\ \htil(\phi) \nabla \cdot {\bf u} 
                + \frac{1}{2} K (\nabla \cdot {\bf u})^2 
 + \mutil \sum_{ij} \left( u_{ij} - \frac{\delta_{ij}}{d} \nabla \cdot {\bf u}
\right)^2 \>. \label{ftotMG}
\end{equation}
The first term is the double well potential.
% The second is similar to $f_{\rm c}$ and shifts the energy at which, for 
%constant elastic coefficients, solid and liquid are at coexsistence.
%Nevertheless, Eq.~(\ref{ftotMG}) corresponds to the case
%$\sigma_{00}=0$ and $g=0$ of (\ref{phasebas}), as we shall see below.
The  second  and third terms are
 due to the particular choice of reference frame, and 
$\eta_0$ is related to the externally applied stress as described by
Eq.~(\ref{etaexpr}).
 Applying the same line of reasoning as in Sec.~\ref{secphase},
 they obtain a system of coupled partial differential
equations:
\begin{eqnarray}
\frac{\partial \phi}{\partial t} & = & - \Gamtil \biggl[\frac1a \gtil'(\phi) 
 - l^2 \nabla^2 \phi \nonumber \\
& + & \left. \frac{\eta_0^2}{K} \htil(\phi) \htil'(\phi) 
 + \eta_0 \htil'(\phi)  \nabla \cdot {\bf u} 
 + \mu_1 \htil'(\phi) 
  \sum_{ij}  \left( u_{ij} - \frac{\delta_{ij}}{d} \nabla \cdot {\bf u} \right)^2 \right]
   \, ,
\label{phitMG}
\end{eqnarray}
and
\begin{equation}
\frac{\partial \sigma_{ij}}{\partial x_j} = \frac{\partial}{\partial x_i} 
  [\eta_0 \htil(\phi) + K \nabla \cdot {\bf u}] 
  + 2 \mu_1 \frac{\partial}{\partial x_j} \left[ \htil(\phi) \left( u_{ij}
  - \frac{\delta_{ij}}{d} \nabla \cdot {\bf u} \right) \right] = 0 \, .
\label{stressMG}
\end{equation}
They show in \cite{Mueller2000} that the
phase field equations of this model also converge to the sharp interface 
equations. By expanding the solution of the 
mechanical equilibrium condition to
first order in the shear modulus they were able to 
 integrate out the elastic
fields, so that they were left with an equation for $\phi$ only. That allowed
them to use a pseudospectral method with which they could study wide periodic
systems and three dimensional systems \cite{Mueller99}.

Whether this expansion is entirely consistent is an open question, since
they consider $\eta_0$ an independent parameter that is $O(1)$,
whereas for fixed $\sigma_0$ we actually have  $\eta_0=O(1/\mu_1)$, and
$\mu_1$ is the small quantity in their expansion. Probably this does not
really matter in the absence of gravity where the precise value of 
$\eta_0$ is immaterial as it only sets the time scale. The problem
may be more awkward in the presence of gravity. 

Equations (\ref{phitMG}) and (\ref{stressMG}) are in a form that allows
direct comparison with Eqs.~(\ref{phasebas}) and (\ref{stressbas}), 
respectively. First note that in two dimensions
$K=\lambda+\mu=\lambda+\mu_1/4$ and that we must set $\lamtil=K$ to
have the same elastic constants in the two sets of equations.
In 2D, we have $[ u_{ij} - (\delta_{ij}/{d}) \nabla \cdot {\bf u}]^2
= u_{ij} u_{ij} - \frac12 u_{ii}^2$ (summations over $i$ and $j$ are implied)
and because of $\lambda-\lamtil=-\mu$, we obtain
\begin{equation}
\mu  u_{ij} u_{ij} + \frac{\lambda-\lamtil}{2} u_{ii}^2
 = \frac14 \mu_1 \left( u_{ij} - \frac{\delta_{ij}}{d} \nabla \cdot {\bf u} \right)^2 \>,
\end{equation}
which shows that on replacing $\htil(\phi)$ with $\frac14 h(\phi)$ the
 term of  (\ref{phitMG}) that is quadratic in the strains 
becomes equal to the corresponding term of (\ref{phasebas}).
 Common prefactors will be discussed below. We then see immediately, that the
choice $\Delta p = \eta_0/4$ will make the linear terms equal. This
choice is the right one as is revealed by 
a quick comparison of Eq.~(\ref{etaexpr})
with Eq.~(\ref{delpsig}),
derived in  appendix \ref{APPSHARPINT}. 
Next we have to compare the constant terms
which are $\eta_0^2/K$ and $\Delta W$. Here we notice that in \cite{Mueller99},
it was assumed that $p_{0\ell}=p_0=0$. If we furthermore set $\sigma_{00}=0$,
then we infer from (\ref{delw}) that $\Delta W = \Delta p^2/2K = \eta_0^2/32K$.
Now the $\phi$-dependent factor of  $\eta_0^2/K$ in (\ref{phitMG}) is
$\htil(\phi) \htil'(\phi) = \frac12 \partial [\htil(\phi)^2]/\partial \phi$.
We have checked by directly performing the sharp-interface limit of
the original MG model, that this limit does not change when $\htil(\phi)^2$
is replaced with $\frac14 \htil(\phi)$ (the solid and liquid phase limits
are obviously unchanged). Doing this replacement first and then substituting
$\frac14 h(\phi)$ for $\htil(\phi)$, we get identity of the constant terms, 
too. 

Finally, the prefactors should be discussed. In order to make the 
prefactor of the elastic expressions  the same in both equations, we must set
$\Gamtil=1/(3 \ktil \epsilon)$, which shows that $l^2=3\gamma\epsilon$.
To determine the factor $1/a$ of the double well potential in (\ref{phitMG}),
 one must actually perform the sharp-interface limit 
[because the potential is not 
the same as in (\ref{phasebas})], which yields $a=3\epsilon/8\gamma$.
With these choices, Eqs.~(\ref{phitMG}) and  (\ref{phasebas}) are 
asymptotically equivalent.

The comparison of the equations describing mechanical equilibrium is
even more straightforward. Inserting the expressions (\ref{hooke1}) and
(\ref{hooke2}) with $p_{0\ell}=0$ and $p_{0s}=-\Delta p = -\eta_0/4$
into (\ref{stressbas}) and using $\lamtil=K$, we get
\begin{eqnarray}
0 &=& \frac{\partial}{\partial x_j} \left\{h(\phi)\left[\frac{\eta_0}{4}\delta_{ij} 
+K u_{kk} \delta_{ij} + 2 \mu \left(u_{ij}-\frac12 u_{kk} \delta_{ij}\right)\right]
+ [1-h(\phi)] K  u_{kk} \delta_{ij}\right\} \nonumber \\
 &=&  \frac{\partial}{\partial x_i} \left[\frac{\eta_0}{4} h(\phi) + K u_{kk}\right] +  2 \mu  \frac{\partial}{\partial x_j} \left[h(\phi)\left(u_{ij}-\frac12 u_{kk} \delta_{ij}\right)\right] \>, 
\end{eqnarray}
which is obviously identical to (\ref{stressMG}), once we replace
 $h(\phi)/4$ by $\htil(\phi)$.

%%%%%%%%%%%%%%%%
\newcommand{\imag}{i}
\section{Analytic solution of the elastic problem for the double cycloid 
\label{APPCONFMAPP}}

In this appendix (only!), we switch back from our notation for geometric 
relations in the plane which was based on a coordinate system spanned by
the $x$ and $z$ axes (thus reminding ourselves that in reality we have
a three-dimensional system and we simply suppress deformations in 
the $y$ direction) to the more conventional use of $x$ and $y$ for the
planar coordinates. This way we ``liberate'' the symbol $z$ for use
as a complex variable: $z=x+\imag y$.

We wish to solve the elastic problem in a half-infinite geometry,
 the top of which is bounded by the double cycloid given by  
Eq.~(\ref{doubcyc1}) with $z$ replaced by $y$. In the complex plane,
this curve is described by
\begin{equation}
z = x+\imag y = \xi - \imag A e^{-\imag k \xi} - \imag B e^{-2 \imag k \xi}
\label{doubcyc2}
\end{equation}
and because the parameter $\xi$ is real, this equation also defines 
a conformal mapping frome the $z$ plane to the $\zeta$ plane, where
$\zeta = \xi + \imag \eta$, mapping the curve $z(x)$ 
to the $\xi$ axis. 
%Hence our solid is mapped to the lower half plane $\eta <0$
%in the  $\zeta$ plane.

\newcommand{\zbar}{{\bar{z}}}
To rephrase the elastic problem in the complex plane, we use the 
Goursat function formalism. Its basic statements can be easily inferred
from the representation of two-dimensional elasticity in terms of a 
single scalar function, the Airy function $\chi(x,y)$.
Setting $\sigma_{xx}=\partial_y^2 \chi$,  $\sigma_{yy}=\partial_x^2 \chi$,
and $\sigma_{xy}=-\partial_x \partial_y\chi$, the mechanical equilibrium
equations $\partial_j \sigma_{ij}=0$ are automatically satisfied.
Hooke's law for isotropic bodies then implies that $\chi$ obeys the
biharmonic equation:
\begin{equation} 
\nabla^4 \chi = 0. 
\label{biharmonic}
\end{equation}
In terms of the complex
variables $z$ and $\zbar=x-\imag y$, the Laplacian becomes
$\nabla^2 = 4 \partial_\zbar \partial_z$ [because $\partial_x =  \partial_z
+ \partial_\zbar$, $\partial_y = \imag (\partial_z-\partial_\zbar)$],
hence the most general form of the solution to (\ref{biharmonic}) is
given by
\begin{equation}
\chi(x,y) \equiv \tilde{\chi}(z,\zbar) = \zbar f_1(z) + g_1(z) + z f_2(\zbar) +
g_2(\zbar) \>,
\end{equation}
where  $f_j$, $g_j$ ($j=1,2$) are analytic functions of their arguments.
Since we are looking for real solutions, we can restrict ourselves to 
two independent complex functions instead of four, i.e., we can write
\begin{equation} 
 \tilde{\chi}(z,\zbar) = \frac12 \left\{ \zbar\, \phi(z) + \int \psi(z) dz
+ z\, \overline{\phi(z)} + \int \overline{\psi(z)} d\zbar \right\} \>,
\label{goursatfunctions}
\end{equation}
where an overbar denotes complex conjugation. In this formula, $\phi$ and 
$\psi$ are the Goursat functions.

{}From (\ref{goursatfunctions}), we obtain by direct differentiation
\begin{eqnarray}
\sigma_{xx}+\sigma_{yy} = \nabla^2 \tilde{\chi} & = & 2 \left[\phi'(z)
+\overline{\phi'(z)}\right] \>, \label{tracesig} \\
\sigma_{yy}-\sigma_{xx}+2\imag \sigma_{xy}  & = & 2  \left[\zbar \phi''(z)+
\psi'(z)\right] \>. \label{otherstress}
\end{eqnarray}
The displacements can also be expressed by the Goursat functions
[using Hooke's law and (\ref{tracesig},\ref{otherstress})], but since
we do not need the corresponding relation, we omit it here.

We have boundary conditions for the stresses on the curve given by
Eq.~(\ref{doubcyc2}), which one could attempt to use directly in 
 (\ref{tracesig},\ref{otherstress}) to obtain equations for the Goursat
functions. However, since we are going to employ conformal mapping, it is
useful to keep the order of derivatives small -- the analytic evaluation
of higher-order derivatives can be quite cumbersome. 
Thus it is desirable to reformulate the boundary conditions in partially
integrated form. The force $(f_x,f_y)$ on the boundary can be condensed
into a single complex number
\begin{equation}
f_x+\imag f_y = \sigma_{xj} n_j +\imag \sigma_{yj} n_j 
              = (\sigma_{xx}+\imag \sigma_{xy}) n_x 
                 +(\sigma_{yy}-\imag \sigma_{xy}) \imag n_y \>,
\label{deffxfy} 
\end{equation}
where $(n_x,n_y)$ is the normal vector to the boundary. Introducing the
arclength $s$, we have (directing $s$ such that  $s\to \infty$ corresponds
to $x\to\infty$)
\begin{equation}
n_x+\imag n_y = -\frac{dy}{ds} + \imag \frac{dx}{ds} = \imag \frac{dz}{ds}
\label{normvec}
\end{equation}
and thus
\begin{eqnarray}
f_x+\imag f_y 
    &=& \frac12 \left(\sigma_{xx}+\sigma_{yy}\right) \imag  \frac{dz}{ds}
     +  \frac12 \left(\sigma_{yy}-\sigma_{xx}-2\imag \sigma_{xy}\right) \imag \frac{d\zbar}{ds} \nonumber \\
&=&  \left[\phi'(z)+\overline{\phi'(z)}\right] \imag  \frac{dz}{ds}
     +  \left[z\overline{\phi''(z)}+\overline{\psi'(z)}\right]\imag \frac{d\zbar}{ds}  \nonumber \\
&=& \imag \frac{d}{ds} \left\{\phi(z) + z \overline{\phi'(z)} + \overline{\psi(z)}\right\} \>.
\label{exprfxfy}
\end{eqnarray}
Integrating this local relation along the boundary, we obtain
\begin{equation}
-\imag f \equiv -\imag \int \left(f_x+\imag f_y\right) ds = \phi(z) + z \overline{\phi'(z)} + \overline{\psi(z)} \>,
\end{equation}
which allows to apply the boundary condition in expressions involving 
first-order derivatives only.

We  substract the stress at infinity from our (linear) 
elastic equations to be able to work with analytic functions that are bounded
at infinity. Hence, we set 
\begin{equation}
\sigma_{ij} = \sigma_{ij}^{(0)} + \sigma_0 \delta_{ix}  \delta_{jx} 
\end{equation}
and replace $\sigma_{ij}$ with $\sigma_{ij}^{(0)}$ in
 Eqs.~(\ref{tracesig},\ref{otherstress}) above.
 Taking the equilibrium pressure in the liquid equal to zero (which we can
do without loss of generality), the boundary conditions at the
liquid-solid interface ($\sigma_{ij} n_j=0$) become
\begin{equation}
\sigma_{ij}^{(0)} n_j = -\sigma_0  \delta_{ix} n_x \>,
\end{equation}
translating into
\begin{equation}
f_x+\imag f_y = \sigma_0 \frac{dy}{ds} \quad \Rightarrow \quad -if = \frac12 (\zbar-z) \sigma_0 \>,
\end{equation}
where we have dropped an arbitrary constant of integration.

Before embarking on the actual calculation, we ought to ponder one more point.
We would like to solve a problem with periodic boundary conditions for
strains and stresses in the $x$ direction. But periodicity of the strains does 
not imply periodicity of the displacements nor does it imply periodicity
of the Goursat functions. On the other hand, the use of periodic functions
greatly facilitates the derivation.
As noted by Spencer and Meiron \cite{Spencer94},
we can express the Goursat functions by periodic functions $\phi_0$
and $\psi_0$ via 
\begin{eqnarray}
\phi(z) &=& \phi_0(z) \>, \nonumber\\
\psi(z) &=& \psi_0(z) - z \phi_0'(z) \>.
\label{pergoursat}
\end{eqnarray}

The mathematical problem is then  to find two periodic 
functions $\phi_0$ and $\psi_0$, analytic in the domain occupied by the
solid, satisfying
\begin{equation}
\phi_0(z) + (z-\zbar) \overline{\phi_0'(z)} + \overline{\psi_0(z)} 
= - \frac12 (z-\zbar) \sigma_0 
\label{interface1}
\end{equation}
at the interface and remaining bounded for $y\to -\infty$.
%\begin{equation}
%\phi_0(z) \to 0\>, \quad\quad \psi_0(z) \to 0 \quad\quad \mbox{for}\quad y\to -\infty\>.
%\end{equation}
%Since we have already disposed of all arbitrary constants, 
The solution to
this problem must be unique apart from 
possible additive constants to the functions $\phi_0(z)$ and  $\psi_0(z)$.

We transform to the $\zeta$ plane, using the analytic continuation of
(\ref{doubcyc2})
\begin{equation}
z  =  \zeta - \imag A e^{-\imag k \zeta} - \imag B e^{-2 \imag k \zeta}
\equiv \omega(\zeta) \>. \label{anadoubcyc} 
\end{equation}
This maps the interface to the real axis and the solid to the half plane
$\eta <0$. To designate functions in the $\zeta$ plane, we put a tilde
on the letter they have in the $z$ plane:
\begin{eqnarray}
\phi_0(z) &=& \phi_0(\omega(\zeta)) \equiv \tilde{\phi}_0(\zeta) \>, \nonumber \\
\psi_0(z) &=& \tilde{\psi}_0(\zeta) \>. \label{transforms}
\end{eqnarray}
The derivative of $\phi_0$ transforms as follows:
\begin{equation}
\phi_0'(z) = \tilde{\phi}_0'(\zeta)\frac{d\zeta}{dz} = \frac{\tilde{\phi}_0'(\zeta)}{\omega'(\zeta)} \>.
\end{equation}

Our task therefore is to construct two analytic functions satisfying
\begin{equation}
 \tilde{\phi}_0(\xi) 
+ [\omega(\xi)-\bar{\omega}(\xi)] \frac{\overline{\tilde{\phi}_0'}(\xi)}{\overline{\omega'}(\xi)} 
+ \overline{\tilde{\psi}}_0(\xi) = -\frac{\sigma_0}{2}[\omega(\xi)-\bar{\omega}(\xi)] 
\label{interface2}
\end{equation}
on the real axis ($\eta=0$) and remaining bounded %approaching zero 
as $\eta\to-\infty$.
$\omega(\xi)$ is given by Eq.~(\ref{anadoubcyc}), hence
\begin{equation}
\omega'(\xi) =  1 -  A k e^{-\imag k \xi} -  2 B k e^{-2 \imag k \xi}
\label{derivom}
\end{equation}
and Eq.~(\ref{interface2}) becomes ($A$ and $B$ are real)
\begin{eqnarray}
&& \tilde{\phi}_0(\xi) 
-\imag \left(A  e^{-\imag k \xi} + B e^{-2 \imag k \xi} + A  e^{\imag k \xi} 
+ B  e^{2 \imag k \xi}\right) 
\frac{\overline{\tilde{\phi}_0'}(\xi)}{1 - A k e^{\imag k \xi} - 2 B k e^{2 \imag k \xi}} + \overline{\tilde{\psi}_0}(\xi) \nonumber \\
&& \quad\quad = \imag\frac{\sigma_0}{2}\left(A  e^{-\imag k \xi} + B e^{-2 \imag k \xi} + A  e^{\imag k \xi} 
+ B  e^{2 \imag k \xi}\right) \>.
\label{interface3}
\end{eqnarray}
Note that $\overline{\tilde{\phi}_0'}(\xi)$ and 
$\overline{\tilde{\psi}_0}(\xi)$
are functions that should be analytically continued to the {\em upper}
half plane, for if  $\overline{\tilde{\phi}_0'}(\bar\zeta)$ 
[$\overline{\tilde{\psi}_0}(\bar\zeta)$] is analytic in the upper half 
plane, $\tilde{\phi}_0'(\zeta)$ [$\tilde{\psi}_0(\zeta)$] is analytic
in the lower half plane which is what we need. The basic idea in
constructing the solution is to divide the terms in (\ref{interface3})
into two groups, one of which corresponds to functions analytic
in the upper half plane, the other to those analytic in the lower half plane.
The equality between these two groups implies that each of them is
equal to a constant, which gives us two equations for the two 
functions sought. It is clear that $ \tilde{\phi}_0(\xi) $
belongs to the terms of (\ref{interface3}) that are analytic in the 
lower half plane (on replacement of $\xi$ by $\zeta$) whereas 
$ \overline{\tilde{\psi}_0}(\xi)$ has to be analytic in the upper half plane.
The difficult term is the middle one on the left hand side as it contains
some
expressions that are analytic and bounded in the upper half plane 
(e.g., $e^{\imag k \xi}$) but also some for which this is the case in the lower
half plane (e.g., $e^{-\imag k \xi}$). One way to proceed is to 
expand both $\tilde\phi_0$ and $\tilde\psi_0$ 
in a series in powers of  $e^{-\imag k \xi}$
(a one-sided Fourier series, so to speak), to multiply
 Eq.~(\ref{interface3}) by the denominator of the middle expression,
 and to separate terms with
plus signs and minus signs in the exponents. This gives a two-termed
recursion for  $\tilde\phi_0$ containing two constants that have to
be determined from the analyticity properties. As it turns out, the
series for  $\tilde\phi_0$ is finite, only the two first terms are nonzero.
This suggests that a close look at Eq.~(\ref{interface3}) would have
revealed this property, allowing to avoid the tedious expansion procedure.
Indeed, there is a more  elegant way leading to this result. Its 
discovery is left as an exercise to the astute reader. Here, we simply take
\begin{equation}
\tilde{\phi}_0(\xi) = a_1 e^{-\imag k \xi} + a_2 e^{-2 \imag k \xi}
\end{equation}
as an ansatz. Inserting this into (\ref{interface3}), we get
\begin{eqnarray}
&& a_1 e^{-\imag k \xi} + a_2 e^{-2 \imag k \xi}
+ \left(A  e^{-\imag k \xi} + B e^{-2 \imag k \xi} + A  e^{\imag k \xi} 
+ B  e^{2 \imag k \xi}\right) 
\frac{\bar{a}_1 k e^{\imag k \xi} + 2 \bar{a}_2 k e^{2 \imag k \xi}}{1 - A k e^{\imag k \xi} - 2 B k e^{2 \imag k \xi}} \nonumber \\
&& \quad\quad - \imag\frac{\sigma_0}{2}\left(A  e^{-\imag k \xi} + B e^{-2 \imag k \xi} + A  e^{\imag k \xi} 
+ B  e^{2 \imag k \xi}\right) = - \overline{\tilde{\psi}_0}(\xi)\>.
\label{interface4}
\end{eqnarray}
Since we have just $\tilde{\psi}_0(\xi)$ on the right-hand side, which must
be analytically continued to the upper half plane and remain bounded there,
the left-hand side must not contain, after substitution of $\zeta$
for $\xi$, any ``dangerous'' terms diverging as
$\eta\to\infty$. %on replacing $\xi$ with $\zeta$. 
Dangerous terms would obviously be 
 the terms $e^{-\imag k \zeta}$ and $e^{-2 \imag k \zeta}$ as well
as the zeros  of the denominator. Now, we have the condition $Ak +2 Bk <1$
and we have $\vert e^{\imag k \zeta}\vert \le 1$ for $\eta\ge 0$. Therefore,
the denominator is always different from zero in the upper half plane.
All we have to do then is to choose $a_1$ and $a_2$ such that the dangerous
terms cancel. This is straightforward for $a_2$, since there are only two
terms containing $e^{-2 \imag k \zeta}$, after the prefactor of the second
term has been multiplied with the numerator. There remains a term
proportional to $e^{-\imag k \zeta}$, however, in the numerator, which 
must also be canceled by the choice of $a_1$. The result is
\begin{eqnarray}
a_1 &=& \frac{\sigma_0}{2}\,\imag\,\frac{A}{1-Bk} \>,\nonumber \\
a_2 &=& \frac{\sigma_0}{2}\, \imag\, B \>.
\label{coeffsphi}
\end{eqnarray}

{}From this, we immediately get $\tilde\phi_0$ as
\begin{equation}
\tilde\phi_0(\zeta) = \frac{\sigma_0}{2} \,\imag \left\{ \frac{A}{1-Bk} e^{-\imag k \zeta} + B e^{-2 \imag k \zeta}\right\} \>.
\label{solphi}
\end{equation}
Once  $\tilde\phi_0$ is known, $\tilde\psi_0$ is obtained from 
(\ref{interface3}) or, using (\ref{coeffsphi}), from (\ref{interface4}).
We give the result for reference purposes, even though it will not 
be needed in the following
\begin{eqnarray}
\tilde\psi_0(\zeta) &=& - \frac{\sigma_0}{2} \imag \frac{1}{1-A k e^{-\imag k \zeta} -2 B k e^{-2 \imag k \zeta}}
\Biggl\{ k\left(2 B^2 + A^2 \frac{1 + B k}{1 - B k}\right)
+ \frac{2 A B k}{1 - B k}  e^{-\imag k \zeta} \nonumber\\
&&\mbox + \left(A e^{-\imag k \zeta} + B e^{-2 \imag k \zeta}\right) 
\left(1+\frac{A B k^2}{1 - B k}e^{-\imag k \zeta}\right)\Biggr\} \>.
\label{solpsi} 
\end{eqnarray}
Note that only exponentials of negative multiples of $\imag k \zeta$
appear; the  numerators  are thus analytic and bounded in the lower half-plane
and  the denominator remains nonzero 
because of $Ak+2Bk<1$, hence $\tilde\psi_0(\zeta)$
satisfies all the required analyticity and boundedness conditions.
In the case $A=0$ or $B=0$, these results reduce to those of 
Gao {\em et al.} \cite{Chiu93} for the simple cycloid (after transforming
back to the nonperiodic Goursat functions).

We need only $\tilde\phi_0$ to compute the tangential
 stresses on the boundary.
Since we know the normal stresses to be equal
to zero from the boundary condition, we can write $\sigma_{tt} =
\mbox{tr} \sigma = \sigma_{xx}^{(0)} + \sigma_{yy}^{(0)} + \sigma_0$.
We have 
\begin{equation}
\sigma_{xx}^{(0)} + \sigma_{yy}^{(0)} = 2 \left[\frac{\tilde\phi_0'(\zeta)}{\omega'(\zeta)} + \frac{\overline{\tilde\phi_0'(\zeta)}}{\overline{\omega'(\zeta)}}\right] \>,
\label{tracesigma0}
\end{equation}
which, when specialized to the boundary, gives us
\begin{equation}
\sigma_{tt} = \frac{\sigma_0}{2} 
\left\{\frac{1+\frac{1+Bk}{1-Bk} Ak e^{-\imag k \xi} + 2Bk   e^{-2 \imag k \xi}}{1- Ak e^{-\imag k \xi} - 2Bk   e^{-2 \imag k \xi}} + c.c.\right\}
\>\label{sigttcyclo}
\end{equation}
In the grooves of the pattern, we have $k\xi=m\pi$ hence $e^{-\imag k \xi} = (-1)^m$, $e^{-2\imag k \xi}=1$, from which we obtain the tangential stress
as
\begin{equation}
\sigma_{tt} = \sigma_0
\frac{1+\frac{1+Bk}{1-Bk} Ak (-1)^m + 2Bk}{1- Ak (-1)^m - 2Bk}
\>.
\label{sigttcyclomin}
\end{equation}

To obtain the normal velocity in the grooves, we need 
the curvature there. The curvature is easily calculated from the parametric
representation of the double cycloid
\begin{eqnarray}
\kappa &=& - \frac{x'(\xi) y''(\xi) - y'(\xi) x''(\xi)}{\left(x'(\xi)^2+y'(\xi)^2\right)^{3/2}} \nonumber\\
 &=& \frac{k^3 \left[A^2+8B^2+6AB \cos k\xi\right] 
- k^2 \left[A \cos k\xi + 4B  \cos 2k\xi\right]}{\left(A^2 k^2+4 B^2 k^2
 + 4 A B k^2 \cos k\xi + 1- 2 A k \cos k\xi - 4B k \cos 2k\xi\right)^{3/2}}
\label{curvcyclo}
\end{eqnarray}
and its value in the bottom of the grooves is ($\cos k\xi = \pm 1$)
\begin{equation}
\kappa = -\frac{k^2\left[ A (-1)^m + 4B\right]}{\left[1- A k (-1)^m - 2 B k\right]^2} \>.
\label{curvcyclomin}
\end{equation}
Note that as a cusp is approached ($A k + 2 B k \to 1$), $\sigma_{tt}^2$ and
the curvature diverge with the same denominator,  for even $m$.

Inserting (\ref{sigttcyclomin}) and (\ref{curvcyclomin}) into
Eq.~(\ref{vntil}) for the nondimensional normal velocity and neglecting the
gravity term we obtain
\begin{equation}
\tilde{v}_n = - \frac{1}{2\left[1- A k (-1)^m - 2 B k\right]^2}
\left\{\left(1+ 2Bk +\frac{1+Bk}{1-Bk} Ak (-1)^m\right)^2 - 2\ell_1 k^2 \left[A(-1)^m +4 B\right]
\right\} \>,
\label{vntilcycl}
\end{equation}  
where we have nondimensionalized the curvature via multiplication by $\ell_1$.
Setting $\alpha=2\ell_1 k$, we generate Eq.~(\ref{vntilcyc}). Note that
if we consider the amplitude $A$ to be a small perturbation of a cycloid
determined by $B$, the basic wave number is $2k$, not $k$. Then $\alpha$
is just the ratio of the basic wave number and the wave number of the 
fastest-growing mode.

%%%%%%%%%%%%%%%%%%%%%%%%%%
%\bibliography{/home/phase4/misbah/bibgroup/sample}

\begin{thebibliography}{10}
\bibitem{footnote2}
These parameters are:
 $E=3 \times 10^8$ dynes/cm$^2$, $\nu = 0.3333$, $\gamma = 0.2$ dynes/cm,
$\rho_s = 0.193$ g/cm$^3$,  $\rho_\ell = 0.170$ g/cm$^3$, hence  
$\Delta \rho = 0.017$ g/cm$^3$.
The value for $k$ is estimated as $k=100$ cm/s.
With these parameters, the critical stress for appearance of
the Grinfeld instability is  $\sigma_{0c} = 2.48\times 10^{4}$ dynes/cm$^2$.
For the purpose of estimating $\ell_1$, we assumed a slightly supercritical
stress   $\sigma_{0} = 3\times 10^{4}$ dynes/cm$^2$ and we took $K\approx E$.
\bibitem{footnote6}
It should however be mentioned that the program at that time still contained an error,
resulting in an effective stress that was wrong by a factor of about 
10$^{1/4}\approx 1.78$.
Therefore, the quantitative descriptions of the figures given in
\protect\cite{Kassner99} are incorrect,
even though the qualitative statements remain valid. That we found  
approximately the right value for the instability threshold was
due to our taking an overly large value for the 
interface thickness $\epsilon$ which
compensated  the effect of the too large stresses.
\bibitem{footnote7}
The precise condition is $(1-2Bk)^2 (1-Bk)/((1+2Bk)^2+3Bk) < Ak$,
where $Ak < 1-2Bk$. The inequality given in the main text implies the
one stated here.
\bibitem{footnote3}Note that the actual orthogonal
coordinates are $r$ and a polar angle $\theta$, connected with
$s$ via the curvature of the interface: $d\theta/ds = \kappa$.
This difference becomes important only when taking second derivatives.
\end{thebibliography}
%\bibliographystyle{prsty}

\end{document}